\documentclass[11pt,a4paper]{article}
\pdfoutput=1
\usepackage{jheppub}

\RequirePackage[utf8]{inputenc}

\usepackage{amsfonts}
\usepackage{amssymb}
\usepackage{comment}
\usepackage{graphicx}
\usepackage{amsmath}
\usepackage{tabu}
\usepackage{dsfont}
\usepackage{float}
\usepackage{amsthm}
\usepackage{slashed}
\usepackage{setspace}
\usepackage[labelformat=simple]{subcaption}

\usepackage{pgfplots}
\usepackage{tikz}
\usepackage{tikz-cd}
\usetikzlibrary{shapes.misc,shadows,fit,decorations.pathmorphing,positioning,trees,decorations.markings,decorations.pathreplacing,calc,shapes,patterns,arrows,positioning,arrows.meta}
\usepackage{xcolor}
\usepackage{pgfplots}
\usepackage{pgfplotstable}
\usepackage{xcolor}
\usepackage[boxsize=0.5em,aligntableaux=center]{ytableau}
\usepackage{makecell}
\usepackage{diagbox}
\usepackage{environ}
\usepackage[utf8]{inputenc}
\usepackage{amsmath, amsthm, amssymb, amsfonts}
\usepackage{amsmath, amsthm, amssymb, amsfonts}
\usepackage{multirow}
\usepackage{graphicx}
\usepackage{float}
\usepackage[numbers,sort&compress]{natbib}
\usepackage{jheppub}
\usepackage{enumerate}
\usepackage{enumitem}
\usepackage[capitalize,sort]{cleveref}
\usepackage{hhline}
\usepackage{mathtools}
\usepackage{longtable}
\usepackage{alphalph}

\crefname{figure}{Figure}{Figures}
\crefname{table}{Table}{Tables}

\newcommand{\ee}{\mathrm{e}}
\newcommand{\kom}{\, ,\quad }

\def\ov{\overline}

\DeclareMathOperator{\SL}{SL}

\DeclareMathOperator{\GCD}{GCD}
\newcommand{\de}{\partial}
\newcommand{\CC}{\mathbb{C}}
\newcommand{\PP}{\mathbb{P}}

\newcommand{\ZZ}{\mathbb{Z}}

\newcommand{\pderr}[1]{\frac{\de}{\de #1}}

\newcommand{\coma}{\, , \quad}
\newcommand{\fstop}{\, .}

\renewcommand{\epsilon}{\varepsilon}

\makeatletter
\newsavebox{\measure@tikzpicture}
\NewEnviron{scaletikzpicturetowidth}[1]{%
  \def\tikz@width{#1}%
  \begin{lrbox}{\measure@tikzpicture}%
  \BODY
  \end{lrbox}%
  \pgfmathparse{#1/\wd\measure@tikzpicture}%
  \BODY
}
\makeatother

\definecolor{col1}{HTML}{f94144}
\definecolor{col2}{HTML}{f3722c}
\definecolor{col3}{HTML}{f8961e}
\definecolor{col4}{HTML}{f9844a}
\definecolor{col5}{HTML}{f9c74f}
\definecolor{col6}{HTML}{90be6d}
\definecolor{col7}{HTML}{43aa8b}
\definecolor{col8}{HTML}{4d908e}
\definecolor{col9}{HTML}{577590}
\definecolor{col10}{HTML}{277da1}

\catcode`\@=11   
\newdimen\@rotdimen
\newbox\@rotbox  

\def\@vspec#1{\special{ps:#1}}
\def\@rotstart#1{\@vspec{gsave currentpoint currentpoint translate
		#1 neg exch neg exch translate}}
\def\@rotfinish{\@vspec{currentpoint grestore moveto}}
\def\@rotr#1{\@rotdimen=\ht#1\advance\@rotdimen by\dp#1%
	\hbox to\@rotdimen{\hskip\ht#1\vbox to\wd#1{\@rotstart{90 rotate}%
			\box#1\vss}\hss}\@rotfinish}
\def\@rotl#1{\@rotdimen=\ht#1\advance\@rotdimen by\dp#1%
	\hbox to\@rotdimen{\vbox to\wd#1{\vskip\wd#1\@rotstart{270 rotate}%
			\box#1\vss}\hss}\@rotfinish}%
\def\@rotu#1{\@rotdimen=\ht#1\advance\@rotdimen by\dp#1%
	\hbox to\wd#1{\hskip\wd#1\vbox to\@rotdimen{\vskip\@rotdimen
			\@rotstart{-1 dup scale}\box#1\vss}\hss}\@rotfinish}%
\def\@rotf#1{\hbox to\wd#1{\hskip\wd#1\@rotstart{-1 1 scale}%
		\box#1\hss}\@rotfinish}%
\def\rotate{\@ifnextchar[{\@rotate}{\@rotate[l]}}
\def\@rotate[#1]#2{\setbox\@rotbox=\hbox{#2}\@nameuse{@rot#1}\@rotbox}

\catcode`\@=12

\tikzset{
    partial ellipse/.style args={#1:#2:#3}{
        insert path={+ (#1:#3) arc (#1:#2:#3)}
    }
}

\allowdisplaybreaks[2]
\numberwithin{equation}{section}

\usepackage[framemethod=TikZ]{mdframed}
\mdfsetup{skipabove=\topskip, skipbelow=\topskip}

\newcounter{equ}[section]

\newcounter{Boxequ}[section]

\makeatletter
\def\user@resume{resume}
\def\user@intermezzo{intermezzo}
\newcounter{previousequation}
\newcounter{lastsubequation}
\newcounter{savedparentequation}
\renewenvironment{subequations}[1][]{%
      \def\user@decides{#1}%
      \setcounter{previousequation}{\value{equation}}%
      \ifx\user@decides\user@resume 
           \setcounter{equation}{\value{savedparentequation}}%
      \else  
      \ifx\user@decides\user@intermezzo
           \refstepcounter{equation}%
      \else
           \setcounter{lastsubequation}{0}%
           \refstepcounter{equation}%
      \fi\fi
      \protected@edef\theHparentequation{%
          \@ifundefined {theHequation}\theequation \theHequation}%
      \protected@edef\theparentequation{\theequation}%
      \setcounter{parentequation}{\value{equation}}%
      \ifx\user@decides\user@resume 
           \setcounter{equation}{\value{lastsubequation}}%
         \else
           \setcounter{equation}{0}%
      \fi
      \def\theequation  {\theparentequation  \alph{equation}}%
      \def\theHequation {\theHparentequation \alph{equation}}%
      \ignorespaces
}{%
  \ifx\user@decides\user@resume
       \setcounter{lastsubequation}{\value{equation}}%
       \setcounter{equation}{\value{previousequation}}%
  \else
  \ifx\user@decides\user@intermezzo
       \setcounter{equation}{\value{parentequation}}%
  \else
       \setcounter{lastsubequation}{\value{equation}}%
       \setcounter{savedparentequation}{\value{parentequation}}%
       \setcounter{equation}{\value{parentequation}}%
  \fi\fi
  \ignorespacesafterend
}
\makeatother

\preprint{ZMP-HH/21-27}

\title{Systematics of perturbatively flat flux vacua}

\author[a]{Federico Carta,}
\author[b]{Alessandro Mininno,}
\author[c]{Pramod Shukla}

\affiliation[a]{Department of Mathematical Sciences, Durham University, \\ Durham, DH1 3LE, United Kingdom}
\affiliation[b]{II. Institut f\"ur Theoretische Physik, Universit\"at Hamburg,\\
Luruper Chaussee 149, 22607 Hamburg, Germany}
\affiliation[c]{\small ICTP, Strada Costiera 11, Trieste 34151, Italy}

\emailAdd{federico.carta@durham.ac.uk}
\emailAdd{alessandro.mininno@desy.de}
\emailAdd{pramodmaths@gmail.com}

\abstract{In this article, we present a systematic analysis of the so-called perturbatively flat flux vacua (PFFV) for the mirror Calabi-Yau (CY) $3$-folds ($\tilde{X}_3$) with $h^{1,1}(\tilde{X}_3) =2$ arising from the Kreuzer-Skarke database of the four-dimensional reflexive polytopes. We consider the divisor topologies of the CY $3$-folds for classifying the subsequent models into three categories; (i) models with the so-called Swiss-cheese structure, (ii) models with the K3-fibered structure, and (iii) the remaining ones which we call as models of ``Hybrid type". In our detailed analysis of PFFV we find that for a given fixed value of the D$3$ tadpole charge $N_{\rm flux}$, the K3-fibered mirror CY $3$-folds have significantly larger number of such PFFV as compared to those which have Swiss-cheese structure, while the Hybrid type models have a mixed behavior. We also compute the Gopakumar-Vafa invariants necessary for fixing the flat valley in the weak string-coupling and large complex-structure regime by using the non-perturbative effects, which subsequently reduce the number of physically trustworthy vacua quite significantly. Moreover, we find that there are some examples in which the PFFV are protected even at the leading orders of the non-perturbative effects due to the some underlying symmetry in the CY geometry, which we call as ``Exponentially flat flux vacua". We also present a new class of PFFV using the ${\cal S}$-duality arguments.
}


\begin{document}


\maketitle

\bigskip

\section{Introduction}
\label{sec_intro}

In the context of type IIB orientifold compactifications, moduli stabilization has attracted huge amount of interests since the past two decades \cite{Kachru:2003aw,Balasubramanian:2005zx,Blumenhagen:2006ci,Curio:2000sc,Grana:2005jc, DeWolfe:2005uu,Lust:2006zg,Douglas:2006es}. This is usually a two-step process where in the first step complex structure moduli and the axio-dilaton are stabilized by turning on background fluxes \cite{Gukov:1999ya,Taylor:1999ii,Giddings:2001yu,Kachru:2002he,Blumenhagen:2003vr} while in the second step K\"ahler moduli are stabilized using sub-leading corrections \cite{Witten:1996bn,Becker:2002nn}. In the meantime, there have been multiple proposals specific to K\"ahler moduli stabilization such as KKLT scheme \cite{Kachru:2003aw}, LVS scheme \cite{Balasubramanian:2005zx}, Racetrack scheme \cite{Blanco-Pillado:2004aap,Denef:2004dm} along with other ones like using string-loop corrections \cite{Berg:2005yu,Cicoli:2007xp,Antoniadis:2018hqy,Basiouris:2020jgp,Basiouris:2021sdf}, however in all these K\"ahler moduli stabilization schemes the central theme regarding complex structure moduli remains the same as what was developed in \cite{Gukov:1999ya,Taylor:1999ii,Giddings:2001yu,Kachru:2002he}. The reason being the fact that the prescription is quite generic and rich in the sense that there exist a landscape of flux vacua which have been extensively studied in the meantime \cite{Ashok:2003gk,Denef:2004cf, Denef:2005mm,Douglas:2004qg,Denef:2004ze, Conlon:2004ds,Misra:2004ky,Giddings:2005ff, Garcia-Etxebarria:2014wla, Cole:2019enn,Hebecker:2020aqr,  Marchesano:2019hfb,Aganagic:2002qg,Blanco-Pillado:2020hbw}. For attempts of exploring flux vacua in the type IIA, M-theory and F-theoretic setups, see also \cite{Douglas:2003um, Marchesano:2020uqz,Honma:2017uzn,Marchesano:2021gyv}.

One of the central ingredients of the type IIB phenomenological models is the dynamically realized VEV for the classical flux superpotential $W_0$ arising from the ${\cal S}$-dual pair of R-R and NS-NS fluxes, usually denoted as $(F_3, H_3)$. This so-called Gukov-Vafa-Witten (GVW) flux superpotential is given as below,
\begin{equation}
\label{eq:W0-value}
W_0 = \sqrt{\frac{2}{\pi}} \left\langle e^{K/2} \int_{\rm CY} \left(F_3 - S H_3\right) \wedge \Omega_3 \right\rangle \coma 
\end{equation}
where $S$ corresponds to the axio-dilaton, while $\Omega_3$ denotes the nowhere vanishing holomorphic $3$-form of the compactifying Calabi-Yau $3$-fold. Here, a normalization factor $e^{K/2}$ from the K\"ahler potential is included, where $K$ denotes the contributions from the complex structure and the axio-dilaton sector. In KKLT scheme \cite{Kachru:2003aw} and its variants, one requires a relatively lower value of the $W_0$ which has been argued to be achievable by tuning the fluxes in the landscape. In this context, the recent proposal of \cite{Demirtas:2019sip, Demirtas:2020ffz} has attracted a great amount of interests as it somehow creates an exception to the usual notion of being ``natural" or ``tuned" values of $W_0$. The proposal presents an elegant recipe of naturally realizing exponentially small value of $W_0$  via analytically invoking some perturbatively flat valleys of supersymmetric flux vacua, which are stabilized by using the non-perturbative effects. In fact, the magnitude of $W_0$ has been always a center of attraction for model builders \cite{Cicoli:2013swa,Louis:2012nb} and some significant amount of efforts has been made in this regard, specially in recent years \cite{Broeckel:2021uty, Bastian:2021hpc, Alvarez-Garcia:2020pxd, Grimm:2021ckh} 

The main motivation for the current work is to implement the recipe of \cite{Demirtas:2019sip} for explicitly constructing the perturbatively flat flux vacua (PFFV) using several CY geometries. The reason behind this goal is to examine if such vacua are unique (or accidental and limited to just few CY geometries)  or there is some (statistical) pattern in the sense of counting the number of PFFV. For that purpose, we consider the CY geometries arising from the triangulation of four-dimensional reflexive polytopes of the Kreuzer-Skarke (KS) database \cite{Kreuzer:2000xy}, and use the relevant topological data from the collection in \cite{Altman:2014bfa,Altman:2021pyc}. Given our current focus being limited to the stabilization of complex structure moduli and the axio-dilaton, we will use 39 CY geometries with $h^{1,1} = 2$ assuming that their respective mirrors would be the ones used to compactify the type IIB superstring theory. To make our study systematic, we divide these 39 CY geometries into three categories based on the underlying geometries of the CY $3$-folds being of the so-called Swiss-cheese type, K3-fibered type and the so-called ``Hybrid" type. In our detailed and systematic analysis, we find that there is a significant number of PFFV, though we observed that the K3-fibered CY geometries can have reasonably large number of PFFV configurations as compared to the other CY geometries. In order to fix the flat direction by using non-perturbative effects in the prepotential we compute the Gopakumar-Vafa invariants \cite{Gopakumar:1998ii,Gopakumar:1998jq} for all the 39 CY geometries and have subsequently utilized them in getting physical vacua in the large complex-structure and weak string-coupling regime. We find that only some of the CYs with a Swiss-cheese structure can result a couple of physical vacua, those with K3-fibered structure can have many physical vacua with $W_0$ being as low as of the order of $10^{-30}$. In addition, we find that there are a couple of CY geometries having some peculiar types of PFFV which are protected against non-perturbative effects as well, and we call such vacua as ``exponentially flat flux vacua" (EFFV). While re-deriving the recipe of PFFV \cite{Demirtas:2019sip}, we realized that there is another class of flat vacua which can be accessed using the ${\cal S}$-duality arguments leading to similar but a different set of requirements. 

The article is organized as follows: we briefly review the relevant preliminaries about type IIB orientifold compactifications, along with a quick re-derivation of the recipe about finding the perturbatively flat flux vacua (PFFV) in Section \ref{sec_basics}. In Section \ref{sec_methodology} we set the methodology and implementation strategy for the search of PFFV configurations where we illustrate the detailed steps in one concrete model which we replicate for all the 39 CY geometries. Section \ref{sec_classification} presents a classification of PFFV in three categories depending on the nature of mirror CY geometry being Swiss-cheese, K3-fibered or ``Hybrid" type. In Section \ref{sec_lowW0} we discuss the stabilization of the flat direction using instanton effects in the prepotential giving rise to low $W_0$, and classify such physical vacua. Section \ref{sec_EFFV} presents some more insights on the PFFV in the sense of finding those PFFV which are naturally protected against non-perturbative corrections due to the presence of some symmetry in the CY geometry. We also present a new class of PFFV configurations based on ${\cal S}$-duality arguments. Finally, in Section \ref{sec_conclusion} we conclude the findings with future plans, along with presenting 3 appendices regarding topological data of the mirror CY $3$-fold and the computation of the Gopakumar-Vafa (GV) invariants.

\section{On perturbatively flat flux vacua (PFFV)}
\label{sec_basics}

\subsection{Type IIB orientifolds and fluxes}

In this section, we present the relevant details about non-geometric type IIB orientifold setup. The allowed orientifold projections can be classified by their action ${\cal O}$ on the
K\"ahler form $J$ and the holomorphic $3$-form $\Omega_3$ of
the Calabi-Yau, which can be explicitly given as \cite{Grimm:2004uq}:
\begin{equation}
\label{eq:orientifold}
{\cal O}= \begin{dcases}
                       \Omega_p\, \sigma  & : \, 
                       \sigma^*(J)=J\,,  \qquad  \sigma^*(\Omega_3)=\Omega_3 \, ,\\[0.1cm]
                       (-)^{F_L}\,\Omega_p\, \sigma & :\, 
        \sigma^*(J)=J\,, \qquad \sigma^*(\Omega_3)=-\Omega_3\,.
\end{dcases}
\end{equation}
Note that $\Omega_p$ is the world-sheet parity, $F_L$ is the left-moving space-time fermion number, and $\sigma$ is a holomorphic, isometric involution. The first choice leads to orientifold with O$5$/O$9$-planes, whereas the second choice to O$3$/O$7$-planes. We further denote the bases of even/odd $2$-forms as $(\mu_\alpha, \, \nu_a)$ while $4$-forms as $(\tilde{\mu}_\alpha, \, \tilde{\nu}_a)$ where $\alpha\in h^{1,1}_+(X_3), \, a\in h^{1,1}_-(X_3)$. Some explicit constructions of type IIB toroidal/CY orientifold setups with $h^{1,1}_-(X_3)\neq0$ can be found in \cite{Lust:2006zg,Lust:2006zh,Blumenhagen:2008zz,Cicoli:2012vw,Gao:2013rra,Gao:2013pra,Gao:2014uha,Carta:2020ohw,Carta:2021sms, Carta:2021uwv,Cicoli:2021tzt,Altman:2021pyc}. The various field ingredients can be expanded in appropriate bases of the equivariant cohomologies. For example, the K\"{a}hler form $J$, the
$2$-forms $B_2$,  $C_2$ and the RR four-form $C_4$ can be expanded as
\begin{equation}
\begin{array}{llllll}
J & = & t^\alpha\, \mu_\alpha \coma & B_2 & = & -\, b^a\, \nu_a\coma \\
C_2 & = & -\, c^a\, \nu_a \coma & C_4 & = & c_{\alpha} \, \tilde\mu^\alpha + \, D_2^{\alpha}\wedge \mu_\alpha + V^{K}\wedge a_K - V_{K}\wedge b^K\fstop    
\end{array}
    \label{eq:fieldExpansions}
\end{equation}
Note that $t^\alpha$ are string-frame two-cycle volume moduli, while $b^a, \, c^a$ and $c_\alpha$ are various axions. Further, ($V^K$, $V_K$) forms a dual pair of space-time $1$-forms and $D_2^{\alpha}$ is a space-time $2$-form dual to the scalar field $c_\alpha$.

In addition, we denote the bases for the even/odd cohomologies $H^3_\pm(X_3)$ of $3$-forms as symplectic pairs $(a_K, b^J)$ and $({\cal A}_\Lambda, {\cal B}^\Delta)$ respectively, where we fix their normalization as,
\begin{equation}
\int_X a_K \wedge b^J = \delta_K{}^J \quad \text{and} \quad \int_X {\cal A}_\Lambda \wedge {\cal B}^\Delta = \delta_\Lambda{}^\Delta \,.    
\end{equation}
Here, for the orientifold choice with O$3$/O$7$-planes, the indices are distributed in the even/odd sector as $K,J\in \{1, ..., h^{2,1}_+(X_3)\}$ and $\Lambda,\Delta\in \{0, \ldots, h^{2,1}_-(X_3)\}$, while for O$5$/O$9$-planes, one has $K,J\in \{0, ..., h^{2,1}_+(X_3)\}$ and $\Lambda,\Delta\in \{1, ..., h^{2,1}_-(X_3)\}$. In this article, our focus will be only in the orientifold involutions leading to the O$3$/O$7$-planes. Also, for this setting, since $\sigma^*$ reflects the holomorphic $3$-form $\Omega_3$, we have $h^{2,1}_-(X_3)$ complex structure moduli appearing as complex scalars. 

\subsubsection{The K\"ahler potential}

The generic form of the type IIB K\"{a}hler potential can be written as a sum of two pieces motivated from their underlying ${\cal N}=2$ special K\"ahler and quaternionic structure, and the same is given as \cite{Grimm:2004uq},
\begin{equation}
\label{eq:KTypeIIB0}
K_{\rm IIB} = K^{\text{c.s}}+ K^{Q} \, , 
\end{equation}
where the $K^{\text{c.s.}}$ piece depends mainly on the complex structure moduli, while the $K^{Q}$ part depends on the volume of the Calabi-Yau $3$-fold and the dilaton. For computing the $K^{\text{c.s.}}$ piece, we consider the involutively-odd holomorphic $3$-form $\Omega_3 \equiv  {\cal X}^\Lambda {\cal A}_\Lambda - {\cal F}_{\Lambda} {\cal B}^\Lambda$ which can be written out using a prepotential of the following form \cite{Hosono:1994av,Arends:2014qca}, 
\begin{equation}
\label{eq:prepotential}
{\cal F} = -\, \frac{\kappa_{ijk}^0 \,{\cal X}^i\, {\cal X}^j \, {\cal X}^k}{3!\,{\cal X}^0} +  \frac{1}{2} \,\tilde{p}_{ij} {\cal X}^i\, {\cal X}^j + \,\tilde{p}_{i} \, {\cal X}^i {\cal X}^0- \frac{i}{2}\,{\tilde{p}}_0 ({\cal X}^0)^2 + \, ({\cal X}^0)^2 {\cal F}_{\rm inst}({\cal X}^i/{\cal X}^0)\fstop    
\end{equation}
This is a homogeneous function of degree $2$ in the symplectic coordinates ${\cal X}^\Lambda$, and one can use ${\cal X}^0 = 1$ and ${\cal X}^i = U^i$ to write it in terms of the non-homogeneous complex variables $U^i$. Further, the parameters $\kappa_{ijk}^0$ are the classical triple intersection numbers on the mirror $3$-fold $\tilde{X}_3$ which, along with the other rational/real parameters, are defined as \cite{Hosono:1994av,Arends:2014qca}
\begin{equation}
\begin{array}{rllrll}
\kappa_{ijk}^0  & = & \displaystyle{\int_{\tilde{X_3}} \, J_i \wedge J_j \wedge J_k}\coma & \tilde{p}_{ij} & = & \displaystyle{\frac{1}{2}\int_{\tilde{X_3}} \, J_i \wedge J_j \wedge J_j \quad ({\rm mod} \, \, {\mathbb Z})\coma}\\
\tilde{p}_j & = & \displaystyle{\frac{1}{4\cdot 3!}\int_{\tilde{X_3}} \,c_2(\tilde{X_3}) \wedge J_j} \coma &  \tilde{p}_0 &= & \displaystyle{-\, \frac{\zeta(3)\, \chi(\tilde{X_3})}{(2\, \pi)^3}}\fstop    
\end{array}
\end{equation}
Let us note that $\tilde{p}_0= - p_0$ as the Euler characteristics satisfy $\chi(\tilde{X}_{3})=-\chi(X_{3})$. Given that $\tilde{p}_{ij}$'s are symmetric only up to an integral difference in the off-diagonal entries, the following form has been also suggested in \cite{Demirtas:2020ffz},
\begin{equation}
\tilde{p}_{ij} = \frac{1}{2} \begin{dcases}
    \kappa_{iij}^0 & \text{for } i \geq j \\
    \kappa_{ijj}^0 & \text{for }  i < j  \\
\end{dcases}
\end{equation}
Subsequently, the prepotential can be expressed as below,
\begin{equation}
\label{eq:Fpre}
{\cal F}(U) = {\cal F}_{\rm poly}(U^i)  + {\cal F}_{\rm inst}(U^i)
\end{equation}
where
\begin{equation}
\label{eq:Fpert}
{\cal F}_{\rm poly} = -\frac{1}{3!} \kappa_{ijk}^0 \, U^i\, U^j\, U^k + \frac{1}{2} \tilde{p}_{ij} U^i\, U^j + \tilde{p}_j \, U^i - \frac{i}{2}\, \tilde{p}_0
\end{equation}
and
\begin{equation}
\label{eq:Finst}
{\cal F}_{\rm inst} = \frac{1}{(2\pi i)^3}\, \sum_{d_i} A_{d_i} e^{2 \pi i d_i\, U^i}\fstop
\end{equation}
Here, the string worldsheet corrections on the mirror dual side give rise to a general form of the instanton corrections given as below \cite{Hosono:1994av,Hosono:1994ax,Hosono:1993qy, Candelas:1994hw, Arends:2014qca}
\begin{equation}
    \begin{split}
        \mathcal{F}_{\text{inst}}(U^{i}) &=  \frac{1}{(2\pi i)^3}\,  \sum_{\beta\in H_{2}^{-}(\tilde{X}_{3},\mathbb{Z})\setminus\lbrace0\rbrace}\, n_{\beta}\, \text{Li}_{3}(q^{\beta})\\
\text{Li}_{3}(x) &=\sum_{m=1}^{\infty}\, \dfrac{x^{m}}{m^{3}}\kom q^{\beta}=\ee^{2\pi i d_{i}U^{i}}\coma
    \end{split}
\end{equation}
in terms of Gopakumar-Vafa invariants $n_{\beta}$ \cite{Gopakumar:1998ii,Gopakumar:1998jq},
which naively count the number of rational (oriented) curves $\Sigma_{g}$ of genus $g$ and of class $\beta=q_{i}\beta^{i}$ that can be holomorphically mapped into $\tilde{X}_{3}$.
Now, the first derivatives of the prepotential ${\cal F}$ are given by
\begin{equation}
\label{eq:Prepder1}
\begin{split}
    {\cal F}_0 &= \, \frac{1}{6}\, \kappa_{ijk}^0 \, U^i \, U^j\, U^k + \tilde{p}_i \, U^i - i\, \tilde{p}_0 +\left(2\,{\cal F}_{\text{inst}} - U^i\, \partial_{i} {\cal F}_{\text{inst}} \right)\coma \\
    {\cal F}_i &= -\,\frac{1}{2}\, \kappa_{ijk}^0  \, U^j\, U^k + \tilde{p}_{ij}\, U^j + \tilde{p}_i + \left(\partial_{i} {\cal F}_{\text{inst}} \right)\fstop
\end{split}
\end{equation}
Subsequently, the components of holomorphic $3$-form $\Omega_3$ can be explicitly rewritten as period vectors in terms of complex structure moduli $U^i$ given as,
\begin{equation}
\label{eq:IIBOmegawithU}
\Pi_{\Omega_3} = \begin{pmatrix} 
{\cal A}_0  \\
U^i \, {\cal A}_i \\
 \left(\frac{1}{2}\, \kappa_{ijk}^0  \, U^j\, U^k \,-  \tilde{p}_{ij}\, U^i\, U^j - \tilde{p}_i - \left(\partial_{i} {\cal F}_{\text{inst}} \right)\right) \, {\cal B}^i \\
- \left(\frac{1}{6} \kappa_{ijk}^0 \, U^i \, U^j\, U^k\, + \tilde{p}_i \, U^i - i\, \tilde{p}_0 +\left(2\,{\cal F}_{\text{inst}} - U^i\, \partial_{i} {\cal F}_{\text{inst}} \right) \right) {\cal B}^0
\end{pmatrix}\fstop
\end{equation}
Now the complex structure moduli dependent part of the K\"ahler potential can be simply given as,
\begin{equation}
\label{eq:KcsSimp}
\begin{split}
    K^{\text{c.s.}} & \equiv -\ln\left(-\, i \, \int_{X} \Pi_{\Omega_3} \wedge \ov\Pi_{\Omega_3} \right)\fstop
\end{split}
\end{equation}
The full type IIB K\"ahler potential can be given by including the quaternionic sector contribution $K^{Q}$, i.e.,
\begin{equation}
\label{eq:KIIB}
K_{\rm IIB} = -\ln\left(-\, i \, \int_{X} \Pi_{\Omega_3} \wedge \ov\Pi_{\Omega_3} \right)\, - \, \ln\left(-i(S - \ov{S}) \right) - 2 \ln\left({\cal V} + \ldots \right)\coma
\end{equation}
where the complexified axio-dilaton $S$ is given as $S = c_0 + i\, e^{-\phi}$, and the complex structure moduli ($U^i$) can be given as $U^i = v^i  - i \, u^i$. Further, the overall internal volume of the CY $3$-fold is written as ${\cal V} = \frac{1}{6} \, \, \ell_{\alpha \beta \gamma} \, t^\alpha\, t^\beta \, t^{\gamma}$ using the Einstein-frame two-cycle volume moduli $(t^\alpha)$ and the triple intersection numbers ($\ell_{\alpha\beta\gamma}$) of the compactifying CY $3$-fold. Note that the last piece can receive  $(\alpha^\prime)^3$-correction \cite{Becker:2002nn} which have been used for naturally realizing the LARGE volume scenarios (LVSs) \cite{Balasubramanian:2005zx}. However, our current interest is mainly to study the dynamics of the complex-structure moduli along with axio-dilaton in the fluxed background, and therefore we limit ourselves to detailing only those quantities which are directly relevant for the current purpose.

\subsubsection{The superpotential}

It is important to note that in a given setup, all flux-components will not be generically allowed under the full orientifold action ${\cal O} = \Omega_p (-)^{F_L} \sigma$. For example, the standard fluxes $(F, H)$ are anti-invariant under the involution \cite{Blumenhagen:2015kja, Robbins:2007yv}. Subsequently, the flux induced superpotential $W(S, U^i)$ can be  written as \cite{Gukov:1999ya},
\begin{equation}
\label{eq:W_gen}
W \equiv W_{\rm R} + W_{\rm NS} 
 = \sqrt{\frac{2}{\pi}}\, \left({F}_{\Lambda} - S \, {H}_{\Lambda} 
\right) \, {\cal X}^\Lambda -\sqrt{\frac{2}{\pi}}\, \left({F}^{\Lambda} - S \, {H}^{\Lambda} 
\right) \, {\cal F}_\Lambda\fstop 
\end{equation} 
Now using Eq. \eqref{eq:Prepder1} leads to the following explicit form of the flux superpotential,
\begin{equation}
W = W_{\rm poly} + W_{\rm exp}\coma
\end{equation}
in which the polynomial piece $W_{\rm poly}$ is induced through polynomial terms of ${\cal F}_\Lambda$ and the remaining pieces, arising from the instanton effects, are collected as $W_{\rm exp}$. To make them explicit we have the following form,
\begin{equation}
\label{eq:WgenIIB}
\begin{split}
    W_{\rm poly}= &\,\sqrt{\frac{2}{\pi}}\left[\ov{F}_0 + \, U^i \, \ov{F}_i + \frac{1}{2} \, \kappa_{ijk}^0  U^i U^j\, F^k - \frac{1}{6} \, \kappa_{ijk}^0  U^i U^j U^k\, F^0 - i\, \tilde{p}_0 \, F^0 \right]+ \\
& - \, \sqrt{\frac{2}{\pi}}\, S \left[\ov{H}_0 + \, U^i \, \ov{H}_i + \frac{1}{2} \, \kappa_{ijk}^0  U^i U^j \, H^k - \frac{1}{6} \, \kappa_{ijk}^0  U^i U^j U^k \, H^0 - i\, \tilde{p}_0 \, H^0 \right] 
\end{split}
\end{equation}
and 
\begin{equation}
\begin{split}
W_{\rm exp} = &\,-\sqrt{\frac{2}{\pi}} \left({F}^{0} - S \, {H}^{0}\right) \left(2\,{\cal F}_{\text{inst}} - U^i\, \partial_{i} {\cal F}_{\text{inst}} \right)+ \\
&\,- \sqrt{\frac{2}{\pi}}\, \left({F}^{i} - S \, {H}^{i}\right) (\partial_{i} {\cal F}_{\text{inst}})\fstop
\end{split}
\end{equation}
Note that because of the $\alpha^\prime$-corrections on the mirror side, the complex structure sector is modified such that to induce rational shifts in the usual flux components given as below,
\begin{equation}
\label{eq:IIB-W-fluxshift}
\begin{array}{lllclll}
\ov{F}_0 & = & F_0 - \tilde{p}_i \, F^i\coma & \text{\hspace{0.5cm}} & \ov{F}_i & = & F_i - \tilde{p}_{ij}\, F^j - \tilde{p}_i\, F^0\,, \\
\ov{H}_0 & = & H_0 - \tilde{p}_i \, H^i\coma & & \ov{H}_i & = & H_i - \tilde{p}_{ij}\, H^i - \tilde{p}_i H^0\,.
\end{array}
\end{equation}
Note that such rationally shifted fluxes appear only in the polynomial piece $W_{\rm poly}$, and have been also observed in \cite{Blumenhagen:2014nba} for type IIB $F_3/H_3$ flux model, which was generalized in \cite{Shukla:2019wfo} in the presence of non-geometric flux. Also, in the type IIA case, it was reported in \cite{Escobar:2018rna}. However, we note that these shifts are absent in the $W_{\rm exp}$ piece, which involves fluxes with upper indices only.

\subsection{Recipe for finding PFFV and low $|W_0|$}

In this section, we briefly review the recipe for dynamically realizing a low value for the GVW flux superpotential $W_0$ as proposed in \cite{Demirtas:2019sip}, and subsequently have been studied in \cite{Demirtas:2020ffz, Alvarez-Garcia:2020pxd, Honma:2021klo, Broeckel:2021uty, Bastian:2021hpc}. Let us begin by considering the derivatives of the polynomial piece $W_{\rm poly}$ of the flux superpotential $W_0$ which are given as,
\begin{equation}
\label{eq:derW}
\begin{split}
\frac{\partial W_{\rm poly}}{\partial U^i} = &\,\sqrt{\frac{2}{\pi}}\left[\ov{F}_i + \, \kappa_{ijk}^0 \,U^j\, F^k - \frac{1}{2} \, \kappa_{ijk}^0 \, U^j \,U^k\, F^0 \right]+ \\
&\,- \, \sqrt{\frac{2}{\pi}}\, S \left[\ov{H}_i +  \, \kappa_{ijk}^0 \, U^j \, H^k - \frac{1}{2} \, \kappa_{ijk}^0 \, U^j\, U^k \, H^0 \right]\coma\\
\frac{\partial W_{\rm poly}}{\partial S} = &- \, \sqrt{\frac{2}{\pi}} \left[\ov{H}_0 + \, U^i \, \ov{H}_i + \frac{1}{2} \, \kappa_{ijk}^0  U^i U^j \, H^k - \frac{1}{6} \, \kappa_{ijk}^0  U^i U^j U^k \, H^0 - i\, \tilde{p}_0 \, H^0 \right]\fstop    
\end{split}
\end{equation}
Now we take the following simplification for the choice of fluxes,
\begin{equation}
\label{eq:flux-cond1}
F_0 = \tilde{p}_i F^i\coma F_i = \tilde{p}_{ij} F^j\coma F^0 = 0\coma H_0 = 0\coma H^i = 0\coma H^0 = 0\coma
\end{equation}
which implies that the shifted fluxes in Eq. \eqref{eq:IIB-W-fluxshift} take the following form,
\begin{equation}
\label{eq:flux-cond2}
\ov{F}_0 = 0\coma \ov{F}_i = 0\coma \ov{H}_0 = 0\coma \ov{H}_i = H_i\fstop
\end{equation}
Moreover, the generic superpotential \eqref{eq:WgenIIB} gets simplified into the following form,
\begin{equation}
\label{eq:Wpolysimp}
W_{\rm poly}=
\sqrt{\frac{2}{\pi}}\left[\frac{1}{2} \, \kappa_{ijk}^0  U^i U^j\, F^k  - \, S \,U^i \, \ov{H}_i \right]\coma 
\end{equation}
while the derivatives of the superpotential are simplified as below,
\begin{equation}
\label{eq:dWpolysimp}
\begin{split}
    \frac{\partial W_{\rm poly}}{\partial U^i} &= \sqrt{\frac{2}{\pi}} \left[\kappa_{ijk}^0 \,U^j\, F^k  - \, S \, {H}_i \right]\coma\\
    \frac{\partial W_{\rm poly}}{\partial S} &= - \, \sqrt{\frac{2}{\pi}} \,U^i \, \ov{H}_i\fstop 
\end{split}
\end{equation}
The perturbative supersymmetric solutions are given as,
\begin{equation}
\begin{split}
    D_{U^i} \, W_{\rm poly} &= W_{\rm poly}\, \frac{\partial K}{\partial U^i}  + \frac{\partial W_{\rm poly}}{\partial U^i} = 0\coma \\
D_{S} \, W_{\rm poly} &= W_{\rm poly}\, \frac{\partial K}{\partial S}  + \frac{\partial W_{\rm poly}}{\partial S} = 0\fstop
\end{split}    
\end{equation}
Subsequently we have a perturbatively flat valley described by $U^i = S\, M^i$ for some flux vector $M^i$ which satisfies the following constraints,
\begin{equation}
    H_i\, M^i = 0 = \kappa_{ijk}^0 \,M^i \, M^j\, F^k\coma \kappa_{ijk}^0 \,M^j\, F^k = H_i\coma
\end{equation}
which lead to flat vacua such that superpotential and the derivatives trivially vanish. Moreover, these constraints suggest the following generic form for the flux vector $M^i$,
\begin{equation}
M^i = (\kappa_{ijk}^0 \, F^k)^{-1}\, H_j\coma    
\end{equation}
which has to lie within the K\"ahler cone. Subsequently, given that $W_{\rm poly} =0$ along the flat valley $U^i = S M^i$, the effective flux superpotential $W_0$ takes the following form,
\begin{equation}
\label{eq:W0eff}
W^{\rm eff}(S) = - \sqrt{\frac{2}{\pi}}\, {F}^{i} \, \partial_{i} {\cal F}_{\text{inst}}(S) \simeq - \sqrt{\frac{2}{\pi}}\, \sum_{d_i} \frac{A_{d_i}\, F^i\, d_i}{(2\pi\,i)^2}\, e^{2 \pi i \, S M^i d_i}\coma
\end{equation}
where we have used the flux constraints in \cref{eq:flux-cond1,eq:flux-cond2}. This effective superpotential helps in fixing the axio-dilaton ($S$) along with the complex structure moduli ($U^i$), and can subsequently lead to an exponentially low value of the flux superpotential $|W_0|$.

\section{Scanning the physical vacua}
\label{sec_methodology}

\subsection{Methodology}

We follow a three-step strategy to find physical vacua in the weak-coupling and large complex structure regime.

\subsubsection{Step 1: finding the PFFV}
\label{sec:step1}

To summarize, the conditions for having the perturbatively flat flux vacua lead to finding a flux vector $M^i$ such that the following conditions hold:
\begin{enumerate}[align=left, leftmargin=*]
\item[{\bf Existence}:] To find a matrix $N_{ij} = (\kappa_{ijk}^0 \, F^k)$ depending on the RR flux $F_3$, which is invertible, i.e. $\det[\kappa_{ijk}^0 \, F^k] \neq 0$, so that to guarantee the existence of another flux vector $M^i$ such that $M^i = N^{ij} H_j = (\kappa_{ijk}^0 \, F^k)^{-1}\, H_j$.
\item[{\bf Orthogonality}:] The flux vector $M^i$ must be orthogonal to the NS-NS flux vector $H_i$, i.e., $H_i\, M^i = 0$ which is also equivalent to the constraint: $M^i \, N_{ij} \, M^j = 0$.
\item[{\bf Tadpole condition}:] For cancellation of the total D$3$ charge, say $Q$, the tadpole constraint $ 0 \leq - F^i\, H_i \leq -2 \, Q$ needs to be satisfied.
\item[{\bf K\"ahler cone condition}:] One has to ensure that the flux vector $M^i$ lies within the K\"ahler cone. So if one is working in a $2$-form basis which involves the K\"ahler cone generators itself, then $M^i > 0, \, \, \forall i \in h^{2,1}_-({X})$ for simplicial cases. For mirror CY $3$-folds with non-simplicial K\"ahler cones, some additional constrains on the components of the $M^i$ flux vector would be needed.
\item[{\bf Integrality condition}:] Recalling that in setting the overall mechanism to work, we have set $F_0 = \tilde{p}_i F^i, \,\, F_i = \tilde{p}_{ij} F^j$ to make some of the shifted fluxes in Eq. (\ref{eq:IIB-W-fluxshift}) vanish, and therefore one needs to ensure that $(\tilde{p}_i F^i)$ and $(\tilde{p}_{ij} F^j)$ are integral valued.
\end{enumerate}

\subsubsection{Step 2: using the Gopakumar-Vafa invariants to generate a dilaton dependent effective superpotential}

Using the details about GV invariants from the Appendix \ref{sec:GVinvarcomputation}, the instanton part of the prepotential ${\cal F}_{\text{inst}}$ can be subsequently expressed in the following form,
\begin{equation}
\label{eq:F-inst-nis}
{\cal F}_{\text{inst}} = -\frac{1}{(2 \pi i)^3} \left(n_1 \, q_1 + n_2 \, q_2 + n_{11} \, q_1^2 + n_{12} \, q_1 \, q_2 + n_{22} \, q_2^2 +\ldots \right)
\end{equation}
where $q_i = e^{2 \pi i\, U^i}$ for $i \in \{1, 2\}$ and we have considered only up to quadratic terms in $q_i$'s in the instanton expansion. The leading order numbers $n_i$'s are collected for all the 39 CY geometries in Table \ref{tab_cydata-h11eq2-2}. For example, the Swiss-cheese case of $\mathbb{W}\CC\PP^4[1,1,1,6,9]$ these are given as below
\begin{equation}
\label{eq:nis-11169}
n_1= 540\coma n_2 = 3\coma \quad n_{11}= \frac{1215}{2}\coma \quad n_{12} = - 1080, \quad n_{22} = - \frac{45}{8}\fstop
\end{equation}
which have been also reported earlier in \cite{Candelas:1994hw}.

Assuming the large complex structure limit, we can ignore the non-perturbative contributions in the K\"ahler potential \eqref{eq:KcsSimp} which can be given as below,
\begin{equation}
\label{eq:KcsSimp1}
\begin{split}
    K^{\text{c.s.}} & \equiv -\ln\left(-\, i \, \int_{X} \Pi_{\Omega_3} \wedge \ov\Pi_{\Omega_3} \right)\\
&= -\ln\left[-\, i\, (\ov {\cal X}^\Lambda \, {\cal F}_\Lambda - {\cal X}^\Lambda \, \ov {\cal F}_\Lambda)\right] \simeq -\ln \left(\frac{4}{3} \, \kappa_{ijk}^0 \, u^i u^j u^k + 2\, \tilde{p}_0 \right) \\
&\simeq -\ln \left[-\, \frac{i}{6} \, \kappa_{ijk}^0 \,(U^i - \ov U^i)\, (U^j - \ov U^j)\, (U^k - \ov U^k) + 2\, \tilde{p}_0 \right]\fstop
\end{split}
\end{equation}
Notice that along the flat valley $U^i = S\, M^i$, the K\"ahler potential $K^{\text{c.s.}}$ solely depends on the axio-dilaton and using \cref{eq:KIIB} the ``effective" tree level axio-dilaton dependent term can be approximated as below,
\begin{equation}
\label{eq:Keff}
K^{\rm eff}(S) \simeq - 4 \, \ln \left[-i \left(S -\ov{S} \right)\right] -\ln \left[\frac{1}{6} \kappa_{ijk}^0 \,M^i M^j M^k \right]\coma 
\end{equation}
where we have neglected the Euler characteristic dependent constant pieces denoted as $\tilde{p}_0$. This is  justified because the intersection numbers $\kappa_{ijk}^0 $ are integral valued while the term with $\tilde{p}_0$ effectively appears with an extra factor of $g_s^{-3}$ and hence in the weak coupling regime, it is not likely to compete with the cubic term in $M^i$ flux vector.

For stabilizing the flat valley $U^i = S \, M^i$, we utilize the non-perturbative effects in the superpotential Eq. \eqref{eq:W0eff} given as,
\begin{equation}
\label{eq:Weff}
\begin{split}
W^{\rm eff}(S) \simeq &\, \frac{-1}{(2 \pi i)^2} \sqrt{\frac{2}{\pi}} \left[F^1 \, \left(n_1 \, e^{2\pi i S \, M^1} +2 \, n_{11} \,e^{4\pi i S \, M^1} + n_{12}\, e^{2\pi i S \, M^1}\, e^{2\pi i S\, M^2} \right)\right.+\\
& + F^2 \left.\left(n_2 \, e^{2\pi i S \, M^2} +2 \, n_{22} \,e^{4\pi i S \, M^2} + n_{12}\, e^{2\pi i S \, M^1}\, e^{2\pi i S\, M^2} \right) \right] + \ldots
\end{split}
\end{equation}
where we have used \cref{eq:F-inst-nis} which defines $n_i$'s and $n_{ij}$ coefficients.

\subsubsection{Step 3: stabilizing the PFFV in a physical regime}
\label{sec:step3}

With all the necessary pieces of information at hand, now we look for the supersymmetric stabilization of the PFFV via imposing the following constraints,
\begin{equation}
\label{eq:SUSY-eff}
D_S W^{\rm eff}(S) = \partial_S W^{\rm eff}(S) + W^{\rm eff}(S) (\partial_S K^{\rm eff}) = 0\fstop
\end{equation}
Note that the second piece in \cref{eq:SUSY-eff} with the K\"ahler derivative is always small in the weak string coupling limit. 
Using $S = c_0 + i\, s$ and the non-perturbative prepotential terms up to quadratics in $q_i$'s, the SUSY constraint \eqref{eq:SUSY-eff} translates into the following condition,
\begin{equation}
\label{eq:SUSY-explicit}
\begin{split}
0 = &\, e^{- 2 \pi \,s \,  {M^1}} \, F^1\, n_1 (1+ \pi\, s\, M^1) + e^{- 2 \pi \,s \,  {M^2}} \, F^2\, n_2 (1+ \pi\, s\, M^2)+ \\
& + 2\, e^{- 4 \pi \,s \,  {M^1}} \, F^1\, n_{11} \, (1+ 2 \pi\, s\, M^1) + 2\, e^{- 4 \pi \,s \,  {M^2}} \, F^2\, n_{22} \, (1+ 2 \pi\, s\, M^2)+\\
& + \, e^{- 2 \pi \,s \,  (M^1 + M^2)} \, (F^1 + F^2)\, n_{12} \, (1+ \pi\, s\, M^1+  \pi\, s\, M^2)\fstop
\end{split}
\end{equation}
It turns out that the universal axions $c_0$ can always be stabilized with a VEV $\langle c_0 \rangle = 0$, and the string coupling is determined by $g_s = \langle s \rangle^{-1}$ by solving \cref{eq:SUSY-explicit} numerically. Finally, we also need to check the following constraints to hold in order to ensure that the solution lies in the large complex structure and weak string coupling regime,
\begin{equation}
\label{eq:su-physical}
\langle s \rangle > 1\coma \langle u^1 \rangle = \langle s \rangle \, M^1 > 1\coma \langle u^2 \rangle = \langle s \rangle \, M^2 > 1\fstop
\end{equation}
Although we will use \cref{eq:SUSY-eff} or equivalently \cref{eq:SUSY-explicit} in our numerical search of physical vacua, let us make some approximations to see through the underlying analytical structure of the vacua. 

\subsubsection*{Case 1: $M^1 \neq M^2$}

For the cases when $M^1 \neq M^2$, if we restrict $W^{\rm eff}(S)$ to only single exponentials by neglecting the terms with coefficients $n_{ij}$, then the SUSY condition in \cref{eq:SUSY-eff} or equivalently \cref{eq:SUSY-explicit} boils down to the following simple relation,
\begin{equation}
\label{eq:gsVEV}
e^{- 2\pi \langle s \rangle (M^1 - M^2)} \simeq -\frac{A\, M^2}{M^1}\coma A = \frac{F^2 \,n_2}{F^1 \,n_1}\fstop
\end{equation}
Given the fact that each of the $M^i$ flux components is positive definite as follows from the K\"ahler cone conditions, the constraint in  \cref{eq:gsVEV}, which solves for $\langle s \rangle$, has physical solutions only for $A < 0$. This subsequently leads to the following VEV for the flux superpotential $W^{\rm eff}$,
\begin{equation}
\label{eq:W0VEV}
\langle W^{\rm eff} \rangle \simeq c \,\left( e^{-2\pi \langle s \rangle \, M^1} + A \, e^{-2\pi \langle s \rangle \, M^2} \right)
\coma c = \frac{1}{(2 \pi)^2} \sqrt{\frac{2}{\pi}} \, F^1 \,n_1\fstop
\end{equation}
 To connect with the claimed values of \cite{Broeckel:2021uty}, we note that $F^i = \{-16, 50\}$ and $n_i =\{540, 3\}$ from \cref{eq:nis-11169} one gets, $c = -\sqrt{\frac{2}{\pi}} \times 8640/{(2 \pi)^2}$ and $A = -5/288$ for the $\mathbb{W}\CC\PP^4[1,1,1,6,9]$ Swiss-cheese model. However, note that this approximate estimation fails for the flux vacua with $M^1 = M^2$.
 
\subsubsection*{Case 2: $M^1 = M^2 = M$}

This case is a bit special one, because neglecting the second order terms in $q_i$'s does not result in physical solutions for $M^1 = M^2 = M > 0$. In such case, the first order  non-perturbative terms will have to complete with the second order terms leading to the following relation arising from \cref{eq:SUSY-eff},
\begin{equation}
\label{eq:SUSY-eff-M1=M2}
\begin{split}
e^{2 \pi  {M} \,\langle s \rangle } \, {(\pi  {M} s+1)\, ({F^1} {n_1}+{F^2} {n_2})}  = -(2 \pi  {M} s+1)\, ({F^1} (2 {n_{11}}+{n_{12}})+{F^2} ({n_{12}}+2 {n_{22}})) \fstop
\end{split}
\end{equation}
This shows that if one does not include the second order terms, i.e., neglecting the pieces containing $n_{ij}$'s, the SUSY solution can only be satisfied for $ \langle s \rangle = - 1/(\pi M)$ if $({F^1} {n_1}+{F^2} {n_2}) \neq 0$. However, given that $M > 0$ follows from the K\"ahler cone condition, such flat vacua would not result in receiving VEVs in the physical domain. Therefore, for all the PFFV configurations resulting in $M^1 = M^2$ one can check the sign of the following quantity,
\begin{equation}
\label{eq:calA}
{\cal A} = \frac{{F^1} (2 {n_{11}}+{n_{12}})+{F^2} ({n_{12}}+2 {n_{22}})}{{F^1} {n_1}+{F^2} {n_2}}\fstop
\end{equation}
If one finds that ${\cal A} \geq 0$, this would imply that there is no physical solution in the weak coupling regime for the corresponding flat vacua. Also, even if ${\cal A} < 0$, one has to check that one gets weak coupling i.e., $\langle s \rangle > 1$ using the corresponding flux values $M^1=M^2=M$. By these criteria, one can discard several flat vacua arising from Step 1 in Section \ref{sec:step1}.

In fact, such a solution would always be suspicious, as it uses two sub-leading corrections in the expansion of the same (non-perturbative) series. However, these can either lead to unphysical solution beyond the weak coupling regime, or some special cases when ${\cal A}$ takes $\frac00$ form leading to what we call as Exponentially flat vacua (EFFV). We will discuss these issues in some detail in the upcoming sections.
 
\subsection{Implementation}

There are a total of 39 distinct CY geometries with $h^{1,1} = 2$ in the KS database which arise from the 36 four-dimensional reflexive polytopes giving 48 triangulations. We consider studying these CYs for complex-structure moduli sector, assuming that the K\"ahler moduli stabilization is performed using the mirror of such CYs as the compactifying $3$-fold for the type IIB superstring theory. In order to perform a general analysis for the two-field models, we consider the following flux vectors for the RR and NS-NS fluxes,
\begin{equation}
F^i = \left(\begin{matrix}
    F^1 \\
    F^2 \\
\end{matrix} \right)\coma  H_i = \left(\begin{matrix}
    H_1 \\
    H_2 \\
\end{matrix} \right)\fstop
\end{equation}
In addition, we consider the basis of divisors on the mirror CY $3$-fold ($\tilde{X}$) to be given by $\{J_1, J_2\}$ which subsequently leads to the following triple intersection numbers ($\kappa_{ijk}^0 $) along with the other ingredients:
\begin{equation}
\label{eq:ingredients-h11eq2}
\begin{split}
 \kappa_{ijk}^0  &= \{\kappa_{111}^0,\, \kappa_{112}^0,\, \kappa_{122}^0,\, \kappa_{222}^0 \}\coma\\
\tilde{p}_{ij} &= \frac{1}{2} \int_{\tilde{X}} J_i J_j^2 = \frac{1}{2} \begin{pmatrix}
    \kappa_{111}^0  & \kappa_{122}^0 \\
    \kappa_{112}^0 & \kappa_{222}^0 
\end{pmatrix} \coma\\
\tilde{p}_i &= \frac{1}{24} \begin{pmatrix}
    c_{11} \, \kappa_{111}^0 + c_{12} \, \kappa_{112}^0 + c_{22} \, \kappa_{122}^0 \\
    c_{11} \, \kappa_{112}^0 + c_{12} \, \kappa_{122}^0 + c_{22} \, \kappa_{222}^0 
\end{pmatrix}\fstop
\end{split}
\end{equation}
Here the coefficients $c_{ij}$ are introduced through the second Chern class of the mirror CY $3$-fold $(\tilde{X})$ which can be generically given as $c_2(\tilde{X}) = c_{11} \, J_1^2 + c_{12}\, J_1\, J_2 + c_{22}\, J_2^2$. These pieces of information can be taken from the database \cite{Altman:2014bfa}. However, one has to appropriately convert these quantities into the divisor basis of K\"ahler cone generators, which we present in Table \ref{tab_cydata-h11eq2}. Therefore, the overall task of finding perturbatively flat flux vacua boils down to the following plan:

Given the following symplectic flux vectors parametrized by a pair of fluxes $\{F^i, \, H_i\}$ along with the quantities $\kappa_{ijk}^0 $ and $\tilde{p}_{ij}$ and $\tilde{p}_i$,
\begin{equation}
F^\Lambda =  \left\{\begin{matrix}
      \tilde{p}_i F^i \\
      \tilde{p}_{ij} F^j \\
      0 \\
      F^i \\
\end{matrix} \right\}\coma  H_\Lambda =  \left\{\begin{matrix}
     0 \\
     H_i \\
     0 \\
     0 \\
\end{matrix} \right\}\coma
\end{equation}
with $i=1,2$, we know how to find another flux vector $M^i$ defined as $M^i = N^{ij}H_j = (\kappa_{ijk}^0 \, F^k)^{-1}\, H_j$ that satisfies
\begin{equation}
H_i\, M^i = 0\coma 0 \leq - F^i\, H_i \leq -2 \, Q\coma
\end{equation}
along with the requirement that the flux vector $M^i$ lies inside the K\"ahler cone, and $(\tilde{p}_{ij}F^j)$ and $(\tilde{p}_i F^i)$ are integral valued. The next step is to stabilize the flat directions using the non-perturbative effects and ensure that the subsequent vacua are physical in the sense of being within weak coupling and large complex structure regime. This is done by solving \cref{eq:SUSY-eff} using \cref{eq:Keff,eq:Weff}. To illustrate the overall recipe, we present all the details in a concrete model, which we will subsequently replicate for all the 39 CY geometries later on.

\subsection{An explicit (Swiss-cheese) example}

As the original proposal \cite{Demirtas:2019sip} was based on the vanilla Swiss-cheese Calabi-Yau realized as a degree-18 hypersurface in $\mathbb{W}\CC\PP^4[1,1,1,6,9]$, we will present the relevant topological data for this example, which we will subsequently use to explore the possibilities for the flat flux vacua. Also, we present the relevant data for this Swiss-cheese model in three formulations separately, where each formulation is useful for manifestation of certain aspects while may not be as appropriate as the other one for some other aspects.

\subsubsection*{AGHJN's basis of divisors}

This Swiss-cheese CY $3$-fold corresponds to the polytope ID $41$ in the CY database of Altman, Gray, He, Jejjala and Nelson (AGHJN) \cite{Altman:2014bfa} and it is defined by the following toric data:
\begin{center}
\begin{tabular}{|c|cccccc|}
\hline
CY & $x_1$  & $x_2$  & $x_3$  & $x_4$  & $x_5$ & $x_6$      \\
\hline
 6 & 0 & 0  & 0 & 2 & 3 & 1  \\
18 & 1 & 1  & 1 & 6 & 9 & 0  \\
\hline
 \multicolumn{1}{c|}{}& SD$_1$  & SD$_1$ & SD$_1$ &  SD$_2$ & SD$_3$ & ${\mathbb P}^2$  \\
\cline{2-7}
\end{tabular}
\end{center}
For this CY $3$-fold the Hodge numbers are $(h^{2,1}, h^{1,1}) = (272, 2)$, the Euler number is $\chi=-540$ and the SR ideal is ${\rm SR} =  \{x_1 \, x_2 \, x_3, \,  \, x_4 \, x_5 \, x_6\}$. Here, what we call the `special deformation' (SD) divisors SD$_1$, SD$_2$ and SD$_3$ (following the nomenclature of \cite{Gao:2013pra, Cicoli:2016xae, Cicoli:2017axo}) are represented by the following Hodge diamond:
\begin{equation}
    \begin{array}{ccccccccccccccc}
                 &   &   &   &  1 &   &         &\text{\hspace{0.5cm}} &           &   &    &   &  1  &   &         \\
                 &   &   & 0 &    & 0 &         & &           &   &    & 0 &     & 0 &         \\
      {\rm SD}_1 & = & 2 &   & 30 &   & 2 \coma & & {\rm SD}_2 & = & 28 &   & 218 &   & 28\coma \\
                 &   &   & 0 &    & 0 &         & &           &   &    & 0 &     & 0 &         \\
                 &   &   &   &  1 &   &         & &           &   &    &   &  1  &   &         \\
                 \\
                 &   &   &   &  1 &   &         & &           &   &    &   &  1  &   &         \\
                 &   &   & 0 &    & 0 &         & &           &   &    & 0 &     & 0 &         \\
      {\rm SD}_3 & = & 65 &   & 417 &   & 65 \coma & & {\mathbb P}^2 & = & 0 &   & 1 &   & 0\fstop \\
                 &   &   & 0 &    & 0 &         &  &          &   &    & 0 &     & 0 &         \\
                 &   &   &   &  1 &   &         & &        &   &    &   &  1  &   &         \\
    \end{array}
 \end{equation}

\noindent
This nomenclature is used in the sense that the simplest non-rigid divisor, i.e., one which can be deformed, has $h^{2,0}(D) = 1$, and K3 is one such example. `Special deformation' (SD) divisors are those which have instead $h^{2,0}(D) > 1$.  The divisor topology analysis has been performed using the package \texttt{cohomCalg} \cite{Blumenhagen:2010pv, Blumenhagen:2011xn}. The triple intersection numbers and the K\"ahler cone in the basis of smooth divisors $J_1\equiv D_1=\{0, 1\}$ and $J_2 \equiv D_6 =\{1, 0\}$ are:
\begin{equation}
\begin{split}
    &\kappa^0_{111} = 0\coma  \kappa^0_{112} = 1\coma  \kappa^0_{122} = -3\coma \kappa^0_{222} = 9\coma\\
    &\text{K\"ahler Cone:} \quad t_1 -3 \, t_2> 0\coma t_2 > 0\fstop 
\end{split}    
\end{equation}
In this case the overall volume and the 4-cycle moduli take the following form:
\begin{equation}
{\cal V} = \frac{1}{2}t_1^2\, t_2 - \frac{3}{2} t_1\, t_2^2 + \frac{3}{2}t_2^3\coma \tau_1 = t_1 t_2 - \frac32\,t_2^2\coma \tau_2 = \frac12 \,(t_1 - 3\, t_2)^2\fstop    
\end{equation}
The parameters $\tilde{p}_{ij}$ appearing in the prepotential ${\cal F}$ are given as below \cite{Candelas:1994hw, Cicoli:2013cha},
\begin{equation}
\tilde{p}_{ij} = \frac{1}{2} \int_{CY} D_i D_j^2 = \frac{1}{2} \left(\begin{matrix}
    0  & 1\\
    -3 & 9 
\end{matrix} \right)\fstop
\end{equation}
Note that these $\tilde{p}_{ij}$ parameters are symmetric only modulo integers. The second Chern-class $c_2(CY)$ is given as,
\begin{equation}
c_2(CY) = 102\, D_1^2 + 69\, D_1 D_6 + 11\, D_6^2\coma
\end{equation}
which subsequently gives the prepotential parameter $\tilde{p}_i$'s as below \cite{Candelas:1994hw, Cicoli:2013cha},
\begin{equation}
\tilde{p}_i = \frac{1}{24} \int_{CY} c_2(CY) D_i = \frac{1}{4} \left\{\begin{matrix}
    6 \\
    -1 
\end{matrix} \right\}\fstop
\end{equation}

\subsubsection*{Diagonal basis of divisors}

Now we switch to work in the so-called ``diagonal" basis of divisors, which turns out to be consisting of $J_1\equiv 3 \, D_1 + D_6 =\{1,3\}$ and $J_2 \equiv D_6 =\{1,0\}$. Subsequently, the triple intersection numbers and the K\"ahler cone are given as below:
\begin{equation}
    \begin{split}
    & \kappa^0_{111} = 9\coma  \kappa^0_{112} = 0\coma  \kappa^0_{122} = 0\coma \kappa^0_{222} = 9\coma  \\
&  \text{K\"ahler Cone:} \quad t_1 + \, t_2> 0\coma t_2 < 0\fstop 
    \end{split}
\end{equation}
In this case the overall volume and the 4-cycle moduli take the following form:
\begin{equation}
{\cal V} = \frac{3}{2}t_1^3 + \frac{3}{2}t_2^3\coma \tau_1 = \frac92 t_1^2\coma \tau_2 =\frac92 t_2^2 \coma
\end{equation}
which ensures that this CY will lead to a Swiss-cheese type volume form after converting the two-cycle volume moduli $(t^i)$ into the four-cycle volume moduli $(\tau_i)$ according to K\"ahler cone conditions. The parameters $\tilde{p}_{ij}$ appearing in the prepotential ${\cal F}$ are given as below,
\begin{equation}
\tilde{p}_{ij} = \frac{1}{2} \int_{CY} D_i D_j^2 = \frac{1}{2} \left(\begin{matrix}
    9  & 0\\
    0 & 9 \\
\end{matrix} \right)\fstop
\end{equation}
while the $\tilde{p}_i$'s parameter are as below,
\begin{equation}
\tilde{p}_i = \frac{1}{24} \int_{CY} c_2(CY) D_i = \frac{1}{4} \left\{\begin{matrix}
    17 \\
    -1 \\
\end{matrix} \right\}\fstop
\end{equation}
Now working with the diagonal basis one may like to explore the recipe of realizing low $W_0$ for the $\mathbb{W}\CC\PP^4[1,1,1,6,9]$ Swiss-cheese model. The diagonal nature of the intersection polynomial gives the various relevant quantities as,
\begin{equation}
\label{eq:diag-Nij}
N_{ij} = \kappa_{ijk}^0 \, F^k = \left(\begin{matrix}
    9 F^1 & 0\\
    0 & 9 F^2 \\
\end{matrix} \right) \quad \implies \quad M^i = N^{ij}\, H_j = \left(\frac{H_1}{9 F^1},\, \frac{H_2}{9F^2} \right)^T\fstop
\end{equation}
Subsequently, the condition $H_i\, M^i = 0$ results in 
\begin{equation}
F^1 = - F^2  \frac{H_1^2}{H_2^2} \quad \implies \quad M^i = \frac{H_2}{9F^2}\left(-\frac{H_2}{H_1},\, 1 \right)^T\fstop
\end{equation}
Given that the generators of the K\"aher cone are given as:
\begin{equation}
\label{eq:11169KCgen}
{\cal K}_1 =  3 \, D_1 + D_6 =\{1, 3\}\,,  \quad \text{and} \quad {\cal K}_2 \equiv D_1 =\{0, 1\}\coma
\end{equation}
imposing the condition that $M^i$ lies inside the K\"ahler cone results in the following two constraints,
\begin{equation}
\label{eq:pi-KC-11169-diag}
-\frac{H_2^2}{9H_1 F^2} + \frac{H_2}{9F^2} > 0 \coma -\frac{H_2}{3F^2}  > 0\fstop
\end{equation}
Also, the tadpole condition is simplified into the following form,
\begin{equation}
N_{\rm flux} =\frac12 \left(\frac{H_1^3\, F^2}{H_2^2} - H_2 F^2\right)\fstop
\end{equation}
Subsequently, the set of constraints for flat vacua can be summarized as below,
\begin{equation}
    \label{eq:final-constraints-diag}
    \begin{dcases}
   -\frac{H_2^2}{9H_1 F^2} + \frac{H_2}{9F^2} > 0\coma\\
   -\frac{H_2}{3F^2} > 0\coma \\
    N_{\rm flux} = \frac12 \left(\frac{H_1^3\, F^2}{H_2^2} - H_2 F^2\right) \in {\mathbb Z} & \text{such that }0 < N_{\rm flux} \leq N_{\rm flux}^{\text{max}}\coma\\
  - \frac{9 H_1^2 F^2}{2 H_2^2} \in {\mathbb Z}\coma\\
   \frac{9 F^2}{2} \in {\mathbb Z}\coma\\
   -\frac{F^2\left(17 H_1^2 + H_2^2\right)}{4H_2^2} \in {\mathbb Z} \coma\\
   F^1 = - F^2  \frac{H_1^2}{H_2^2} \in {\mathbb Z} \fstop
    \end{dcases}
\end{equation}
Note that the last requirement of $F^1$ to be an integer has to be also imposed because we have solved $M^i\, H_i = 0 $ for $F^1$ which has helped to eliminate it from other expressions, and we will be subsequently providing an input of integer flux configuration consisting of only three fluxes, namely $\{H_1, H_2, F^2\}$. 

Let us point out that while working in this basis, there can be a couple of benefits for analyzing the so-called `strong' Swiss-cheese examples,
\begin{itemize}

\item
It is easy to figure out the presence of the so-called Swiss-cheese structure in the CY.

\item 
The rigid del Pezzo divisor(s) are part of the basis, and hence the consistency checks for the superpotential contributions would be manifestly doable for the K\"ahler moduli sector.

\item 
Being diagonal, the $\tilde{p}_{ij}$ matrix will be symmetric.

\item
In the diagonal basis of divisors, there are fewer terms in the intersection polynomial and one can expect to have a relatively simpler expressions for various intermediate flux dependent quantities, e.g., $N_{ij}$ given in \cref{eq:diag-Nij} or any other constraint in \cref{eq:final-constraints-diag}.

\end{itemize}
However, there are some limitations working in the diagonal basis, and it turns out that using the basis of K\"ahler cone generators can be more suitable because of the following reasons,
\begin{itemize}
\item 
The very first obstacle is the fact that not every Calabi-Yau has Swiss-cheese type structure and hence this diagonal basis choice cannot be applied to generic CY cases in an algorithmic manner. As opposed to this, a basis of K\"ahler cone generators can be applied to all the CY $3$-folds.

\item
While working in the K\"ahler cone basis, the condition for the $M^i$ flux vector to lie within the K\"ahler cone simply boils down to demanding $\{M^1 > 0 , M^2 > 0 \}$ whereas while using the diagonal basis (or any other basis for that matter) one has to take extra care of the K\"ahler cone conditions on the flux vector $M^i$.

\item
Given that the \texttt{INSTANTON} code \cite{Klemm:2001aaa} used for computing the Gopakumar-Vafa invariants works efficiently for the K\"ahler cone basis, this choice of diagonal basis may not be as encouraging to use as it has been for LVS model building. 
\end{itemize}

\subsubsection*{CFKM's basis of divisors}

With the motivation of using the divisor basis of K\"ahler cone generators, now we present the same Swiss-cheese Calabi-Yau example in the notations of  Candelas, Font, Katz and Morrison (CFKM) \cite{Candelas:1994hw} which we refer as CFKM's notation from now onwards. In order to identify AGHJN's notations with those of CFKM, one needs to consider the divisor basis $\{{\cal H}, {\cal L}\}$ where ${\cal H} = 3 \, D_1 + D_6$ and ${\cal L} = D_1$, which leads to triple intersection numbers being given as
\begin{equation}
{\cal H}^3 =9\coma {\cal H}^2 \, {\cal L} = 3\coma {\cal H} \, {\cal L}^2 =1\coma {\cal L}^3 = 0\coma
\end{equation}
while $\{{\cal H}, {\cal L}\}$ also turns out to be the generators of the K\"ahler cone in \cite{Candelas:1994hw}. This means that the divisor basis of \cite{Candelas:1994hw} does not contain the rigid ${\mathbb P}^2$ divisor, which turns out to be useful in realizing the LVS models. Subsequently, working in the CFKM's basis of divisors turns out to be $J_1\equiv 3 \, D_1 + D_6 =\{1, 3\}$ and $J_2 \equiv D_1 =\{0, 1\}$ for which the triple intersection numbers and the K\"ahler cone conditions are given as below:
\begin{equation}
    \begin{split}
    & \kappa^0_{111} = 9\coma  \kappa^0_{112} = 3\coma  \kappa^0_{122} = 1\coma \kappa^0_{222} = 0\coma  \\
&  \text{K\"ahler Cone:} \quad t_1 > 0\coma t_2 > 0\fstop 
    \end{split}
\end{equation}
In this case the overall volume and the 4-cycle moduli take the following form:
\begin{equation}
{\cal V} = \frac{3}{2}t_1^3 + \frac{3}{2}t_1^2\, t_2 + \frac{1}{2} t_1\, t_2^2\coma \tau_1 = \frac12 \,(3\, t_1 + t_2)^2\coma \tau_2 = t_1 t_2 + \frac32\,t_1^2\fstop    
\end{equation}
The parameters $\tilde{p}_{ij}$ appearing in the prepotential ${\cal F}$ are given as below,
\begin{equation}
\tilde{p}_{ij} = \frac{1}{2} \int_{CY} D_i D_j^2 = \frac{1}{2} \left(\begin{matrix}
    9  & 3\\
    1 & 0 \\
\end{matrix} \right)\coma
\end{equation}
while the $\tilde{p}_i$'s parameter are as below,
\begin{equation}
\tilde{p}_i = \frac{1}{24} \int_{CY} c_2(CY) D_i = \frac{1}{4} \left\{\begin{matrix}
    17 \\
    6 
\end{matrix} \right\}\fstop
\end{equation}
The instanton corrections are known in the CFKM notations, and take the following form,
\begin{equation}
    \begin{split}
        (2 \pi i)^3 {\cal F}_{\text{inst}} &= -540 \, q_1 - 3 \, q_2 -\frac{1215}{2} \, q_1^2 + 1080 \, q_1 \, q_2 + \frac{45}{8} q_2^2\coma
    \end{split}
\end{equation}
where we have defined $q_i = e^{2\pi i \, U^i}, \, \, \forall i \in \{1, 2\}$.

\subsubsection*{Numerics of the flat flux vacua}

Now, working with the CFKM basis (which can be also called as the K\"ahler cone basis), we will explore the recipe of realizing low $W_0$ for the $\mathbb{W}\CC\PP^4[1,1,1,6,9]$ Swiss-cheese model. The $N_{ij}$ matrix is given as below,
\begin{equation}
N_{ij} = \kappa_{ijk}^0 \, F^k = \left(\begin{matrix}
    9F^1 + 3F^2 & 3F^1 + F^2\\
    3F^1 + F^2 & F^1 
\end{matrix} \right)\coma
\end{equation}
which being invertible guarantees the existence of the flux vector $M^i$ given by $M^i = N^{ij}\, H_j$. Moreover, the condition $H_i\, M^i = 0$ results in the following solution,
\begin{equation}
F^1 = - \frac{F^2 H_2 (2 H_1 - 3 H_2)}{(H_1-3H_2)^2}\coma
\end{equation}
which gives a simplified form of the $M^i$ flux vector as below,
\begin{equation}
M^i = \frac{H_1-3H_2}{F^2}\left(-\frac{H_2}{H_1},\, 1 \right)^T\fstop
\end{equation}
Now, imposing the condition that $M^i$ lies inside the K\"ahler cone can simply be imposed through the following constraints,
\begin{equation}
\label{eq:pi-KC-11169}
\frac{H_1-3H_2}{F^2} > 0\coma -\frac{H_2}{H_1} > 0\coma
\end{equation}
which can result in two possible solutions,
\begin{equation}
\{H_1 > 0,\,H_2 < 0,\, F^2 > 0\} \quad \text{or} \quad \{H_1 < 0,\, H_2 > 0,\, F^2 < 0\}\fstop
\end{equation}
In fact it has been also pointed out in \cite{Broeckel:2021uty} that the number of flat vacua obtained by demanding the five set of requirements in the recipe can be significantly reduced by considering their equivalence under the following ${\cal S}$-duality transformations,
\begin{equation}
F^i \to - F^i\coma H_i \to - H_i\coma
\end{equation}
which subsequently turns out to be more general than the current particular Swiss-cheese example as we will see later on. So, without loss of any generality, one can choose any of the two solutions, and we take the second one. Further, the simplified tadpole condition after using $H_i\, M^i = 0$ turns out to take the following form,
\begin{equation}
\label{eq:tadpole-11169}
N_{\rm flux} = -\frac{3 {F^2} {H_2} \left({H_1}^2-3 {H_1} {H_2}+3 {H_2}^2\right)}{2 ({H_1}-3 {H_2})^2}\fstop
\end{equation}
Demanding $N_{\rm flux} > 0$ does not put any extra constraints and gets trivially satisfied by the previous constraint $\{H_1 < 0,\, H_2 > 0,\, F^2 < 0\}$. Now the desired flat flux vacua can be obtained by scanning through the possible values of fluxes $\{F^2, H_1, H_2\}$ satisfying the following constraints simultaneously,
\begin{equation}
    \label{eq:final-constraints}
    \begin{dcases}
    \frac{H_1-3H_2}{F^2} > 0\coma\\
    -\frac{H_2}{H_1} > 0\coma \\
    N_{\rm flux} = -\frac{3 {F^2} {H_2} \left({H_1}^2-3 {H_1} {H_2}+3 {H_2}^2\right)}{2 ({H_1}-3 {H_2})^2} \in {\mathbb Z} & \text{such that }0 < N_{\rm flux} \leq N_{\rm flux}^{\text{max}}\coma\\
   \frac{3 {F^2} {H_1}^2}{2 ({H_1}-3 {H_2})^2} \in {\mathbb Z}\coma\\
   \frac{3 {F^2} {H_2} (2 {H_1}-3 {H_2})}{2 ({H_1}-3 {H_2})^2} \in {\mathbb Z}\coma\\
    \frac{1}{4} {F^2} \left(\frac{17 {H_2} (2 {H_1}-3 {H_2})}{({H_1}-3 {H_2})^2}+6\right) \in {\mathbb Z} \coma \\
    F^1 = - \frac{F^2 H_2 (2 H_1 - 3 H_2)}{(H_1-3H_2)^2} \in {\mathbb Z}\fstop
    \end{dcases}
\end{equation}
Note that the last requirement of $F^1$ to be an integer has to be also imposed because we have solved $M^i\, H_i = 0 $ for $F^1$ which has helped to eliminate it from other expressions, and we will be subsequently providing an input of integer flux configuration consisting of only three fluxes, namely $\{H_1, H_2, F^2\}$. Here, the exact value of $N_{\rm flux}^{\text{max}}$ depends on the specifics of the orientifold actions and the brane setting. However, for some estimates, a range for these values can be considered to be, say $N_{\rm flux}^{\text{max}} = 150$ or 200, which appear to be more than the maximum values known for the models built using this CY $3$-fold in the literature; for example $N_{\rm flux}^{\text{max}} = 138$ \cite{Louis:2012nb}. 

To have some estimates, let us quickly recall some scenarios; in the case of orientifold involutions where four D$7$-brane and its image are put on top of the O$7$ components, the divisor with highest GLSM charges can usually lead to larger value of $N_{\rm flux}$. For this purpose, one can take the involution $\sigma: x_5 \to - x_5$ which results in the following two components of the O$7$-planes
\begin{equation}
O7_1 = D_5\coma O7_2 = D_6 \coma
\end{equation}
while there are no O$3$-planes present in the fixed point set. This leads to the following condition for cancellation of the D$3$-brane tadpole \cite{Blumenhagen:2008zz},
\begin{equation}
N_{\rm flux} +N_{\rm gauge} =\frac{\chi(O7)}{4} =\frac{\chi(O7_1) + \chi(O7_2)}{4} = \frac{549+3}{4} = 138\fstop
\end{equation}
On the other hand if we consider the involution to be $\sigma: x_4 \to - x_4$ which reflects the divisor with lower GLSM charge, then the fixed point set has only one O$7$-plane denoted as O$7 = D_4$, and there are no O$3$-points present. Subsequently, one has the following estimates for the $N_{\rm flux}$ value corresponding to the D$7$-brane being on top of O$7$ scenario,
\begin{equation}
N_{\rm flux} +N_{\rm gauge} =\frac{\chi(O7)}{4} =\frac{\chi(O7_4)}{4} = \frac{276}{4} = 69\fstop
\end{equation}
We note that this value is relatively lower than the previous value however it is still not too low as compared to what one often encounters during explicit construction for model building. 

Considering a range of values for $N_{\rm flux}$ we scanned for possible flat flux vacua. In scanning these vacua we have set the magnitude of the independent fluxes, namely $\{F^2, H_1, H_2\}$, to lie within a range $\{1, 300\}$. This means that we have scanned through $27000000$ flux configurations to test the criteria mentioned in Eq. \eqref{eq:final-constraints}. The results are summarized in Table~\ref{tab_vacua-11169}. 

\begin{table}[H]
\centering
\renewcommand\arraystretch{1.2}
\begin{tabular}{|c || *{9}{>{\centering}p{0.7cm}|}c|}
\hline
\# & \multicolumn{10}{c|}{Number of flat flux vacua for $50 \leq N_{\rm flux} \leq 500$}   \\
\hline
   & 50 & 100 & 150 & 200 & 250 & 300 & 350 & 400 & 450 & 500  \\
\hhline{|=#=|=|=|=|=|=|=|=|=|=|}
Vacua &  6 & 21 & 38 & 60 & 86 & 107 & 130 & 153 & 173 & 198 \\
$M^1$=$M^2$ &  1 & 3 & 5 & 8 & 9 & 13 & 16 & 18 & 20 & 21 \\
Physical &  0 & 0 & 1 & 1 & 2 & 2 & 2 & 3 & 3 & 4 \\
\hline  
\end{tabular}
\caption{Statistics of PFFV for Swiss-Cheese CY defined in $\mathbb{W}\CC\PP^4[1, 1, 1, 6, 9]$. }
\label{tab_vacua-11169}
\end{table}
\noindent
In order to zoom a bit more into the flux vacua, we present the details of the 38 PFFV which we find by demanding $0 < N_{\rm flux} \leq 150$.

\begin{center}
\renewcommand*{\arraystretch}{1.0}
\begin{longtable}{|c||c|cc|cc|cc|cc|c|}
 \caption{Details of 38 PFFV obtained by demanding $0 < N_{\rm flux} \leq 150$.}\\
 \hline
 \# &  $N_{\rm flux} $& $F^1 $& $F^2  $& $H_1 $& $H_2 $&  $M^1  $& $M^2  $& $\tilde{p}_{1j}F^j $& $\tilde{p}_{2j}F^j $& $\tilde{p}_{i}F^i$ \\
 \hhline{|=#=|==|==|==|==|=|}
 \endhead
 \label{tab:38vacuaNflux150}1 &  $14$ & $4   $& $-16  $& $-3  $& $1   $&  $\frac{1}{8}  $& $\frac{3}{8}       $& $-6    $& $6     $& $-7     $ \\
2 &  $28$ & $4   $& $-16  $& $-6  $& $2   $&  $\frac{1}{4}  $& $\frac{3}{4}       $& $-6    $& $6     $& $-7     $ \\
3 &  $28$ & $8   $& $-32  $& $-3  $& $1   $&  $\frac{1}{16} $& $\frac{3}{16}      $& $-12   $& $12    $& $-14    $ \\
4 &  $42$ & $4   $& $-16  $& $-9  $& $3   $&  $\frac{3}{8}  $& $\frac{9}{8}       $& $-6    $& $6     $& $-7     $ \\
5 &  $42$ & $12  $& $-48  $& $-3  $& $1   $&  $\frac{1}{24} $& $\frac{1}{8}       $& $-18   $& $18    $& $-21    $ \\
6 &  $42$ & $20  $& $-64  $& $-1  $& $1   $&  $\frac{1}{16} $& $\frac{1}{16}      $& $-6    $& $30    $& $-11    $ \\
7 &  $56$ & $4   $& $-16  $& $-12 $& $4   $&  $\frac{1}{2}  $& $\frac{3}{2}       $& $-6    $& $6     $& $-7     $ \\
8 &  $56$ & $8   $& $-32  $& $-6  $& $2   $&  $\frac{1}{8}  $& $\frac{3}{8}       $& $-12   $& $12    $& $-14    $ \\
9 &  $56$ & $16  $& $-64  $& $-3  $& $1   $&  $\frac{1}{32} $& $\frac{3}{32}      $& $-24   $& $24    $& $-28    $ \\
10 &  $70$ & $4   $& $-16  $& $-15 $& $5   $&  $\frac{5}{8}  $& $\frac{15}{8}      $& $-6    $& $6     $& $-7     $ \\
11 &  $70$ & $20  $& $-80  $& $-3  $& $1   $&  $\frac{1}{40} $& $\frac{3}{40}      $& $-30   $& $30    $& $-35    $ \\
12 &  $78$ & $16  $& $-54  $& $-3  $& $2   $&  $\frac{1}{9}  $& $\frac{1}{6}       $& $-9    $& $24    $& $-13    $ \\
13 &  $78$ & $28  $& $-100 $& $-2  $& $1   $&  $\frac{1}{40} $& $\frac{1}{20}      $& $-24   $& $42    $& $-31    $ \\
14 &  $84$ & $4   $& $-16  $& $-18 $& $6   $&  $\frac{3}{4}  $& $\frac{9}{4}       $& $-6    $& $6     $& $-7     $ \\
15 &  $84$ & $8   $& $-32  $& $-9  $& $3   $&  $\frac{3}{16} $& $\frac{9}{16}      $& $-12   $& $12    $& $-14    $ \\
16 &  $84$ & $12  $& $-48  $& $-6  $& $2   $&  $\frac{1}{12} $& $\frac{1}{4}       $& $-18   $& $18    $& $-21    $ \\
17 &  $84$ & $20  $& $-64  $& $-2  $& $2   $&  $\frac{1}{8}  $& $\frac{1}{8}       $& $-6    $& $30    $& $-11    $ \\
18 &  $84$ & $24  $& $-96  $& $-3  $& $1   $&  $\frac{1}{48} $& $\frac{1}{16}      $& $-36   $& $36    $& $-42    $ \\
19 &  $84$ & $40  $& $-128 $& $-1  $& $1   $&  $\frac{1}{32} $& $\frac{1}{32}      $& $-12   $& $60    $& $-22    $ \\
20 &  $98$ & $4   $& $-16  $& $-21 $& $7   $&  $\frac{7}{8}  $& $\frac{21}{8}      $& $-6    $& $6     $& $-7     $ \\
21 &  $98$ & $28  $& $-112 $& $-3  $& $1   $&  $\frac{1}{56} $& $\frac{3}{56}      $& $-42   $& $42    $& $-49    $ \\
22 &  $112$ & $4   $& $-16  $& $-24 $& $8   $&  $1            $& $3                 $& $-6    $& $6     $& $-7     $ \\
23 &  $112$ & $8   $& $-32  $& $-12 $& $4   $&  $\frac{1}{4}  $& $\frac{3}{4}       $& $-12   $& $12    $& $-14    $ \\
24 &  $112$ & $16  $& $-64  $& $-6  $& $2   $&  $\frac{1}{16} $& $\frac{3}{16}      $& $-24   $& $24    $& $-28    $ \\
25 &  $112$ & $32  $& $-128 $& $-3  $& $1   $&  $\frac{1}{64} $& $\frac{3}{64}      $& $-48   $& $48    $& $-56    $ \\
26 & $114$ & $20  $& $-108 $& $-6  $& $1   $&  $\frac{1}{72} $& $\frac{1}{12}      $& $-72   $& $30    $& $-77    $ \\
27 &  $114$ & $32  $& $-98  $& $-1  $& $2   $&  $\frac{1}{7}  $& $\frac{1}{14}      $& $-3    $& $48    $& $-11    $ \\
28 &  $122$ & $12  $& $-100 $& $-12 $& $1   $&  $\frac{1}{80} $& $\frac{3}{20}      $& $-96   $& $18    $& $-99    $ \\
29 &  $124$ & $16  $& $-50  $& $-3  $& $4   $&  $\frac{2}{5}  $& $\frac{3}{10}      $& $-3    $& $24    $& $-7     $ \\
30 &  $126$ & $4   $& $-16  $& $-27 $& $9   $&  $\frac{9}{8}  $& $\frac{27}{8}      $& $-6    $& $6     $& $-7     $ \\
31 &  $126$ & $12  $& $-48  $& $-9  $& $3   $&  $\frac{1}{8}  $& $\frac{3}{8}       $& $-18   $& $18    $& $-21    $ \\
32 &  $126$ & $20  $& $-64  $& $-3  $& $3   $&  $\frac{3}{16} $& $\frac{3}{16}      $& $-6    $& $30    $& $-11    $ \\
33 &  $126$ & $36  $& $-144 $& $-3  $& $1   $&  $\frac{1}{72} $& $\frac{1}{24}      $& $-54   $& $54    $& $-63    $ \\
34 &  $126$ & $60  $& $-192 $& $-1  $& $1   $&  $\frac{1}{48} $& $\frac{1}{48}      $& $-18   $& $90    $& $-33    $ \\
35 &  $140$ & $4   $& $-16  $& $-30 $& $10  $&  $\frac{5}{4}  $& $\frac{15}{4}      $& $-6    $& $6     $& $-7     $ \\
36 &  $140$ & $8   $& $-32  $& $-15 $& $5   $&  $\frac{5}{16} $& $\frac{15}{16}     $& $-12   $& $12    $& $-14    $ \\
37 &  $140$ & $20  $& $-80  $& $-6  $& $2   $&  $\frac{1}{20} $& $\frac{3}{20}      $& $-30   $& $30    $& $-35    $ \\
38 &  $140$ & $40  $& $-160 $& $-3  $& $1   $&  $\frac{1}{80} $& $\frac{3}{80}      $& $-60   $& $60    $& $-70    $ \\
 \hline
    \end{longtable}     
    \end{center}

\noindent
We note that the flux vacua numbered as 29 in Table \ref{tab:38vacuaNflux150} is the one claimed in \cite{Demirtas:2019sip}. These vacua have the following details on the moduli/dilaton VEVs, 
\begin{equation}
 \begin{array}{rllrllrll}
   F^i                      & = & \{16, -50\}\coma & H_i                 & = & \{-3, 4\}\coma              & M^i          & = & \{2/5, \, 3/10\}\coma\\
   \tilde{p}_{ij} F^j       & = & \{-3, 24\}\coma  & \, \, \tilde{p}_i F^i     & = & -7\coma                     & N_{\rm flux} & = & 124\coma\\
   \langle g_s \rangle^{-1} & = & 6.85505\coma     & \langle U^i \rangle & = & \{2.74202, 2.05652\} \coma  & |W_0|        & = & 2.0482 \cdot 10^{-8}\fstop\\
\end{array}  
\end{equation}
In addition, there was a claim of another physical vacua in \cite{Broeckel:2021uty} with the following flux configuration,
\begin{equation}
\begin{array}{rllrllrll}
   F^i                      & = & \{32, -98\}\coma & H_i                 & = & \{ -1, 2\}\coma            & M^i          & = & \{1/7, \, 1/14\}\coma \\
   \tilde{p}_{ij} F^j       & = & \{-3, 48\}\coma  & \tilde{p}_i F^i     & = & -11\coma                   & N_{\rm flux} & = & 114\coma \\
   \langle g_s \rangle^{-1} & = & 10.5073\coma     & \langle U^i \rangle & = & \{1.50104, 0.750521\}\coma & |W_0|        & = & 0.00060898\fstop  
\end{array}  
\end{equation}
We observe that this solution, which appears at number 27 in Table \ref{tab:38vacuaNflux150} of our collection, lies in the weak coupling regime, however it is not a physical solution. Besides, having differences in the string coupling $(g_s)$ and the $|W_0|$ values as compared to the claim of \cite{Broeckel:2021uty}, we observe that complex structure moduli are not within the physical regime.


\section{Classification of the PFFV for mirror CY with $h^{1,1} =2$}
\label{sec_classification}

Before discussing the statistics of the flat vacua, let us mention some interesting observations which appear from the generic analysis of 39 CY geometries. The very first observation to make is the fact that imposing the orthogonality condition $M^i H_i = 0$ always results in the following simplified form of the $M^i$ flux vector,
\begin{equation}
\label{eq:Mi-via-f}
M^i = f(H_1, H_2, F^2)\left(-\frac{H_2}{H_1},\, 1 \right)^T
\end{equation}
where $f(H_1, H_2, F^2)$ is some function depending on the fluxes $H_1, H_2$ and $F^2$. Therefore, demanding that $M^i$ lies in the K\"ahler cone gives,
\begin{equation}
f(H_1, H_2, F^2) > 0\coma -\frac{H_2}{H_1} > 0\fstop
\end{equation}
This shows that $H_1$ and $H_2$ always need to be of opposite sign. In fact, using the explicit expressions of the function $f(H_1, H_2, F^2)$ for each of the 39 CY geometries, we observe that $M^i H_i = 0$ along with demanding $M^i$'s to be within K\"ahler cone always result in only two choices which are as follows,
\begin{equation}
\{H_1 > 0,\,H_2 < 0,\, F^2 > 0\} \quad \text{or} \quad \{H_1 < 0,\, H_2 > 0,\, F^2 < 0\}\coma
\end{equation}
which we have witnessed in the explicit $\mathbb{W}\CC\PP^4[1,1,1,6,9]$ example as well. This observation is quite helpful in the sense that the same set of 27000000 flux configurations which we have used for the previous model can be imposed for other 38 CY geometries as well. By this we mean that we have the following flux configurations as input:
\begin{equation}
H_1 \in \{-300, -1\}\coma \quad H_2 \in \{1, 300\}\coma \quad F^2 \in \{-300, -1\}\fstop
\end{equation}
Going by this approach is a quite simple test, and does not require developing any algorithm which in the process need to solve Diophantine equations. The search for PFFV boils down to impose these flux configurations on the set of 39 constraints similar to \cref{eq:final-constraints}, which is derived for each of the examples. Also, K\"ahler cone conditions being already imposed in constructing the input data helps us to avoid any unwanted situation in the numerical search which could arise, for example due to a flux configuration where (some) conditions become singular. In other words, none of the quantities $M^i, \, \tilde{p}_{ij}F^j$ and $\tilde{p}_i F^i$ or even the tadpole expression $N_{\rm flux}$ for that matter (like \cref{eq:tadpole-11169}) gets ill-defined within this choice of $\{F_3, H_3\}$ fluxes.

\begin{center}
\renewcommand\arraystretch{1}
   \begin{longtable}{|c||c|c|c|c|c|c|c|c|c|c|}
\caption{Number of perturbatively flat flux vacua for $h^{1,1}({\rm CY}) =2$ with a fixed value of $N_{\rm flux}$. These flat vacua are obtained by choosing the three independent fluxes $|F^2|, |H_1|, |H_2|$ in the range $\{1, 300\}$ while the remaining flux $F^1$ is obtained by solving the orthogonality condition $M^i H_i  = 0$. }\\
\hline
CY & \multicolumn{10}{c|}{Number of flat flux vacua for $50 \leq N_{\rm flux} \leq 500$} \\
\cline{2-11}
 \# & 50 & 100 & 150 & 200 & 250 & 300 & 350 & 400 & 450 & 500 \\
\hhline{|=#=|=|=|=|=|=|=|=|=|=|}
		\endhead
		\label{tab_fluxdata}1 & 101 & 261 & 446 & 648 & 845 & 1090 & 1275 & 1509 & 1725 & 1905  \\
 2 & 6 & 21 & 38 & 60 & 86 & 107 & 132 & 153 & 173 & 198  \\
 3 & 1 & 4 & 7 & 12 & 16 & 22 & 28 & 35 & 39 & 45  \\
 4 & 1 & 4 & 7 & 12 & 16 & 22 & 28 & 35 & 38 & 44  \\
 5 & 1 & 4 & 7 & 12 & 16 & 22 & 28 & 35 & 38 & 44  \\
 6 & 101 & 261 & 446 & 648 & 845 & 1090 & 1275 & 1509 & 1725 & 1905  \\
 7 & 1 & 3 & 5 & 10 & 14 & 17 & 24 & 30 & 31 & 41  \\
 8 & 42 & 159 & 332 & 534 & 735 & 981 & 1200 & 1473 & 1768 & 2017  \\
 9 & 42 & 159 & 332 & 534 & 735 & 981 & 1200 & 1473 & 1768 & 2017  \\
 10 & 42 & 159 & 332 & 535 & 757 & 1041 & 1294 & 1609 & 1956 & 2276  \\
 11 & 32 & 118 & 246 & 393 & 552 & 765 & 946 & 1176 & 1426 & 1648  \\
 12 & 2 & 18 & 41 & 79 & 119 & 165 & 220 & 278 & 341 & 411  \\
 13 & 3 & 11 & 21 & 37 & 47 & 60 & 74 & 97 & 105 & 120  \\
 14 & 0 & 1 & 5 & 8 & 13 & 16 & 22 & 28 & 28 & 38  \\
 15 & 0 & 1 & 4 & 9 & 11 & 14 & 21 & 31 & 32 & 39  \\
 16 & 1 & 3 & 5 & 10 & 14 & 16 & 20 & 27 & 28 & 33  \\
 17 & 0 & 0 & 0 & 0 & 0 & 1 & 1 & 1 & 1 & 2 \\
 18 & 0 & 1 & 4 & 4 & 7 & 14 & 14 & 17 & 22 & 23  \\
 19 & 7 & 26 & 49 & 77 & 107 & 141 & 176 & 213 & 244 & 278  \\
 20 & 1 & 4 & 9 & 15 & 17 & 29 & 35 & 46 & 50 & 57 \\
 21 & 4 & 14 & 23 & 41 & 52 & 71 & 84 & 106 & 121 & 139  \\
 22 & 0 & 2 & 3 & 6 & 7 & 15 & 17 & 21 & 23 & 29 \\
 23 & 0 & 2 & 3 & 6 & 7 & 15 & 17 & 21 & 23 & 29 \\
 24 & 0 & 2 & 3 & 6 & 7 & 15 & 17 & 21 & 23 & 29 \\
 25 & 5 & 17 & 32 & 48 & 70 & 91 & 115 & 134 & 152 & 176  \\
 26 & 5 & 14 & 29 & 42 & 58 & 78 & 90 & 106 & 124 & 141  \\
 27 & 0 & 0 & 0 & 0 & 0 & 0 & 1 & 1 & 1 & 1 \\
 28 & 0 & 0 & 0 & 0 & 0 & 0 & 1 & 1 & 1 & 1 \\
 29 & 1 & 4 & 9 & 15 & 17 & 29 & 35 & 46 & 50 & 57  \\
 30 & 18 & 76 & 164 & 270 & 377 & 507 & 627 & 772 & 930 & 1065  \\
 31 & 18 & 76 & 164 & 270 & 377 & 507 & 627 & 772 & 930 & 1065  \\
 32 & 18 & 76 & 164 & 271 & 391 & 544 & 685 & 856 & 1047 & 1226  \\
 33 & 376 & 1135 & 2141 & 3256 & 4350 & 5766 & 6976 & 8516 & 10114 & 11208 \\
 34 & 31 & 115 & 242 & 387 & 547 & 757 & 942 & 1172 & 1421 & 1646  \\
 35 & 0 & 2 & 9 & 11 & 21 & 24 & 33 & 41 & 50 & 57  \\
 36 & 16 & 47 & 88 & 132 & 180 & 228 & 291 & 345 & 386 & 434  \\
 37 & 6 & 21 & 38 & 60 & 87 & 113 & 145 & 172 & 200 & 234  \\
 38 & 6 & 20 & 40 & 62 & 88 & 112 & 142 & 171 & 201 & 233  \\
 39 & 6 & 21 & 38 & 60 & 86 & 107 & 132 & 153 & 173 & 198  \\
\hline  
\end{longtable}
\end{center}

\subsection{Class 1: Swiss-cheese type}

For CY $3$-folds with $h^{1,1}=2$, there are 22 geometries which are of Swiss-cheese type. These are numbered as the second column of Table \ref{tab_swiss-cheese-PFFV}. 
Following the strategy and steps we demonstrated for the $\mathbb{W}\CC\PP^4[1,1,1,6,9]$ model, we compute the PFFV for all the 22 examples of Swiss-cheese type, and the corresponding number of vacua for a given value of $N_{\rm flux}$ is given in the following Table \ref{tab_swiss-cheese-PFFV}. 

\begin{center}
\renewcommand\arraystretch{1}
   \begin{longtable}{|c||c|c|c|c|c|c|c|c|c|c||c|}
   \caption{Number of PFFV with a fixed value of $N_{\rm flux}$ for the Swiss-cheese type mirror CY with $h^{1,1}({\rm CY}) =2$. These flat vacua are obtained by choosing the three independent fluxes $|F^2|, |H_1|, |H_2|$ in the range $\{1, 300\}$ while the remaining flux $F^1$ is obtained by solving the orthogonality condition $M^i H_i  = 0$. }\\
\hline
Sr. & CY & \multicolumn{10}{c|}{Number of flat flux vacua for $50 \leq N_{\rm flux} \leq 500$} \\
\cline{3-12}
 \# & \# & 50 & 100 & 150 & 200 & 250 & 300 & 350 & 400 & 450 & 500 \\
\hhline{|=#=|=|=|=|=|=|=|=|=|=#=|}
		\endhead
\label{tab_swiss-cheese-PFFV}1 & 2 & 6 & 21 & 38 & 60 & 86 & 107 & 132 & 153 & 173 & 198 \\
 2 & 3 & 1 & 4 & 7 & 12 & 16 & 22 & 28 & 35 & 39 & 45 \\
 3 & 4 & 1 & 4 & 7 & 12 & 16 & 22 & 28 & 35 & 38 & 44 \\
 4 & 5 & 1 & 4 & 7 & 12 & 16 & 22 & 28 & 35 & 38 & 44 \\
 5 & 7 & 1 & 3 & 5 & 10 & 14 & 17 & 24 & 30 & 31 & 41 \\
 6 & 13 & 3 & 11 & 21 & 37 & 47 & 60 & 74 & 97 & 105 & 120 \\
 7 & 18 & 0 & 1 & 4 & 4 & 7 & 14 & 14 & 17 & 22 & 23 \\
 8 & 19 & 7 & 26 & 49 & 77 & 107 & 141 & 176 & 213 & 244 & 278 \\
 9 & 21 & 4 & 14 & 23 & 41 & 52 & 71 & 84 & 106 & 121 & 139 \\
 10 & 22 & 0 & 2 & 3 & 6 & 7 & 15 & 17 & 21 & 23 & 29 \\
 11 & 23 & 0 & 2 & 3 & 6 & 7 & 15 & 17 & 21 & 23 & 29 \\
 12 & 24 & 0 & 2 & 3 & 6 & 7 & 15 & 17 & 21 & 23 & 29 \\
 13 & 25 & 5 & 17 & 32 & 48 & 70 & 91 & 115 & 134 & 152 & 176 \\
 14 & 26 & 5 & 14 & 29 & 42 & 58 & 78 & 90 & 106 & 124 & 141 \\
 15 & 27 & 0 & 0 & 0 & 0 & 0 & 0 & 1 & 1 & 1 & 1 \\
 16 & 28 & 0 & 0 & 0 & 0 & 0 & 0 & 1 & 1 & 1 & 1 \\
 17 & 29 & 1 & 4 & 9 & 15 & 17 & 29 & 35 & 46 & 50 & 57 \\
 18 & 35 & 0 & 2 & 9 & 11 & 21 & 24 & 33 & 41 & 50 & 57 \\
 19 & 36 & 16 & 47 & 88 & 132 & 180 & 228 & 291 & 345 & 386 & 434 \\
 20 & 37 & 6 & 21 & 38 & 60 & 87 & 113 & 145 & 172 & 200 & 234 \\
 21 & 38 & 6 & 20 & 40 & 62 & 88 & 112 & 142 & 171 & 201 & 233 \\
 22 & 39 & 6 & 21 & 38 & 60 & 86 & 107 & 132 & 153 & 173 & 198 \\
\hline  
\end{longtable}
\end{center}

\begin{figure}[!htp]
 \centering
      \pgfplotstableread{
Label N50 N100 N150 N200 N250 N300 N350 N400 N450 N500 topper
2  6 15 17 22 26 21 25 21 20 25 0
3   1 3 3 5 4 6 6 7 4 6 0
4   1 3 3 5 4 6 6 7 3 6 0
5   1 3 3 5 4 6 6 7 3 6 0
7   1 2 2 5 4 3 7 6 1 10 0
13  3 8 10 16 10 13 14 23 8 15 0
18  0 1 3 0 3 7 0 3 5 1 0
19  7 19 23 28 30 34 35 37 31 34 0
21  4 10 9 18 11 19 13 22 15 18 0
22  0 2 1 3 1 8 2 4 2 6 0
23  0 2 1 3 1 8 2 4 2 6 0
24  0 2 1 3 1 8 2 4 2 6 0
25  5 12 15 16 22 21 24 19 18 24 0
26  5 9 15 13 16 20 12 16 18 17 0
27  0 0 0 0 0 0 1 0 0 0 0
28  0 0 0 0 0 0 1 0 0 0 0
29  1 3 5 6 2 12 6 11 4 7 0
35  0 2 7 2 10 3 9 8 9 7 0
36  16 31 41 44 48 48 63 54 41 48 0
37  6 15 17 22 27 26 32 27 28 34 0
38  6 14 20 22 26 24 30 29 30 32 0
39  6 15 17 22 26 21 25 21 20 25 0
    }\testdata
         \begin{tikzpicture}[scale=1]
    \begin{axis}[
       width=\textwidth,
        ybar stacked,
        ymin=0,
        ymax=450,
        xtick={data},
        legend style={cells={anchor=west}, legend pos=north east},
        reverse legend=false, 
        xticklabels from table={\testdata}{Label},
        xticklabel style={text width=2cm,align=center},
      bar width=0.032\textwidth,
      xlabel={CY \#},
		ylabel={\# flat vacua},
    ]
    \addplot [fill=col1,ybar]
            table [y=N50, meta=Label, x expr=\coordindex]
                {\testdata};
                    \addlegendentry{50}
        \addplot [fill=col2,ybar]
            table [y=N100, meta=Label, x expr=\coordindex]
                {\testdata};
                    \addlegendentry{100}
        \addplot [fill=col3]
            table [y=N150, meta=Label, x expr=\coordindex]
                {\testdata};
                    \addlegendentry{150}
    \addplot [fill=col4,ybar]
            table [y=N200, meta=Label, x expr=\coordindex]
                {\testdata};
                    \addlegendentry{200}
        \addplot [fill=col5,ybar]
            table [y=N250, meta=Label, x expr=\coordindex]
                {\testdata};
                    \addlegendentry{250}
        \addplot [fill=col6]
            table [y=N300, meta=Label, x expr=\coordindex]
                {\testdata};
                    \addlegendentry{300}
    \addplot [fill=col7,ybar]
            table [y=N350, meta=Label, x expr=\coordindex]
                {\testdata};
                    \addlegendentry{350}
        \addplot [fill=col8,ybar]
            table [y=N400, meta=Label, x expr=\coordindex]
                {\testdata};
                    \addlegendentry{400}
        \addplot [fill=col9]
            table [y=N450, meta=Label, x expr=\coordindex]
                {\testdata};
                    \addlegendentry{450}
    \addplot [fill=col10
    ]
            table [y=N500, meta=Label, x expr=\coordindex]
                {\testdata};
                    \addlegendentry{500}
    \end{axis}
    \end{tikzpicture}
 \caption{Statistics of PFFV for Swiss-cheese models.}
\label{fig_upKKLT}
\end{figure}
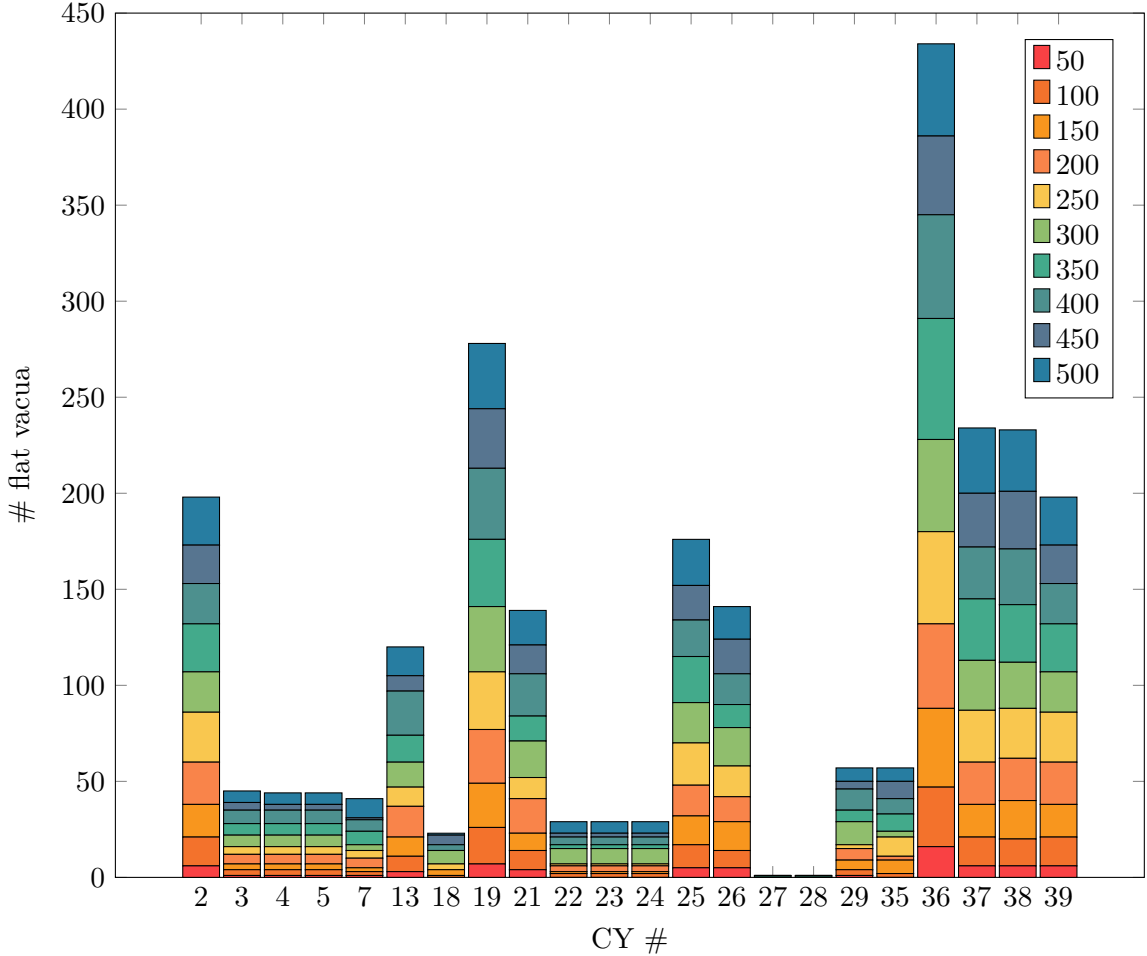

\subsection{Class 2: K3-fibered type}
For CY $3$-fold with $h^{1,1}=2$, there are 10 geometries which are K3-fibered. These are numbered as the second column of Table \ref{tab_K3-fibered-PFFV} where we also present the corresponding number of PFFV configurations. 

\begin{center}
\renewcommand\arraystretch{1}
   \begin{longtable}{|c||c|c|c|c|c|c|c|c|c|c||c|}
\caption{Number of PFFV with a fixed value of $N_{\rm flux}$ for the K3-fibered mirror CY with $h^{1,1}({\rm CY}) =2$. These flat vacua are obtained by choosing the three independent fluxes $|F^2|, |H_1|, |H_2|$ in the range $\{1, 300\}$ while the remaining flux $F^1$ is obtained by solving the orthogonality condition $M^i H_i  = 0$. }\\
\hline
Sr. & CY & \multicolumn{10}{c|}{Number of flat flux vacua for $50 \leq N_{\rm flux} \leq 500$} \\
\cline{3-12}
 \# & \# & 50 & 100 & 150 & 200 & 250 & 300 & 350 & 400 & 450 & 500 \\
\hhline{|=#=|=|=|=|=|=|=|=|=|=|=|}
		\endhead
\label{tab_K3-fibered-PFFV}1 & 8  & 42 & 159 & 332 & 534 & 735 & 981 & 1200 & 1473 & 1768 & 2017 \\
 2 & 9  & 42 & 159 & 332 & 534 & 735 & 981 & 1200 & 1473 & 1768 & 2017 \\
 3 & 10 & 42 & 159 & 332 & 535 & 757 & 1041 & 1294 & 1609 & 1956 & 2276 \\
 4 & 11 & 32 & 118 & 246 & 393 & 552 & 765 & 946 & 1176 & 1426 & 1648 \\
 5 & 12 & 2 & 18 & 41 & 79 & 119 & 165 & 220 & 278 & 341 & 411 \\
 6 & 30 & 18 & 76 & 164 & 270 & 377 & 507 & 627 & 772 & 930 & 1065 \\
 7 & 31 & 18 & 76 & 164 & 270 & 377 & 507 & 627 & 772 & 930 & 1065  \\
 8 & 32 & 18 & 76 & 164 & 271 & 391 & 544 & 685 & 856 & 1047 & 1226  \\
 9 & 33 & 376 & 1135 & 2141 & 3256 & 4350 & 5766 & 6976 & 8516 & 10114 & 11208 \\
 10 & 34 & 31 & 115 & 242 & 387 & 547 & 757 & 942 & 1172 & 1421 & 1646  \\
\hline  
\end{longtable}
\end{center}

\begin{figure}[!htp]
 \centering
      \pgfplotstableread{
Label N50 N100 N150 N200 N250 N300 N350 N400 N450 N500 topper
8    42  117  173  202  201  246  219  273  295  249 0
9    42  117  173  202  201  246  219  273  295  249 0
10   42  117  173  203  222  284  253  315  347  320 0
11   32  86  128  147  159  213  181  230  250  222 0
12   2  16  23  38  40  46  55  58  63  70 0
30   18  58  88  106  107  130  120  145  158  135 0
31   18  58  88  106  107  130  120  145  158  135 0
32   18  58  88  107  120  153  141  171  191  179 0
33   376  759  1006  1115  1094  1416  1210  1540  1598  1094 0
34   31  84  127  145  160  210  185  230  249  225 0
    }\testdata
         \begin{tikzpicture}[scale=1]
    \begin{axis}[
       width=0.8\textwidth,
        ybar stacked,
        ymin=0,
        ymax=11600,
        xtick={data},
        legend style={cells={anchor=west}, legend pos=north east},
        reverse legend=false, 
        xticklabels from table={\testdata}{Label},
        xticklabel style={text width=2cm,align=center},
      bar width=0.032\textwidth,
      xlabel={CY \#},
		ylabel={\# flat vacua},
    ]
    \addplot [fill=col1,ybar]
            table [y=N50, meta=Label, x expr=\coordindex]
                {\testdata};
                    \addlegendentry{50}
        \addplot [fill=col2,ybar]
            table [y=N100, meta=Label, x expr=\coordindex]
                {\testdata};
                    \addlegendentry{100}
        \addplot [fill=col3]
            table [y=N150, meta=Label, x expr=\coordindex]
                {\testdata};
                    \addlegendentry{150}
    \addplot [fill=col4,ybar]
            table [y=N200, meta=Label, x expr=\coordindex]
                {\testdata};
                    \addlegendentry{200}
        \addplot [fill=col5,ybar]
            table [y=N250, meta=Label, x expr=\coordindex]
                {\testdata};
                    \addlegendentry{250}
        \addplot [fill=col6]
            table [y=N300, meta=Label, x expr=\coordindex]
                {\testdata};
                    \addlegendentry{300}
    \addplot [fill=col7,ybar]
            table [y=N350, meta=Label, x expr=\coordindex]
                {\testdata};
                    \addlegendentry{350}
        \addplot [fill=col8,ybar]
            table [y=N400, meta=Label, x expr=\coordindex]
                {\testdata};
                    \addlegendentry{400}
        \addplot [fill=col9]
            table [y=N450, meta=Label, x expr=\coordindex]
                {\testdata};
                    \addlegendentry{450}
    \addplot [fill=col10
    ]
            table [y=N500, meta=Label, x expr=\coordindex]
                {\testdata};
                    \addlegendentry{500}
    \end{axis}
    \end{tikzpicture}
\caption{Statistics of PFFV for K3-fibered models.}
\label{fig:PFFV-K3fibered}
\end{figure}
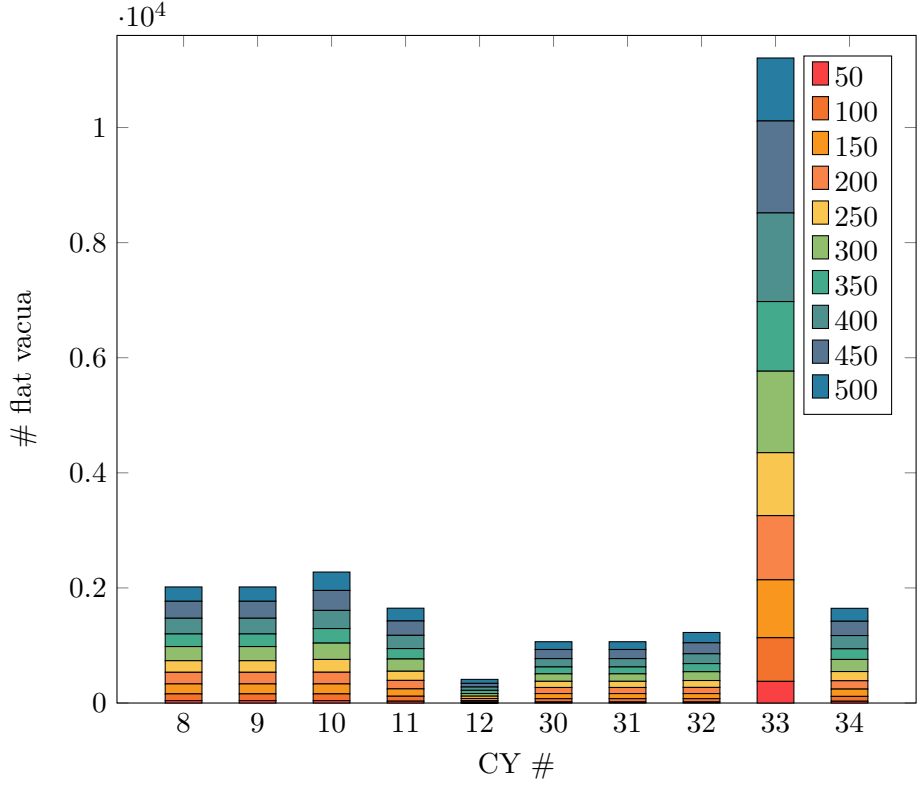

\noindent
Let us observe that there is one example for which the number of flat vacua is much larger than other models. This corresponds to the CY geometry numbered as 33, i.e., model ${\cal M}_{2, 33}$ in Table \ref{tab_cydata-h11eq2}. We will take this to study the subsequent analysis of the physical vacua.

\subsection{Class 3: Hybrid type}

For CY $3$-folds with $h^{1,1}=2$, there are 7 geometries which are neither K3-fibered type nor the Swiss-cheese type. However, we find that these are examples which share properties from both types and that is why we call them as ``Hybrid type". These are numbered as the second column in Table \ref{tab_hybrid-fibered-PFFV} which also presents their number of PFFV configurations.

\begin{center}
\renewcommand\arraystretch{1}
   \begin{longtable}{|c||c|c|c|c|c|c|c|c|c|c||c|}
   \caption{Number of PFFV with a fixed value of $N_{\rm flux}$ for the ``Hybrid type" mirror CY with $h^{1,1}({\rm CY}) =2$. These flat vacua are obtained by choosing the three independent fluxes $|F^2|, |H_1|, |H_2|$ in the range $\{1, 300\}$ while the remaining flux $F^1$ is obtained by solving the orthogonality condition $M^i H_i  = 0$. }\\
\hline
Sr. & CY & \multicolumn{10}{c|}{Number of flat flux vacua for $50 \leq N_{\rm flux} \leq 500$} \\
\cline{3-12}
 \# & \# & 50 & 100 & 150 & 200 & 250 & 300 & 350 & 400 & 450 & 500 \\
\hhline{|=#=|=|=|=|=|=|=|=|=|=|=|}
		\endhead
\label{tab_hybrid-fibered-PFFV}1 & 1 & 101 & 261 & 446 & 648 & 845 & 1090 & 1275 & 1509 & 1725 & 1905 \\
2 & 6 & 101 & 261 & 446 & 648 & 845 & 1090 & 1275 & 1509 & 1725 & 1905 \\
3 & 14 & 0 & 1 & 5 & 8 & 13 & 16 & 22 & 28 & 28 & 38  \\
4 & 15 & 0 & 1 & 4 & 9 & 11 & 14 & 21 & 31 & 32 & 39 \\
5 & 16 & 1 & 3 & 5 & 10 & 14 & 16 & 20 & 27 & 28 & 33 \\
6 & 17 & 0 & 0 & 0 & 0 & 0 & 1 & 1 & 1 & 1 & 2 \\
7 & 20 & 1 & 4 & 9 & 15 & 17 & 29 & 35 & 46 & 50 & 57 \\
\hline  
\end{longtable}
\end{center}

\newpage

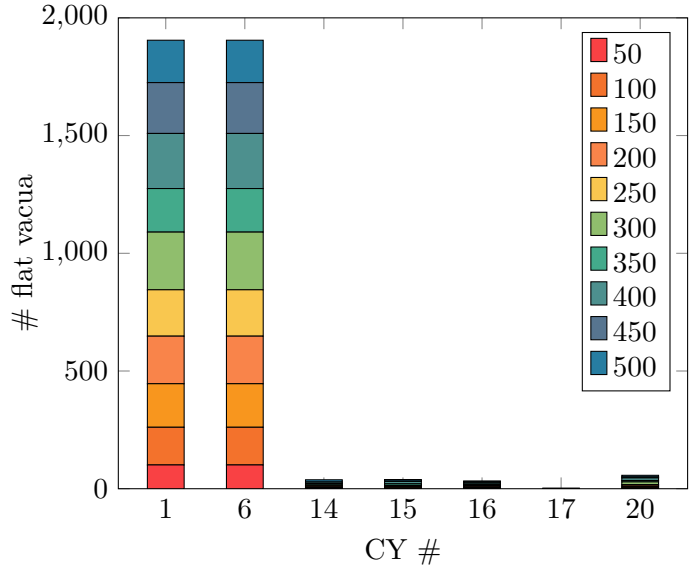
\begin{figure}[!htp]
 \centering
      \pgfplotstableread{
Label N50 N100 N150 N200 N250 N300 N350 N400 N450 N500 topper
1  101  160  185  202  197  245  185  234  216  180 0
6   101  160  185  202  197  245  185  234  216  180 0
14  0  1  4  3  5  3  6  6  0  10 0
15  0  1  3  5  2  3  7  10  1  7 0
16  1  2  2  5  4  2  4  7  1  5 0
17  0  0  0  0  0  1  0  0  0  1 0
20  1  3  5  6  2  12  6  11  4  7 0
    }\testdata
         \begin{tikzpicture}[scale=1]
    \begin{axis}[
       width=0.6\textwidth,
        ybar stacked,
        ymin=0,
        ymax=2000,
        xtick={data},
        legend style={cells={anchor=west}, legend pos=north east},
        reverse legend=false, 
        xticklabels from table={\testdata}{Label},
        xticklabel style={text width=2cm,align=center},
      bar width=0.032\textwidth,
      xlabel={CY \#},
		ylabel={\# flat vacua},
    ]
    \addplot [fill=col1,ybar]
            table [y=N50, meta=Label, x expr=\coordindex]
                {\testdata};
                    \addlegendentry{50}
        \addplot [fill=col2,ybar]
            table [y=N100, meta=Label, x expr=\coordindex]
                {\testdata};
                    \addlegendentry{100}
        \addplot [fill=col3]
            table [y=N150, meta=Label, x expr=\coordindex]
                {\testdata};
                    \addlegendentry{150}
    \addplot [fill=col4,ybar]
            table [y=N200, meta=Label, x expr=\coordindex]
                {\testdata};
                    \addlegendentry{200}
        \addplot [fill=col5,ybar]
            table [y=N250, meta=Label, x expr=\coordindex]
                {\testdata};
                    \addlegendentry{250}
        \addplot [fill=col6]
            table [y=N300, meta=Label, x expr=\coordindex]
                {\testdata};
                    \addlegendentry{300}
    \addplot [fill=col7,ybar]
            table [y=N350, meta=Label, x expr=\coordindex]
                {\testdata};
                    \addlegendentry{350}
        \addplot [fill=col8,ybar]
            table [y=N400, meta=Label, x expr=\coordindex]
                {\testdata};
                    \addlegendentry{400}
        \addplot [fill=col9]
            table [y=N450, meta=Label, x expr=\coordindex]
                {\testdata};
                    \addlegendentry{450}
    \addplot [fill=col10
    ]
            table [y=N500, meta=Label, x expr=\coordindex]
                {\testdata};
                    \addlegendentry{500}
    \end{axis}
    \end{tikzpicture}
\caption{Statistics of PFFV for remaining models.}
\label{fig:PFFVremainmodel}
\end{figure}

\noindent
To begin with, we find that this category of examples has a mixed behavior of number of PFFV as compared to the K3-fibered and the Swiss-cheese type examples. Some examples show many flat vacua like K3-fibered cases, while some examples have relatively less number of flat vacua like the Swiss-cheese cases. Similar types of hybrid nature of such examples was also observed in \cite{AbdusSalam:2020ywo} where they have been classified as ``Hard" examples. However, some detailed investigation shows that most of the PFFV found for CY number 1 and 6 are such that the flux vector $M^i$ satisfies $M^1 = M^2$. This happens due to the presence of an underlying symmetry in the CY $3$-fold, and such vacua have some interesting features of being what we call exponentially flat. We will get back to this aspect later on.

\section{Stabilizing the PFFV for low $|W_0|$}
\label{sec_lowW0}

\subsection{Removing the redundancy in PFFV counts}
\label{sec:step2}
By following Step 1 in Section \ref{sec:step1} in the methodology, which is basically the recipe proposed in \cite{Demirtas:2019sip}, it turns out that the number of PFFV is quite large. However, there is the possibility of a significant amount of over counting in the number of equivalent flat vacua. 

\subsubsection{Flux scaling redundancies}
Given that the underlying requirement for finding the PFFV configurations is to impose the vanishing of $W_{\rm poly}$ in \cref{eq:Wpolysimp} and its derivatives in \cref{eq:dWpolysimp}, we observe that one can demand the following relations to avoid several equivalent flux vacua related up to scaling of $F^i$ and $H_i$ fluxes,
\begin{subequations}
\begin{align} 
 \GCD\left(\{H_i\}\right) &= 1\coma  \label{eq:gcdC1}\\
 \GCD\left(\{F^i, \, H_i\}\right) &= 1\fstop \label{eq:gcdC2}
\end{align}
\label{eq:gcd1}
\end{subequations}
\noindent
The above two requirements are generic and for our particular case of two field models, we need $H_1$ and $H_2$ to be coprime. However, note the fact that there is no such constraint for the $F^1$ and $F^2$ fluxes.

\noindent
To illustrate about these conditions, let us look at the PFFV configurations corresponding to $\mathbb{W}\CC\PP^4[1,1,1,6,9]$ model listed in Table \ref{tab:38vacuaNflux150}. There are in total 38 PFFV configurations corresponding to $0 < N_{\rm flux} \leq 150$, however, we find that 19 of those satisfy \eqref{eq:gcdC1} while 26 of those satisfy \eqref{eq:gcdC2}. 

We find that imposing these two constraints in \cref{eq:gcdC1,eq:gcdC2} significantly reduces the number of PFFV configurations, and we summarize the same in the Table \ref{tab_reducedfluxdata1}. 

\begin{center}
\renewcommand\arraystretch{1}
   \begin{longtable}{|c||c|c|c|c|c|c|c|c|c|c|}
\caption{Reduced number of perturbatively flat flux vacua (listed in Table \ref{tab_fluxdata}) after being subjected to the conditions of \cref{eq:gcdC1,eq:gcdC2}. }\\
\hline
CY & \multicolumn{10}{c|}{Reduced number of flat flux vacua for $50 \leq N_{\rm flux} \leq 500$} \\
\cline{2-11}
 \# & 50 & 100 & 150 & 200 & 250 & 300 & 350 & 400 & 450 & 500\\
\hhline{|=#=|=|=|=|=|=|=|=|=|=|}
		\endhead
		\label{tab_reducedfluxdata1}1 & 35 & 82 & 129 & 176 & 219 & 278 & 297 & 325 & 339 & 348\\
 2 & 4 & 11 & 19 & 27 & 38 & 41 & 44 & 47 & 49 & 50\\
 3 & 1 & 3 & 5 & 7 & 9 & 11 & 14 & 17 & 18 & 18\\
 4 & 1 & 3 & 5 & 7 & 9 & 11 & 14 & 17 & 17 & 17\\
 5 & 1 & 3 & 5 & 7 & 9 & 11 & 14 & 17 & 17 & 17\\
 6 & 35 & 82 & 129 & 176 & 219 & 278 & 297 & 325 & 339 & 348\\
 7 & 1 & 2 & 3 & 5 & 6 & 8 & 10 & 12 & 12 & 15\\
 8 & 23 & 75 & 145 & 220 & 282 & 342 & 394 & 459 & 526 & 585\\
 9 & 23 & 75 & 145 & 220 & 282 & 342 & 394 & 459 & 526 & 585\\
 10 & 23 & 75 & 145 & 221 & 304 & 402 & 488 & 594 & 705 & 811\\
 11 & 19 & 59 & 112 & 170 & 231 & 307 & 372 & 451 & 532 & 610\\
 12 & 2 & 12 & 26 & 45 & 65 & 87 & 111 & 136 & 165 & 192\\
 13 & 2 & 6 & 10 & 16 & 20 & 23 & 26 & 31 & 32 & 33\\
 14 & 0 & 1 & 4 & 6 & 9 & 10 & 13 & 15 & 15 & 19\\
 15 & 0 & 1 & 3 & 7 & 8 & 9 & 13 & 18 & 19 & 23\\
 16 & 1 & 2 & 3 & 5 & 6 & 7 & 8 & 10 & 10 & 10\\
 17 & 0 & 0 & 0 & 0 & 0 & 1 & 1 & 1 & 1 & 2\\
 18 & 0 & 1 & 3 & 3 & 5 & 9 & 9 & 10 & 11 & 12\\
 19 & 5 & 14 & 24 & 34 & 45 & 56 & 67 & 77 & 84 & 91\\
 20 & 1 & 3 & 6 & 9 & 10 & 16 & 19 & 23 & 25 & 28\\
 21 & 3 & 8 & 12 & 18 & 22 & 29 & 34 & 40 & 45 & 49\\
 22 & 0 & 1 & 2 & 3 & 4 & 7 & 8 & 9 & 10 & 12\\
 23 & 0 & 1 & 2 & 3 & 4 & 7 & 8 & 9 & 10 & 12\\
 24 & 0 & 1 & 2 & 3 & 4 & 7 & 8 & 9 & 10 & 12\\
 25 & 3 & 8 & 15 & 20 & 30 & 36 & 44 & 48 & 50 & 55\\
 26 & 3 & 6 & 12 & 17 & 22 & 28 & 31 & 33 & 34 & 36\\
 27 & 0 & 0 & 0 & 0 & 0 & 0 & 1 & 1 & 1 & 1\\
 28 & 0 & 0 & 0 & 0 & 0 & 0 & 1 & 1 & 1 & 1\\
 29 & 1 & 3 & 6 & 9 & 10 & 16 & 19 & 23 & 25 & 28\\
 30 & 14 & 50 & 98 & 152 & 198 & 245 & 286 & 337 & 390 & 434\\
 31 & 14 & 50 & 98 & 152 & 198 & 245 & 286 & 337 & 390 & 434\\
 32 & 14 & 50 & 98 & 153 & 212 & 282 & 344 & 420 & 503 & 581\\
 33 & 188 & 505 & 886 & 1302 & 1705 & 2201 & 2645 & 3167 & 3697 & 3995\\
 34 & 22 & 69 & 132 & 200 & 274 & 363 & 440 & 535 & 631 & 726\\
 35 & 0 & 2 & 6 & 7 & 11 & 12 & 16 & 19 & 21 & 23\\
 36 & 7 & 18 & 31 & 43 & 57 & 69 & 86 & 94 & 98 & 103\\
 37 & 4 & 11 & 19 & 27 & 39 & 47 & 57 & 66 & 75 & 85\\
 38 & 4 & 10 & 20 & 28 & 38 & 45 & 55 & 63 & 73 & 83\\
 39 & 4 & 11 & 19 & 27 & 38 & 41 & 44 & 47 & 49 & 50\\
\hline  
\end{longtable}
\end{center}
In addition to the flux scaling redundancies, we have found in our numerical analysis that there are PFFV configurations with $M^1 = M^2$ which do not result in physical vacua, and these should also be isolated.

\subsubsection{Removing the PFFV with $M^1 = M^2$}

\begin{center}
\renewcommand{\arraystretch}{1.1}
  \begin{longtable}{|c|cccc||c|cccc|}
  \caption{Number of PFFV configurations with $M^1 =M^2$ along with the sign($\cal A$) where ${\cal A}$ is defined in Eq. \eqref{eq:calA}. }\\
    \hline
CY \# &$a$ & $M^1 = M^2$ & ${\cal A} > 0$ & ${\cal A} < 0$ & CY \# &$a$ & $M^1 = M^2$ & ${\cal A} > 0$ & ${\cal A} < 0$ \\   
\hhline{|=|====#=|====|}
  \endhead
  \label{tab:M1M2values}$1  $&$1$             & $1321 $& $0   $& $0   $& $21  $&$\frac{13}{18}$ & $85  $& $0   $& $85  $\\
 $2  $&$4$             & $21   $& $21  $& $0   $& $22  $&$\frac{21}{16}$ & $2   $& $2   $& $0   $\\
 $3  $&$\frac{11}{6}$  & $23   $& $0   $& $23  $& $23  $&$\frac{21}{16}$ & $2   $& $2   $& $0   $\\
 $4  $&$2$             & $23   $& $0   $& $23  $& $24  $&$\frac{21}{16}$ & $2   $& $2   $& $0   $\\
 $5  $&$2$             & $23   $& $0   $& $23  $& $25  $&$\frac{3}{2}$   & $121 $& $121 $& $0   $\\
 $6  $&$\frac{1}{3}$   & $1321 $& $0   $& $0   $& $26  $&$\frac{7}{4}$   & $111 $& $0   $& $111 $\\
 $7  $&$\frac{9}{10}$  & $32   $& $0   $& $32  $& $27  $&$\frac{37}{28}$ & $0   $& $0   $& $0   $\\
 $8  $&$1$             & $167  $& $167 $& $0   $& $28  $&$\frac{37}{28}$ & $0   $& $0   $& $0   $\\
 $9  $&$1$             & $167  $& $167 $& $0   $& $29  $&$\frac{37}{28}$ & $5   $& $5   $& $0   $\\
 $10 $&$\frac{1}{4}$   & $196  $& $196 $& $0   $& $30  $&$2$             & $98  $& $98  $& $0   $\\
 $11 $&$\frac{1}{4}$   & $123  $& $123 $& $0   $& $31  $&$2$             & $98  $& $98  $& $0   $\\
 $12 $&$\frac{1}{4}$   & $38   $& $38  $& $0   $& $32  $&$\frac{1}{2}$   & $123 $& $123 $& $0   $\\
 $13 $&$1$             & $64   $& $64  $& $0   $& $33  $&$\frac{1}{2}$   & $860 $& $860 $& $0   $\\
 $14 $&$\frac{13}{46}$ & $2    $& $0   $& $2   $& $34  $&$\frac{1}{2}$   & $160 $& $160 $& $0   $\\
 $15 $&$\frac{11}{34}$ & $3    $& $0   $& $3   $& $35  $&$\frac{3}{2}$   & $13  $& $0   $& $13  $\\
 $16 $&$\frac{5}{8}$   & $33   $& $0   $& $33  $& $36  $&$\frac{3}{4}$   & $293 $& $0   $& $293 $\\
 $17 $&$\frac{37}{34}$ & $0    $& $0   $& $0   $& $37  $&$\frac{5}{4}$   & $29  $& $29  $& $0   $\\
 $18 $&$\frac{17}{30}$ & $12   $& $0   $& $12  $& $38  $&$\frac{2}{3}$   & $35  $& $35  $& $0   $\\
 $19 $&$\frac{1}{2}$   & $121  $& $0   $& $121 $& $39  $&$4$             & $21  $& $21  $& $0   $\\  
 \cline{6-10}
 $20 $&$\frac{13}{37}$ & $5    $& $5   $& $0   $ \\
 \cline{1-5}
  \end{longtable}
\end{center}

As we have learned from the previous discussion in the Step 3 of Section \ref{sec:step3}, the case of $M^1 = M^2$ has quite peculiar features inherited from the SUSY flatness condition in \cref{eq:SUSY-explicit}. In fact, this case corresponds to setting $H_2 = - H_1$ and the function $f(H_1, H_2, F^2)$ used to express the flux vector $M^i$ in \cref{eq:Mi-via-f} has the following form,
\begin{equation}
f(H_1, F^2) = a \, \frac{H_1}{F^2}\coma
\end{equation}
where we list the values of the parameter $a$ in the second column of Table \ref{tab:M1M2values} for the 39 CY geometries. 
This ensures that $f > 0$ as $H_1$ and $F^2$ both are negative in our choice of convention. Recall that this is necessary for the $M^i$ flux vector to lie within the K\"ahler cone. The number of such flat vacua having $M^1 = M^2$ and $0 < N_{\rm flux} \leq 500$ are given in the third column of Table \ref{tab:M1M2values}.

This shows that the CY geometries numbered as 1 and 6 have a huge number of such vacua while there are a couple of them which have none. To connect with an explicit example, we again get back to the 38 PFFV configurations of the $\mathbb{W}\CC\PP^4[1,1,1,6,9]$ model corresponding to $0 < N_{\rm flux} \leq 150$, as listed in Table \ref{tab:38vacuaNflux150}. After imposing conditions in \cref{eq:gcdC1,eq:gcdC2} along with $M^1 \neq M^2$, we find that there are only 16 PFFV configurations, as can be read off from the Table \ref{tab_reducedfluxdata2}.

Note that as far as the existence is concerned, these flat vacua do satisfy the necessary requirements listed in the recipe of finding the PFFV. However, we find in our numerical analysis that using ${\cal F}_{\rm inst}$ effects, these vacua cannot be lifted into the physically trustworthy regime of large complex-structure moduli and small string coupling values. To elaborate more on it, let us consider two cases by considering the sign of the parameter ${\cal A}$ as defined in \cref{eq:calA}. For a given set of PFFV configurations the cases for which ${\cal A}$ is positive (resp. negative) are listed in the fourth (resp. fifth) column of Table \ref{tab:M1M2values}, while we find that for the CY geometry 1 and 6, there are 1321 PFFV configurations satisfying $M^1 = M^2$, but the parameter ${\cal A}$ turns out to be of the undefined $\frac00$ form. This has some special implications, as we will elaborate later on.

Thus, comparing with the total number of PFFV as listed in the third column of Table \ref{tab:M1M2values} to the direct sum of numbers listed in the fourth and fifth columns of Table \ref{tab:M1M2values}, we find that given a CY $3$-fold, the sign of ${\cal A}$ is fixed for all the PFFV configurations with $M^1 = M^2$. Also using \cref{eq:SUSY-eff-M1=M2} one can see that all the vacua with ${\cal A} \geq 0$ can never result in physical solution using non-perturbative ${\cal F}_{\rm inst}$ effects, while for $A < 0$ models, one has to investigate it case by case to see if any of the PFFV results in weak coupling solution of \cref{eq:SUSY-eff-M1=M2}. 

\begin{center}
\renewcommand\arraystretch{1}
   \begin{longtable}{|c||c|c|c|c|c|c|c|c|c|c|}
\caption{Reduced number of perturbatively flat flux vacua (listed in Table \ref{tab_fluxdata}) after being subjected to the conditions of \cref{eq:gcdC1,eq:gcdC2} and $M^1 \neq M^2$.}\\
\hline
CY & \multicolumn{10}{c|}{Reduced number of flat flux vacua for $50 \leq N_{\rm flux} \leq 500$} \\
\cline{2-11}
 \# & 50 & 100 & 150 & 200 & 250 & 300 & 350 & 400 & 450 & 500\\
\hhline{|=#=|=|=|=|=|=|=|=|=|=|}
		\endhead
		\label{tab_reducedfluxdata2}1 & 10 & 32 & 54 & 76 & 94 & 128 & 147 & 175 & 189 & 198 \\
 2 & 3 & 9 & 16 & 23 & 34 & 37 & 40 & 43 & 45 & 46 \\
 3 & 0 & 1 & 2 & 3 & 4 & 5 & 8 & 11 & 12 & 12 \\
 4 & 0 & 1 & 2 & 3 & 4 & 5 & 8 & 11 & 11 & 11 \\
 5 & 0 & 1 & 2 & 3 & 4 & 5 & 8 & 11 & 11 & 11 \\
 6 & 10 & 32 & 54 & 76 & 94 & 128 & 147 & 175 & 189 & 198 \\
 7 & 0 & 0 & 0 & 0 & 0 & 1 & 2 & 4 & 4 & 7 \\
 8 & 20 & 69 & 135 & 208 & 270 & 330 & 382 & 447 & 514 & 573 \\
 9 & 20 & 69 & 135 & 208 & 270 & 330 & 382 & 447 & 514 & 573 \\
 10 & 20 & 69 & 135 & 208 & 288 & 382 & 465 & 568 & 675 & 778 \\
 11 & 17 & 55 & 105 & 161 & 220 & 293 & 356 & 432 & 511 & 587 \\
 12 & 2 & 11 & 24 & 42 & 61 & 82 & 105 & 129 & 157 & 183 \\
 13 & 0 & 2 & 4 & 7 & 9 & 11 & 14 & 19 & 20 & 21 \\
 14 & 0 & 1 & 4 & 6 & 8 & 9 & 12 & 14 & 14 & 18 \\
 15 & 0 & 1 & 3 & 6 & 7 & 8 & 11 & 16 & 17 & 21 \\
 16 & 0 & 0 & 0 & 0 & 0 & 0 & 0 & 0 & 0 & 0 \\
 17 & 0 & 0 & 0 & 0 & 0 & 1 & 1 & 1 & 1 & 2 \\
 18 & 0 & 0 & 1 & 1 & 2 & 5 & 5 & 6 & 7 & 8 \\
 19 & 2 & 7 & 14 & 20 & 28 & 35 & 42 & 52 & 59 & 66 \\
 20 & 1 & 3 & 5 & 8 & 9 & 14 & 17 & 20 & 22 & 25 \\
 21 & 1 & 3 & 5 & 8 & 10 & 14 & 17 & 20 & 23 & 26 \\
 22 & 0 & 1 & 2 & 3 & 4 & 6 & 7 & 8 & 9 & 11 \\
 23 & 0 & 1 & 2 & 3 & 4 & 6 & 7 & 8 & 9 & 11 \\
 24 & 0 & 1 & 2 & 3 & 4 & 6 & 7 & 8 & 9 & 11 \\
 25 & 0 & 1 & 5 & 6 & 13 & 15 & 19 & 23 & 25 & 30 \\
 26 & 0 & 0 & 2 & 4 & 6 & 8 & 10 & 12 & 13 & 15 \\
 27 & 0 & 0 & 0 & 0 & 0 & 0 & 1 & 1 & 1 & 1 \\
 28 & 0 & 0 & 0 & 0 & 0 & 0 & 1 & 1 & 1 & 1 \\
 29 & 1 & 3 & 5 & 8 & 9 & 14 & 17 & 20 & 22 & 25 \\
 30 & 11 & 44 & 88 & 140 & 186 & 233 & 274 & 325 & 378 & 422 \\
 31 & 11 & 44 & 88 & 140 & 186 & 233 & 274 & 325 & 378 & 422 \\
 32 & 11 & 44 & 88 & 140 & 196 & 262 & 321 & 394 & 473 & 548 \\
 33 & 172 & 472 & 836 & 1236 & 1622 & 2101 & 2529 & 3034 & 3547 & 3845 \\
 34 & 18 & 61 & 120 & 184 & 254 & 338 & 411 & 502 & 594 & 685 \\
 35 & 0 & 1 & 4 & 5 & 8 & 8 & 12 & 14 & 16 & 18 \\
 36 & 0 & 4 & 10 & 15 & 22 & 27 & 36 & 44 & 48 & 53 \\
 37 & 3 & 9 & 16 & 23 & 34 & 40 & 49 & 57 & 65 & 74 \\
 38 & 3 & 8 & 17 & 23 & 32 & 38 & 47 & 53 & 62 & 71 \\
 39 & 3 & 9 & 16 & 23 & 34 & 37 & 40 & 43 & 45 & 46 \\
\hline  
\end{longtable}
\end{center}

\noindent
From Table \ref{tab_fluxdata} and the third column of Table \ref{tab:M1M2values}, we make an observation that CY geometry number 16  have all its PFFV configurations with $M^1 = M^2$, and therefore we see no PFFV left for this model in Table \ref{tab_reducedfluxdata2}. For $N_{\rm flux} \leq 100$, there are no PFFV configurations for 7 CY geometries out of 39. Similarly, for $N_{\rm flux} \leq 250$, there are no PFFV configurations for 5 CY geometries. These values of $N_{\rm flux}$ are quite large and reasonable values in typical models and still having no PFFV, shows the rarity of these vacua.

\subsubsection{More multiplicities through $F^i$ fluxes}

By analyzing the list of PFFV configurations, we make an empirical observation that there could still be some multiplicities of equivalent vacua presented in Table \ref{tab_reducedfluxdata2}. These are PFFV vacua such that for a given coprime value of $\{H_1, H_2\}$, there are the following set of quantities which appear with  a common multiplication factor denoted as $m$,
\begin{equation}
\label{eq:gcd4}
{\cal C} : \left[\{H_1, H_2\}, \, \, m \, \left\{N_{\rm flux}, \, \, F^1, \, \, F^2, \, \, \frac{1}{M^1}, \, \, \frac{1}{M^2}, \, \,  \tilde{p}_{1j} F^j, \, \,  \tilde{p}_{2j} F^j,\, \, \tilde{p}_i F^i\right\}\right]\coma  m \in {\mathbb N}.
\end{equation}
For example, looking at the 38 PFFV configurations of WCP$^4[1,1,1,6,9]$ model listed in Table \ref{tab:38vacuaNflux150} we find that after imposing the conditions \cref{eq:gcdC1,eq:gcdC2} and $M^1 \neq M^2$, the number 38 (which corresponds to $N_{\rm flux} = 150$) reduces to 16 as it can be read-off from the Table \ref{tab_reducedfluxdata2}. Moreover, it can be shown that those 16 PFFV configurations can be further collected into the following seven classes,
\begin{equation}
\label{eq:calC-11169}
\renewcommand{\arraystretch}{1.2}
\begin{array}{rclrlrcl}
{\cal C}_1 & : & \left[\left\{-12,1\right\}, \right. &    & \left.\left\{122,12,-100,80,20/3,-96,18,-99\right\}\right]\,; \\
{\cal C}_2 & : & \left[\left\{-6,1\right\} , \right. &    & \left.\left\{114,20,-108,72,12,-72,30,-77\right\}\right]\,; \\
{\cal C}_3 & : & \left[\left\{-3,1\right\} , \right. &  m & \left.\left\{14,4,-16,8,8/3,-6,6,-7\right\}\right]\,, & m &\in & \{1, 2,\ldots, 10\}\,;\\
{\cal C}_4 & : & \left[\left\{-3,2\right\} , \right. &    & \left.\left\{78,16,-54,9,6,-9,24,-13\right\}\right]\,; \\
{\cal C}_5 & : & \left[\left\{-3,4\right\} , \right. &    & \left.\left\{124,16,-50,5/2,10/3,-3,24,-7\right\}\right]\,; \\
{\cal C}_6 & : & \left[\left\{-2,1\right\} , \right. &    & \left.\left\{78,28,-100,40,20,-24,42,-31\right\}\right]\,; \\
{\cal C}_7 & : & \left[\left\{-1,2\right\} , \right. &    & \left.\left\{114,32,-98,7,14,-3,48,-11\right\}\right]\fstop \\
\end{array}
\end{equation}
Depending on the values of $m$ in a given collection ${\cal C}$, the D3 tadpole value $N_{\rm flux}$ changes. Also, note that the configuration ${\cal C}_5$ in \cref{eq:calC-11169} is the one reported in \cite{Demirtas:2019sip} while ${\cal C}_7$ is the additional one reported in \cite{Broeckel:2021uty}. This way of clubbing the PFFV configurations can apparently reduce the number and hence can make the task of finding physical vacua a bit simpler. However, there is a trick/risk in this kind of filtering the vacua.  Let us recall that the leading order SUSY flatness condition is given in \cref{eq:gsVEV}, which is rewritten as below,
\begin{equation}
\label{eq:SUSY-simple}
e^{- 2\pi \langle s \rangle (M^1 - M^2)} \simeq -\frac{n_2\,F^2 \, M^2}{n_1\,F^1 \,M^1}\fstop
\end{equation}
Therefore, for a given CY geometry, i.e., for a fixed set of values of $\{n_1, n_2\}$, the RHS of the above \cref{eq:SUSY-simple} remains the same for all the multiplicities appearing in a given collection ${\cal C}$ while the LHS changes. This creates the possibility of getting the largest $\langle s \rangle$ corresponding to the lowest value of $|M^1 - M^2|$ within a given collection ${\cal C}$, which should preferably be a fraction, i.e. $|M^1 - M^2| < 1$. However, given that the complex structure moduli are stabilized through $\langle u^i \rangle = \langle s \rangle \, M^i$, and therefore for too small fractional values of $M^i$, the requirements of large complex structure may get challenging. This shows that having weak coupling and large complex structure moduli VEVs appear to be contradictory. One has to take this fact into account while looking for physically allowed values and need to figure out which available value of $m \in {\mathbb N}$ in a given collection ${\cal C}$ of multiple PFFV configurations is the most suitable one. Due to this subtlety, such multiplicities should not be even called as over counting or redundancy in PFFV. For the same reason, we consider all the PFFV configurations as available in Table \ref{tab_reducedfluxdata2} for the dilaton minimization process.

\subsection{Finding the physical vacua}

In order to stabilize the flat valley encoded in the PFFV configurations, now we explicitly utilize the non-perturbative effects for each of the 39 CY geometries, and look for the possible physical vacua (if any) with the lowest possible value of the flux superpotential $|W_0|$. 

Let us, first, rule out four CY geometries numbered as $\{17, 20, 21, 38\}$ which cannot result in physical vacua from any of their corresponding PFFV configurations. The reason being the fact that the GV invariant dependent quantities $n_1$ and $n_2$ as listed in Table \ref{tab_cydata-h11eq2-2} have opposite signs for these four examples, and hence the RHS of \cref{eq:SUSY-simple} is always negative given that $\{F^1 > 0, F^2 < 0, M^1 > 0, M^2 > 0\}$ follows from the K\"ahler cone conditions. This subsequently cannot give real values of string coupling $\langle g_s \rangle = \langle s \rangle^{-1}$. Also recall that there is no PFFV configuration for CY geometry number 16 as seen from Table \ref{tab_reducedfluxdata2} and so at this stage effectively we are left with analyzing 34 CY geometries.

In this regard, we performed a detailed analysis of the physicality of the PFFV configurations and to our surprise, we find that most of the vacua does not result in weak coupling and large complex structure limit in the sense of merely imposing $\{\langle g_s \rangle < 1, \langle u^1 \rangle > 1, \langle u^2 \rangle > 1 \}$. In fact, we find that only 16 CY geometries out of 39 can result in physical vacua, however all of the 10 CY geometries with K3-fibered structure have the physical solutions. Moreover, 5 Swiss-cheese examples out of 22 have physical solutions, while one Hybrid type example out of 7 has a physical solution. 

\subsubsection{Swiss-cheese type}
We find that there are only five CY geometries of the Swiss-cheese type which can result in physical PFFV configurations. These five CY geometries can result in a total of 11 PFFV configurations, which turn out to be physical vacua after stabilizing the flat directions using ${\cal F}_{\rm inst}$. Moreover, these 11 PFFV configurations can be further clubbed into five classes corresponding to each of the five CY geometries. These are presented in Table \ref{tab_physical-vacua-sc-vevs-1} and the subsequent details about the VEVs of various quantities are given in Table \ref{tab_physical-vacua-sc-vevs-2}.

\begin{table}[!htp]
\begin{center}
\renewcommand{\arraystretch}{1.2}
\begin{tabular}{|c|c||c|cc|cc|cc|ccc|}
\hline
\# & CY \# & $N_{\rm flux}$ & $H_1$ & $H_2$ & $F^1$ & $F^2$ & $M^1$ & $M^2$ & $\tilde{p}_{1i}F^i$ & $\tilde{p}_{2i}F^i$ & $\tilde{p}_{i}F^i$ \\
\hhline{|=|=#=|==|==|==|===|}
$1  $& $3 $& $348  $& $-2  $& $3  $& $150  $& $-132 $& $\frac{1}{30}$   & $\frac{1}{45}$ & $-48 $& $27   $& $-31 $\\
\hline
$2  $& $4 $& $348  $& $-3  $& $2  $& $132  $& $-150 $& $\frac{1}{45}$ & $\frac{1}{30}$  & $-27 $& $48   $& $31 $\\
\hline
$3  $& $5 $& $348  $& $-3  $& $2  $& $132  $& $-150 $& $\frac{1}{45}$ & $\frac{1}{30}$    & $-27 $& $48   $& $31 $\\
\hline
\multirow{4}{*}{$4$} & \multirow{4}{*}{$37$} & $496  $& $-4 $& $3 $& $200 $& $-64  $& $\frac{3}{40}$ & $\frac{1}{10}$            & $-32   $& $-188   $& $24  $\\
& & $372  $& $-4 $& $3 $& $150 $& $-48  $& $\frac{1}{10}$ & $\frac{2}{15}$            & $-24   $& $-141   $& $18  $\\
& & $248  $& $-4 $& $3 $& $100 $& $-32  $& $\frac{3}{20}$ & $\frac{1}{5}$            & $-16   $& $-94   $& $12  $\\
& & $124  $& $-4 $& $3 $& $50 $& $-16  $& $\frac{3}{10}$ & $\frac{2}{5}$            & $-8   $& $-47   $& $6  $\\
\hline
\multirow{4}{*}{$5$} & \multirow{4}{*}{$39$} & $ 496 $& $-3 $& $4 $& $64 $& $-200  $& $\frac{1}{10}$ & $\frac{3}{40}$            & $-12   $& $96  $& $-28  $\\
& & $ 372 $& $-3 $& $4 $& $48 $& $-150  $& $\frac{2}{15}$ & $\frac{1}{10}$            & $-9   $& $72  $& $-21  $\\
& & $ 248 $& $-3 $& $4 $& $32 $& $-100  $& $\frac{1}{5}$ & $\frac{3}{20}$            & $-6   $& $48  $& $-14  $\\
& & $ 124 $& $-3 $& $4 $& $16 $& $-50  $& $\frac{2}{5}$ & $\frac{3}{10}$            & $-3   $& $24  $& $-7  $\\
 \hline
\end{tabular}
\end{center}
\caption{All the PFFV configurations of the Swiss-cheese CY geometries which result in physical vacua.}
\label{tab_physical-vacua-sc-vevs-1}
\end{table}

\begin{table}[!htp]
\begin{center}
\renewcommand{\arraystretch}{1.2}
\begin{tabular}{|c|c||c|c|c|c|c|}
\hline
\# & CY \# &  $\langle s \rangle$ & $\langle u^1 \rangle$ & $\langle u^2 \rangle$ & $|W_0|$ & $\langle K(s, u^i) \rangle$ \\
\hhline{|=|=#=|=|=|=|=|}
$1$                  & $3  $&  $60.2393 $& $2.00798 $& $1.33865 $& $7.9161 \cdot 10^{-6}$ & $-9.7269  $\\
\hline
$2$                  & $4  $&  $60.2393 $& $1.33865 $& $2.00798 $& $7.9161 \cdot 10^{-6}$ & $-9.7269  $\\
\hline
$3$                  & $5  $&  $60.2393 $& $1.33865 $& $2.00798 $& $7.9161 \cdot 10^{-6}$ & $-9.7269  $\\
\hline
\multirow{4}{*}{$4$} & \multirow{4}{*}{$37$} &  $21.6575 $& $1.62431 $& $2.16575 $& $9.6629 \cdot 10^{-7}$ & $-9.23311 $\\
                     &                       &  $16.2431 $& $1.62431 $& $2.16575 $& $8.3683 \cdot 10^{-7}$ & $-8.94543 $\\
                     &                       &  $10.8288 $& $1.62431 $& $2.16575 $& $6.8327 \cdot 10^{-7}$ & $-8.53996 $\\
                     &                       &  $5.41438 $& $1.62431 $& $2.16575 $& $4.8314 \cdot 10^{-7}$ & $-7.84682 $\\
\hline
\multirow{4}{*}{$5$} & \multirow{4}{*}{$39$} &  $27.4202 $& $2.74202 $& $2.05652 $& $4.0964 \cdot 10^{-8}$ & $-10.1768 $\\
                     &                       &  $20.5652 $& $2.74202 $& $2.05652 $& $3.5476 \cdot 10^{-8}$ & $-9.88914 $\\
                     &                       &  $13.7101 $& $2.74202 $& $2.05652 $& $2.8966 \cdot 10^{-8}$ & $-9.48367 $\\
                     &                       &  $6.85505 $& $2.74202 $& $2.05652 $& $2.0482 \cdot 10^{-8}$ & $-8.79052 $\\
 \hline
\end{tabular}
\end{center}
\caption{VEVs of moduli and $|W_0|$ for PFFV configurations in Table \ref{tab_physical-vacua-sc-vevs-1}.}
\label{tab_physical-vacua-sc-vevs-2}
\end{table}

\noindent
We observe that three CY geometries ${\cal M}_{2,3}$, ${\cal M}_{2,4}$ and ${\cal M}_{2,5}$ listed in Table  \ref{tab_cydata-h11eq2} result in a unique physical vacuum with same value of $|W_0| = 7.9161 \cdot 10^{-6}$ and $\langle g_s \rangle = 0.0166$ while slightly different values for the $\langle u^i \rangle$ moduli. However, we find that CY geometry numbered as ${\cal M}_{2,37}$ and ${\cal M}_{2,39}$ have four solutions each, which are the same in the sense of clubbing into collection like $\cal C$ which we defined in \cref{eq:gcd4}. It turns out that $|W_0| \simeq 10^{-7}$ for model ${\cal M}_{2, 37}$ with the lowest value being $4.8314 \cdot 10^{-7}$ while for model ${\cal M}_{2, 39}$ we have $|W_0| \simeq 10^{-8}$ with the lowest value being $2.0482 \cdot 10^{-8}$ as already reported in \cite{Demirtas:2019sip}. We also note that the VEVs of the complex structure moduli remain the same for all the PFFV configurations clubbed into a given ${\cal C}_i^j$. This could be attributed to the fact that the SUSY flatness condition in \cref{eq:SUSY-explicit} is a function of $\{u^1 = s M^1, u^2 = s M^2\}$ and no explicit dependence on dilaton appears without $M^1$ and $M^2$. However, given that such vacua correspond to different VEVs for string coupling as well as slightly different values of $|W_0|$, we preferred to keep them distinct and avoid concluding that all the Swiss-cheese CY geometries can result in to a unique  PFFV configuration which remain physical after stabilizing through the non-perturbative prepotential ${\cal F}_{\rm inst}$.

\subsubsection*{A Swiss-cheese example which supports anti-D3 uplifting}

Let us mention that apart from the vanilla Swiss-cheese model in $\mathbb{W}\CC\PP^4[1,1,1,6,9]$ which have been studied several times in the literature, there has been another useful CY geometry realized as a degree 14 hypersurfaces in $\mathbb{W}\CC\PP^4[1,1,2,3,7]$. This is a Swiss-cheese type model and supports LVS in the K\"ahler moduli stabilization sector. In addition to that it also has the necessary ingredients for generating an uplifting from the anti-D$3$ brane due to the presence of an involution giving a Whitney-brane realization with two O$3$-planes and a possibility of relatively large $N_{\rm flux} = 149$ \cite{Crino:2020qwk}. This model is ${\cal M}_{2, 37}$ in our collection of 39 CY geometries in Table \ref{tab_cydata-h11eq2}. In our detailed analysis, we find that this CY has a set of four PFFV configurations with string coupling values $\{0.185, 0.092, 0.062, 0.046\}$ with $|W_0| \simeq 10^{-7}$. However, for $N_{\rm flux} \leq 150$, there is a unique physical vacuum.

\subsubsection{K3-fibered type}
The K3-fibered CY geometries have turned out to be the most interesting ones in our search of physical PFFV configurations. It turns out that there are a total of 499 PFFV configurations which are physical. These are distributed as $\{56, 56, 56, 20, 12, 58, 58, 65, 63, 55\}$ across the 10 CY geometries, numbered as $\{8, 9, 10, 11, 12, 30, 31, 32, 33, 34\}$ respectively. It does not appear to be illuminating to present all the 499 physical vacua, and therefore in Table \ref{tab_physical-vacua-K3-vevs-1} we present two PFFV configurations for each of the 10 K3-fibered models according to $|W_0|_{\max}$ and $|W_0|_{\min}$ values attainable in those model.

\begin{table}[!htp]
\begin{center}
\renewcommand{\arraystretch}{1.2}
\begin{tabular}{|c|c||c|cc|cc|cc|ccc|}
\hline
\# & CY \# & $N_{\rm flux}$ & $H_1$ & $H_2$ & $F^1$ & $F^2$ & $M^1$ & $M^2$ & $\tilde{p}_{1i}F^i$ & $\tilde{p}_{2i}F^i$ & $\tilde{p}_{i}F^i$ \\
\hhline{|=|=#=|==|==|==|===|}
\multirow{2}{*}{$1$}  & \multirow{2}{*}{$8$}  &  $171$ & $-11$ & $12 $& $6   $& $-23  $&  $\frac{1}{2}$   &  $\frac{11}{24}$ & $-22 $& $12   $& $-9 $ \\
                      &                       &  $285$ & $-3 $ & $5  $& $30  $& $-96  $&  $\frac{1}{24}$  &  $\frac{1}{40}$  & $-72 $& $60   $& $-26$ \\
\hline
\multirow{2}{*}{$2$}  & \multirow{2}{*}{$9$}  &  $171$ & $-11$ & $12 $& $6   $& $-23  $&  $\frac{1}{2}$   &  $\frac{11}{24}$ & $-22 $& $12   $& $-9 $ \\
                      &                       &  $285$ & $-3 $ & $5  $& $30  $& $-96  $&  $\frac{1}{24}$  &  $\frac{1}{40}$  & $-72 $& $60   $& $-26$ \\
\hline
\multirow{2}{*}{$3$}  & \multirow{2}{*}{$10$} &  $171$ & $-12$ & $11 $& $23  $& $-6   $&  $\frac{11}{24}$ &  $\frac{1}{2}$   & $0   $& $-24  $& $9  $ \\
                      &                       &  $285$ & $-5 $ & $3  $& $96  $& $-30  $&  $\frac{1}{40}$  &  $\frac{1}{24}$  & $0   $& $-120 $& $26 $ \\
\hline
\multirow{2}{*}{$4$}  & \multirow{2}{*}{$11$} &  $486$ & $-24$ & $23 $& $29  $& $-12  $&  $\frac{23}{48}$ &  $\frac{1}{2}$   & $0   $& $-12  $& $7  $ \\
                      &                       &  $414$ & $-24$ & $19 $& $25  $& $-12  $&  $\frac{19}{48}$ &  $\frac{1}{2}$   & $0   $& $-12  $& $3  $ \\
\hline
\multirow{2}{*}{$5$}  & \multirow{2}{*}{$12$} &  $288$ & $-12$ & $11 $& $37  $& $-12  $&  $\frac{11}{48}$ &  $\frac{1}{4}$   & $0   $& $-30  $& $12 $ \\
                      &                       &  $486$ & $-24$ & $17 $& $32  $& $-12  $&  $\frac{17}{48}$ &  $\frac{1}{2}$   & $0   $& $-30  $& $7  $ \\
\hline
\multirow{2}{*}{$6$}  & \multirow{2}{*}{$30$} &  $171$ & $-11$ & $12 $& $6   $& $-23  $&  $1$             &  $\frac{11}{12}$ & $-11 $& $6    $& $-10$ \\
                      &                       &  $336$ & $-1 $ & $2  $& $96  $& $-288 $&  $\frac{1}{96}$  &  $\frac{1}{192}$ & $-96 $& $96   $& $-80$ \\
\hline
\multirow{2}{*}{$7$}  & \multirow{2}{*}{$31$} &  $171$ & $-11$ & $12 $& $6   $& $-23  $&  $1$             &  $\frac{11}{12}$ & $-11 $& $6    $& $-10$ \\
                      &                       &  $336$ & $-1 $ & $2  $& $96  $& $-288 $&  $\frac{1}{96}$  &  $\frac{1}{192}$ & $-96 $& $96   $& $-80$ \\
\hline
\multirow{2}{*}{$8$}  & \multirow{2}{*}{$32$} &  $171$ & $-12$ & $11 $& $23  $& $-6   $&  $\frac{11}{12}$ & 1                & $0   $& $-12  $& $10 $ \\
                      &                       &  $483$ & $-2 $ & $1  $& $414 $& $-138 $&  $\frac{1}{276}$ &  $\frac{1}{138}$ & $0   $& $-276 $& $115$ \\
\hline
\multirow{2}{*}{$9$}  & \multirow{2}{*}{$33$} &  $414$ & $-24$ & $23 $& $23  $& $-12  $&  $\frac{23}{24}$ & 1                & $0   $& $0    $& $5  $ \\
                      &                       &  $378$ & $-9 $ & $7  $& $56  $& $-36  $&  $\frac{7}{72}$  &  $\frac{1}{8}$   & $0   $& $0    $& $2  $ \\
\hline
\multirow{2}{*}{$10$} & \multirow{2}{*}{$34$} &  $135$ & $-12$ & $11 $& $17  $& $-6   $&  $\frac{11}{12}$ & 1                & $0   $& $-6   $& $6  $ \\
                      &                       &  $420$ & $-5 $ & $3  $& $132 $& $-60  $&  $\frac{1}{40}$  &  $\frac{1}{24}$  & $0   $& $-60  $& $22 $ \\
\hline
\end{tabular}
\end{center}
\caption{PFFV configurations for K3-fibered CY geometries. For a given example, the upper row corresponds to $|W_0|_{\rm min}$ while the lower row corresponds to $|W_0|_{\rm max}$ which is attainable within the available PFFV.}
\label{tab_physical-vacua-K3-vevs-1}
\end{table}

\begin{table}[!htp]
\begin{center}
\renewcommand{\arraystretch}{1.2}
\begin{tabular}{|c|c||c|c|c|c|c|}
\hline
\# & CY \#  & $\langle s \rangle$ & $\langle u^1 \rangle$ & $\langle u^2 \rangle$ & $|W_0|$ & $\langle K(s, u^i) \rangle$ \\
\hhline{|=|=#=|=|=|=|=|}
\multirow{2}{*}{$1$}  & \multirow{2}{*}{$8$}  & $14.5709 $& $7.28545 $& $6.67833 $& $1.6608 \cdot 10^{-22}$ & $-12.5619$  \\
                      &                       & $41.2449 $& $1.71854 $& $1.03112 $& $0.000043559            $& $-9.04608$  \\
\hline
\multirow{2}{*}{$2$}  & \multirow{2}{*}{$9$}  & $14.5709 $& $7.28545$ & $6.67833$ & $1.6608 \cdot 10^{-22}$ & $-12.5619$  \\
                      &                       & $41.2449 $& $1.71854$ & $1.03112$ & $0.000043559$              & $-9.04608$  \\
\hline
\multirow{2}{*}{$3$}  & \multirow{2}{*}{$10$} & $14.5709$ & $6.67833$ & $7.28545$ & $1.6608 \cdot 10^{-22}$ &   $-12.5619$  \\
                      &                       & $41.2449$ & $1.03112$ & $1.71854$ & $0.000043559$              & $-9.04608$  \\
\hline
\multirow{2}{*}{$4$}  & \multirow{2}{*}{$11$} & $11.1567$ & $5.34593$ & $5.57836$ & $1.4469 \cdot 10^{-17}$ & $-11.1522$  \\
                      &                       & $2.65799$ & $1.05212$ & $1.32899$ & $0.000517668$              & $-5.25401$  \\
\hline
\multirow{2}{*}{$5$}  & \multirow{2}{*}{$12$} & $20.2025$ & $4.62975$ & $5.05063$ & $6.5461 \cdot 10^{-16}$ & $-11.6178$  \\
                      &                       & $3.25564$ & $1.15304$ & $1.62782$ & $0.0000797843$             & $-6.22563$  \\
\hline
\multirow{2}{*}{$6$}  & \multirow{2}{*}{$30$} & $11.211 $ & $11.211 $ & $10.2768$ & $1.0793 \cdot 10^{-32}$ & $-12.8997$  \\
                      &                       & $201.567$ & $2.09966$ & $1.04983$ & $0.0000369383$             & $-10.4582$  \\
\hline
\multirow{2}{*}{$7$}  & \multirow{2}{*}{$31$} & $11.211 $ & $11.211 $ & $10.2768$ & $1.0793 \cdot 10^{-32}$ & $-12.8997$  \\
                      &                       & $201.567$ & $2.09966$ & $1.04983$ & $0.0000369383$             & $-10.4582$  \\
\hline
\multirow{2}{*}{$8$}  & \multirow{2}{*}{$32$} & $11.211 $ & $10.2768$ & $11.211 $ & $1.0793 \cdot 10^{-32}$ & $-12.8997$  \\
                      &                       & $289.752$ & $1.04983$ & $2.09966$ & $0.0000442875$             &$ -10.8211$  \\
\hline
\multirow{2}{*}{$9$}  & \multirow{2}{*}{$33$} & $5.91766$ & $5.67109$ & $5.91766$ & $1.2931 \cdot 10^{-17}$ & $-9.84179$  \\
                      &                       & $10.706 $ & $1.04086$ & $1.33825$ & $0.00443788$               & $-5.76616$  \\
\hline
\multirow{2}{*}{$10$} & \multirow{2}{*}{$34$} & $7.03979 $& $6.45314 $& $7.03979$ & $7.2980 \cdot 10^{-21}$ & $-10.802 $  \\
                      &                       & $40.5279 $& $1.0132  $& $1.68866$ & $0.000677371$              & $-7.97739$  \\
\hline
\end{tabular}
\end{center}
\caption{VEVs of moduli and $|W_0|$ for PFFV configurations of Table \ref{tab_physical-vacua-K3-vevs-1}.}
\label{tab_physical-vacua-K3-vevs-2}
\end{table}

\noindent
From Table \ref{tab_physical-vacua-K3-vevs-2}, we can  observe that it is indeed possible to achieve a range of values for $|W_0|$, and it can be as low as $1.1 \cdot 10^{-32}$ and as high as $4.4 \cdot 10^{-3}$ in explicit K3-fibered CY geometries. However, the mentioned configurations with lowest and highest values of $|W_0|$ typically falls in the region of larger tadpole charge requirement as seen from Table \ref{tab_physical-vacua-K3-vevs-1}. Given that one usually does not have too large $N_{\rm flux}$ in a typical model, we select the best possible PFFV configurations for each of the 10 K3-fibered models which have lowest $|W_0|$ within $N_{\rm flux} \leq 150$ in \cref{tab_physical-vacua-K3-vevs-3,tab_physical-vacua-K3-vevs-4}.

\begin{table}[!htp]
\begin{center}
\renewcommand{\arraystretch}{1.2}
\begin{tabular}{|c|c||c|cc|cc|cc|ccc|}
\hline
\# & CY \# & $N_{\rm flux}$ & $H_1$ & $H_2$ & $F^1$ & $F^2$ & $M^1$ & $M^2$ & $\tilde{p}_{1i}F^i$ & $\tilde{p}_{2i}F^i$ & $\tilde{p}_{i}F^i$ \\
\hhline{|=|=#=|==|==|==|===|}
$1  $& $8  $& $81  $& $-5  $& $6  $& $6  $& $-22 $& $\frac{1}{4}$   & $\frac{5}{24}$ & $-20 $& $12  $& $-8 $\\
$2  $& $9  $& $81  $& $-5  $& $6  $& $6  $& $-22 $& $\frac{1}{4}$   & $\frac{5}{24}$ & $-20 $& $12  $& $-8 $\\
$3  $& $10 $& $81  $& $-6  $& $5  $& $22 $& $-6  $& $\frac{5}{24}$  & $\frac{1}{4}$  & $0   $& $-24 $& $8  $\\
$4  $& $11 $& $117 $& $-12 $& $11 $& $14 $& $-6  $& $\frac{11}{24}$ & $\frac{1}{2}$  & $0   $& $-6  $& $3  $\\
$5  $& $12 $& $135 $& $-6  $& $5  $& $35 $& $-12 $& $\frac{5}{48}$  & $\frac{1}{8}$  & $0   $& $-30 $& $10 $\\
$6  $& $30 $& $81  $& $-5  $& $6  $& $6  $& $-22 $& $\frac{1}{2}$   & $\frac{5}{12}$ & $-10 $& $6   $& $-9 $\\
$7  $& $31 $& $81  $& $-5  $& $6  $& $6  $& $-22 $& $\frac{1}{2}$   & $\frac{5}{12}$ & $-10 $& $6   $& $-9 $\\
$8  $& $32 $& $81  $& $-6  $& $5  $& $22 $& $-6  $& $\frac{5}{12}$  & $\frac{1}{2}$  & $0   $& $-12 $& $9  $\\
$9  $& $33 $& $99  $& $-12 $& $11 $& $11 $& $-6  $& $\frac{11}{12}$ & $1$            & $0   $& $0   $& $2  $\\
$10 $& $34 $& $135 $& $-12 $& $11 $& $17 $& $-6  $& $\frac{11}{12}$ & $1$            & $0   $& $-6  $& $6  $\\
 \hline
\end{tabular}
\end{center}
\caption{Selected PFFV configurations with $N_{\rm flux} \leq 150$ for physical vacua using K3-fibered CY geometries.}
\label{tab_physical-vacua-K3-vevs-3}
\end{table}

\begin{table}[!htp]
\begin{center}
\renewcommand{\arraystretch}{1.2}
\begin{tabular}{|c|c||c|c|c|c|c|}
\hline
\# & CY \#  & $\langle s \rangle$ & $\langle u^1 \rangle$ & $\langle u^2 \rangle$ & $|W_0|$ & $\langle K(s, u^i) \rangle$ \\
\hhline{|=|=#=|=|=|=|=}
$1  $& $8  $& $15.0602 $& $3.76504 $& $3.13753 $& $3.8187 \cdot 10^{-12}$ & $-10.5605 $\\
$2  $& $9  $& $15.0602 $& $3.76504 $& $3.13753 $& $3.8187 \cdot 10^{-12}$ & $-10.5605 $\\
$3  $& $10 $& $15.0602 $& $3.13753 $& $3.76504 $& $3.8187 \cdot 10^{-12}$ & $-10.5605 $\\
$4  $& $11 $& $5.85725 $& $2.68457 $& $2.92863 $& $9.0065 \cdot 10^{-10}$ & $-8.53706 $\\
$5  $& $12 $& $21.2343 $& $2.21191 $& $2.65429 $& $1.2309 \cdot 10^{-8}$  & $-9.67302 $\\
$6  $& $30 $& $11.4627 $& $5.73136 $& $4.77613 $& $5.7198 \cdot 10^{-17}$ & $-10.855  $\\
$7  $& $31 $& $11.4627 $& $5.73136 $& $4.77613 $& $5.7198 \cdot 10^{-17}$ & $-10.855  $\\
$8  $& $32 $& $11.4627 $& $4.77613 $& $5.73136 $& $5.7198 \cdot 10^{-17}$ & $-10.855  $\\
$9  $& $33 $& $3.11683 $& $2.8571  $& $3.11683 $& $2.0799 \cdot 10^{-9}$  & $-7.23285 $\\
$10 $& $34 $& $7.03979 $& $6.45314 $& $7.03979 $& $7.2980 \cdot 10^{-21}$ & $-10.802  $\\
 \hline
\end{tabular}
\end{center}
\caption{VEVs of moduli and $|W_0|$ for PFFV configurations of Table \ref{tab_physical-vacua-K3-vevs-3}.}
\label{tab_physical-vacua-K3-vevs-4}
\end{table}

\noindent
Thus, we indeed find that $|W_0| \simeq 10^{-20}$ is well attainable with weak string coupling and large complex structure VEVs, which can be reasonably good for KKLT like models.

\subsubsection{Hybrid type}
We find that there remains a unique CY geometry of the Hybrid type, which possesses a physical PFFV configuration. This is model ${\cal M}_{2,15}$ in the list of 39 CY geometries given in Table \ref{tab_cydata-h11eq2}. The relevant details about physical vacuum are given as below,
\begin{equation}
\renewcommand{\arraystretch}{1.2}
\begin{array}{c}
    {\cal C}_1^{15} : \left[\{-6, 5\}, \{342, 74, -48, 138/5, 23, -72, -9, 11 \} \right]\,; \\
\langle s \rangle  =  45.177\coma  \langle u^1 \rangle  =  1.63685\coma \langle u^2 \rangle =  1.96422\coma \\
\langle |W_0| \rangle  =  8.94674 \cdot 10^{-7}\coma  \langle K(s, u^i) \rangle  =  -9.98428\fstop
\end{array}
\end{equation}
However, notice that for having this PFFV one needs quite a large value of $N_{\rm flux} = 342$. Regarding the other five models in this Hybrid type, we find that ${\cal M}_{2,1}$, ${\cal M}_{2,6}$, ${\cal M}_{2,14}$, ${\cal M}_{2,16}$ and ${\cal M}_{2,17}$ there are no physical vacua, despite the fact that the first two models were among the top ones with 1905 PFFV configurations each in Table \ref{tab_fluxdata}. However, they are useful for some other perspective, as we will discuss in the next section.


\section{Some more insights on flat vacua}
\label{sec_EFFV}

\subsection{Exponentially flat flux vacua}
In our detailed analysis, we have observed that there are some PFFV configurations which remain flat even after including the non-perturbative effects, and we call them as ``exponentially flat flux vacua". This is a special class of PFFV configurations which arises due to symmetry in the CY $3$-fold. In order to investigate the origin of such vacua, let us get back to the SUSY flatness solution as given in \cref{eq:SUSY-explicit} which we write again as below,
\begin{equation}
\begin{split}
0 = &\, e^{- 2 \pi \,s \,  {M^1}} \, F^1\, n_1 (1+ \pi\, s\, M^1) + e^{- 2 \pi \,s \,  {M^2}} \, F^2\, n_2 (1+ \pi\, s\, M^2)+ \\
& + 2\, e^{- 4 \pi \,s \,  {M^1}} \, F^1\, n_{11} \, (1+ 2 \pi\, s\, M^1) + 2\, e^{- 4 \pi \,s \,  {M^2}} \, F^2\, n_{22} \, (1+ 2 \pi\, s\, M^2)+\\
& + \, e^{- 2 \pi \,s \,  (M^1 + M^2)} \, (F^1 + F^2)\, n_{12} \, (1+ \pi\, s\, M^1+  \pi\, s\, M^2)\fstop
\end{split}
\end{equation}
This shows that the PFFV configurations are protected at the first order in the non-perturbative series if the following relations hold,
\begin{equation}
\label{eq:cond-EFFVs-1}
M^1 = M^2\coma F^1 \, n_1 + F^2 \, n_2 = 0\fstop
\end{equation}
Moreover, the protection for the flatness will be extended to even the second order if the following relations hold,
\begin{equation}
\label{eq:cond-EFFVs-2}
M^1 = M^2\coma F^1 \, n_1 + F^2 \, n_2 = 0\coma 2 F^1 n_{11} + (F^1 +F^2) \, n_{12} + 2 F^2\, n_{22} = 0 \fstop
\end{equation}
Note that these solutions are a subclass of those which satisfy $M^1 = M^2$, which generically can also have solutions that do not satisfy the additional relations of being EFFV in \cref{eq:cond-EFFVs-1,eq:cond-EFFVs-2}.

One would wonder why such a peculiar set of conditions as \cref{eq:cond-EFFVs-1,eq:cond-EFFVs-2} would be satisfied at all, however we find that depending on the symmetries in the CY geometries, some examples do have PFFV configurations which satisfy this requirement. It turns out that the CY $3$-folds numbered as $\{1, 6\}$ corresponding to models ${\cal M}_{2,1}$ and ${\cal M}_{2,6}$ of Table \ref{tab_cydata-h11eq2} have such properties. We find that out of 1905 flat vacua of PFFV type, there are 1321 number of EFFV type flat vacua which satisfy both these constraints. The underlying reason for this to happen is the fact that both the CY $3$-folds are defined by bi-cubics in ${\mathbb P}^2 \times {\mathbb P}^2$ which have an $ 1 \leftrightarrow 2$ exchange symmetry with the various topological data, like triple intersection number, $c_2({\rm CY})$, K\"ahler cone generators, GV invariants etc. Actually, the symmetry in $\kappa_{ijk}^0 , \, \tilde{p}_{ij}, \, \tilde{p}_i$ and $\chi({\rm CY})$ along with the Mori cone enforces the PFFVs to have $F^1 = - F^2$ and on top of this GV invariants are same as well. Just for illustration purpose we present one of the 1321 EFFV configurations for model ${\cal M}_{2,6}$ as below,
\begin{equation}
    \label{eq:EFFV-model6}
    \renewcommand{\arraystretch}{1.2}
    \begin{array}{rclrclrcl}
     F^i & = & \{10, -10\}\coma & H^i & = & \{-8, 8\}\coma & M^i & = & \{4/15, 4/15\}\coma \\
     N_{\rm flux} & = &80\coma & \tilde{p}_{ij}F^j & =& \{-15, 15\}\coma  & \tilde{p}_i \, F^i &= & 0\coma\\
    \multicolumn{9}{c}{\displaystyle{n_1 = 189 = n_2\coma n_{11} = \frac{1701}{8} =  n_{22}\coma n_{12}= 8662\fstop}}
    \end{array}
\end{equation}
One can see that the PFFV configuration in \cref{eq:EFFV-model6} satisfies the criteria of being exponentially flat as given in \cref{eq:cond-EFFVs-2}. Moreover, let us mention that due to the presence of $ 1 \leftrightarrow 2$ exchange symmetry, this CY model ${\cal M}_{2,6}$ is also useful for constructing orientifolds with non-trivial odd-sector \cite{Gao:2013pra, Carta:2020ohw, Altman:2021pyc}.

The first example ${\cal M}_{2, 1}$ has a non-trivial fundamental group and hence may not be suitable for type IIB ${\cal N} = 1$ orientifold model building. However, the second model ${\cal M}_{2, 6}$ with $\{h^{1,1}, h^{2,1}\} = \{2, 83\}$ is quite unique in the sense that it is the only one with $h^{1,1} =2$ which appears in the collection of KS CY database \cite{Kreuzer:2000xy, Altman:2014bfa} and in the projective complete intersection Calabi-Yau (pCICY) database \cite{Anderson:2017aux} as well. It corresponds to the polytope ID 7884 in the pCICY dataset of \cite{Anderson:2017aux} which have been recently analyzed for the divisor topologies in \cite{Carta:2022web}. It would be interesting to explore about the generality of such EFFV configurations using the pCICY dataset, which we plan in a future work.

Let us mention that apart from the models ${\cal M}_{2,1}$ and ${\cal M}_{2,6}$, there is another model, namely ${\cal M}_{2, 25}$, which has the following GV dependent quantities,
\begin{equation}
\label{eq:model25}
n_1 = 252\coma n_2 = 252\coma n_{11} = -\frac{18441}{2}\coma n_{12} = 216080\coma n_{22}=\frac{519}{2}\fstop
\end{equation}
Notice that the leading order GV invariant dependent quantities $n_1$ and $n_2$ are equal, however they differ in $n_{ij}$ at the sub-leading order making this case different from the previous two models. In fact, for this model, there are 121 PFFV configurations which satisfy $M^1 = M^2$. However, it turns out that none of these satisfy $F^1 = - F^2$ and so do not satisfy the condition \eqref{eq:cond-EFFVs-2}. Therefore, there are no exponentially flat vacua in this model.

\subsection{A new class of PFFV via ${\cal S}$-duality}
Let us mention that the under the two $\SL(2,{\mathbb Z})$ transformations, the ${\cal S}$-dual pair of $(F_3, H_3)$ fluxes transform in the following manner:
\begin{equation}
    \begin{array}{rclcccl}
        S & \to & \displaystyle{- \frac{1}{S}} & \implies & (F_3, H_3) &\to &(-H_3, F_3)\coma\\
        S & \to & S+1 & \implies & (F_3, H_3) &\to &(F_3 - H_3, H_3)\fstop
    \end{array}
\end{equation}
Note that the second transformation is simply shifting the universal axion $C_0$ by a unit, i.e. $C_0 \to C_0 + 1$, and hence it does not play a crucial role at the level of scalar potential while the first one can be thought of as ``strong-weak duality" given that it changes $g_s \to g_s^{-1}$ for $\langle C_0 \rangle = 0$. It also exchanges the $F_3$ fluxes into $H_3$ fluxes and vice versa. Now we take the following simplification for the choice of fluxes in the superpotential given in \cref{eq:WgenIIB},
\begin{equation}
H_0 = p_i H^i\coma H_i = p_{ij} H^j\coma H^0 = 0\coma F_0 = 0\coma F^i = 0\coma F^0 = 0\fstop
\end{equation}
which implies that the shifted fluxes in \cref{eq:IIB-W-fluxshift} take the following form,
\begin{equation}
\ov{H}_0 = 0\coma\ov{H}_i = 0\coma \ov{F}_0 = 0\coma \ov{F}_i = F_i\fstop
\end{equation}
Subsequently, the generic superpotential \eqref{eq:WgenIIB} gets simplified into the following form,
\begin{equation}
    W_{\rm poly}= \frac{1}{\sqrt{2}}\left[U^i \, {F}_i - \,\frac{1}{2} \, S\, \kappa_{ijk}^0  U^i U^j\, H^k\right]\coma
\end{equation}
while the derivatives of the perturbative piece of the superpotential are simplified as below,
\begin{equation}
    \begin{split}
        \frac{\partial W_{\rm poly}}{\partial U^i} &= \frac{1}{\sqrt{2}}\left[{F}_i  - \, S \, \kappa_{ijk}^0 \, U^j \, H^k \right]\coma\\
        \frac{\partial W_{\rm poly}}{\partial S} &= - \, \frac{1}{\sqrt{2}} \left[\frac{1}{2} \, \kappa_{ijk}^0  U^i U^j \, H^k \right]\fstop
    \end{split}
\end{equation}
Subsequently, we have a flat valley $U^i = \frac{1}{S}\, \tilde{M}^i$ for some flux vector $\tilde{M}^i$ satisfying the following constraints,
\begin{equation}
\label{eq:Sdual-PFFV}
    F_i\, \tilde{M}^i = 0 = \kappa_{ijk}^0 \,\tilde{M}^i \, \tilde{M}^j\, H^k\coma \kappa_{ijk}^0 \,\tilde{M}^j\, H^k = F_i\coma
\end{equation}
which leads to flat vacua such that superpotential and the derivatives trivially vanish. Moreover, these constraints suggest the following generic form for the flux vector $\tilde{M}^i$,
\begin{equation}
\tilde{M}^i = (\kappa_{ijk}^0 \, H^k)^{-1}\, F_j\coma
\end{equation}
which has to lie within the K\"ahler cone. Notice that this flat valley is along $U^i = M^i/S$ and therefore weak-coupling and large complex structure limits look incompatible, however depending on the flux vector $M^i$, it might be possible to have $\langle g_s \rangle \ll 1$ and $u^i \gg 1$ for $i \in \{1, 2\}$. We leave the analysis of such ${\cal S}$-dual PFFV configurations for a future work.


\section{Conclusions}
\label{sec_conclusion}

In this article, we have performed a detailed analysis regarding the dynamical realization of the low values for the GVW flux superpotential $|W_0|$, which has been at the center of the type IIB orientifold models such as KKLT. 

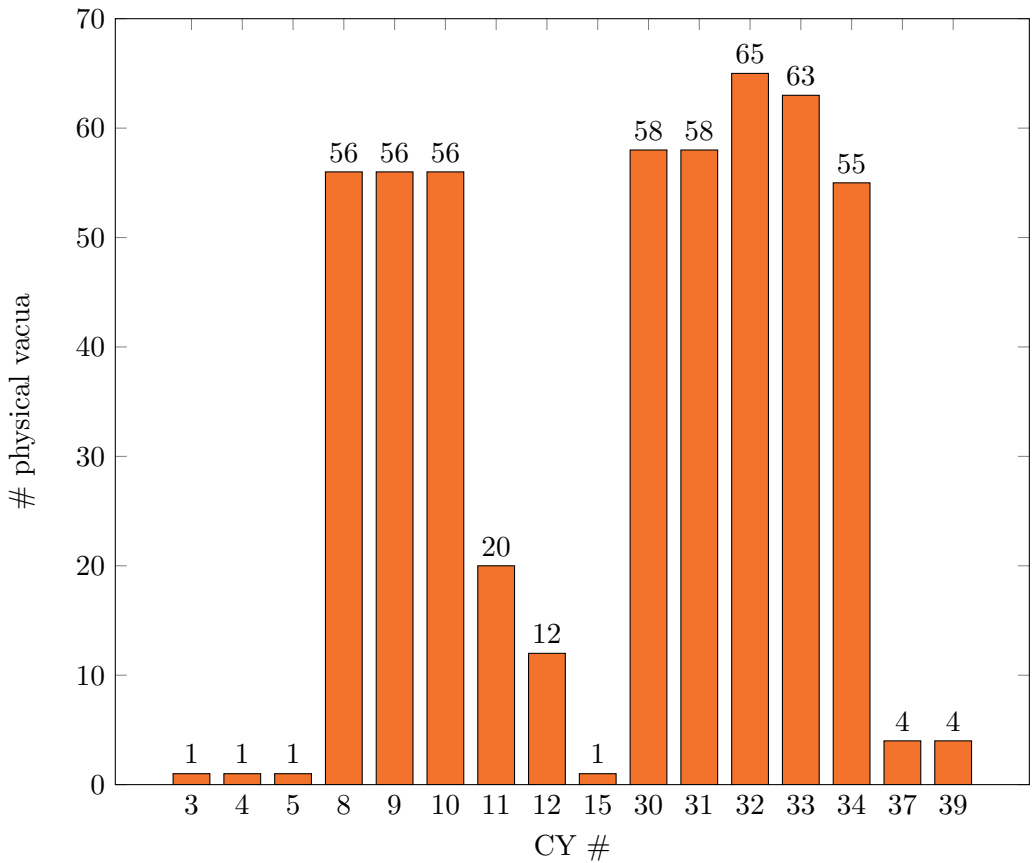
\begin{figure}[H]
 \centering
      \pgfplotstableread{
Label Nf  topper
3 1 0
4 1  0
5 1 0
8 56 0
9 56 0
10 56 0
11 20 0
12 12 0
15 1 0
30 58 0
31 58  0
32 65 0
33 63 0
34 55 0
37 4 0
39 4 0
    }\testdata
         \begin{tikzpicture}[scale=1]
    \begin{axis}[
       width=0.9\textwidth,
        ybar stacked,
        ymin=0,
        ymax=70,
        xtick={data},
        legend style={cells={anchor=west}, legend pos=north east},
        reverse legend=false, 
        xticklabels from table={\testdata}{Label},
        xticklabel style={text width=2cm,align=center},
      bar width=0.032\textwidth,
      xlabel={CY \#},
		ylabel={\# physical vacua},
    ]
    \addplot [fill=col2,ybar,nodes near coords,point meta=y]
            table [y=Nf, meta=Label, x expr=\coordindex]
                {\testdata};
    \end{axis}
    \end{tikzpicture}
\caption{PFFV configurations with physical vacua for all the 16 CY geometries.}
\label{fig:PFFVphysvacua}
\end{figure}

We, first, re-derived the recipe of constructing the perturbatively flat flux vacua proposed in \cite{Demirtas:2019sip}. This process has also led us to find the new class of perturbatively flat valley along $U^i = S^{-1} M^i$ motivated by the ${\cal S}$-duality arguments. However, for these vacua a different set of conditions (as given in \cref{eq:Sdual-PFFV}) is required. Considering the 39 CY geometries with $h^{1,1} =2$ and relevant topological data from \cite{Altman:2014bfa} we have created an input of all the necessary quantities such as $\kappa_{ijk}^0$, $\tilde{p}_{ij}$, $\tilde{p}_i$, conditions $M^i H_i = 0$ along with tadpole equation for $N_{\rm flux}$ for all the geometries and subsequently have imposed the generic flux configurations to test for the PFFV. Let us briefly summarize the journey of scanning the flat vacua in the following point:

\begin{itemize}

\item 
We have checked 27000000 flux configurations for each of the 39 CY geometries which results in a total of 31109 PFFV configurations. These are presented in Table \ref{tab_fluxdata}.

\item
We classify these flat vacua based on the CY geometry being of Swiss-cheese type, K3-fibered type and the so-called ``Hybrid" type and observe that most of the PFFV configurations arise from the K3-fibered CY geometries.

\item
Moreover, we observed that there are some redundancies in counting of the inequivalent PFFV due to the presence of overall flux scalings, and we remove these overcounting by imposing \cref{eq:gcd1} leading to a new set of number of PFFV presented in Table \ref{tab_reducedfluxdata1}.

\item
In our subsequent analysis we have observed that there are 5753 PFFV configurations which satisfy a condition $M^1 = M^2$ on the $M^i$ flux vector, and these are not suitable for giving physical vacua using non-perturbative prepotential contributions. After taking care of such vacua, a new set of number of PFFV is presented in Table \ref{tab_reducedfluxdata2}.

\item
Finally we find that there are only 16 CY geometries out of 39 which have PFFV configurations resulting in physical vacua after including the non-perturbative effects in the prepotential. 

\item
The number of physical vacua is presented in Figure \ref{fig:PFFVphysvacua} which shows 10 peaks corresponding to the K3-fibered CY geometries while the other 6 examples falling in the Swiss-cheese and the ``Hybrid" category have quite less number of physical solutions.

\item
Figure \ref{fig:PFFVphysvacua} also shows that for K3-fibered CY geometries, the physical vacua are quite good in number, and so, one may argue that there is the possibility of ``tuning" fluxes to get a ``suitable" value of $W_0$ depending on the requirement in the model. 

\end{itemize}

\noindent
We have observed that there are some PFFV configurations which are protected against the non-perturbative prepotential contributions, and we call them as exponentially flat flux vacua (EFFV). The underlying reason for the existence of such vacua is the presence of some symmetries in their CY geometry. Moreover, we find that the CY geometry in which such vacua appear is a bi-cubic in ${\mathbb P}^2 \times {\mathbb P}^2$ which appears in the CICY database as well \cite{Anderson:2017aux}. Motivated by this observation and the existence of such symmetries quite frequently as seen in \cite{Carta:2022web} we expect that such EFFV will be more naturally found in abundance in the CICY geometries, and we plan to present a systematic analysis of those in the near future.

Given the fact that the percentage of Swiss-cheese geometries decreases \cite{Cicoli:2018tcq,Cicoli:2021dhg} while the percentage of the K3-fibered geometries increases with increasing the $h^{1,1}$ of the CY in the Kreuzer-Skarke database \cite{Cicoli:aaaaaaa}, our observations about finding more PFFV configurations for K3-fibered geometries are quite encouraging. In fact, it opens a new window of opportunities for using a lot of CY geometries which initially appeared to be of less use for model builders, specially within LVS framework. In this regard, an immediate extension of our work for the $h^{1,1}(\tilde{X}) \geq 3$ cases, which can have K3-fibered CY geometries of Swiss-cheese type, is necessary, and we hope to get back to this in the near future \cite{Carta:2202ccccc}.


\acknowledgments

F.C. is supported by STFC consolidated grant ST/T000708/1. A. M. is supported in part by Deutsche Forschungsgemeinschaft under Germany's Excellence Strategy EXC 2121 Quantum Universe 390833306. P. S. is thankful to Paolo Creminelli, Atish Dabholkar and Fernando Quevedo for their support.


\appendix

\section{Relevant data for CY geometries with $h^{1,1} = 2$}
\label{AppE}

In this appendix, we collect in \cref{tab_cydata-h11eq2,tab_cydata-h11eq2-2} the relevant data for the CY geometries that have been used in the main text. We denote the column $c_2(J)$ for $\int_{\rm CY}c_2 \wedge J_i$, where $J_i$'s correspond to the K\"ahler cone generators. Here ``Swiss-cheese" means CY examples which have one `diagonal' del Pezzo divisor to support a Swiss-cheese like structure, ``K3-fibered" means CY examples which are K3-fibered, and the remaining ones are denoted as ``Hard" following the nomenclature of \cite{AbdusSalam:2020ywo}.

\begin{center}
\begin{longtable}{|c||c|c|c|c|c|c|c|}
\caption{Topological data of the 39 CY geometries corresponding to $h^{1,1}=2$.}\\
\hline
\# & Type & $-\chi$ & $\kappa^0_{111}$ & $\kappa^0_{112}$ & $\kappa^0_{122}$ & $\kappa^0_{222}$ & $c_2(J)$ \\
\hhline{|=#=|=|=|=|=|=|=|}
\endhead
\label{tab_cydata-h11eq2}${\cal M}_{2,1}$ & Hybrid & 54  & 0    & 1    & 1 & 0     & $\{12, 12\}$  \\
${\cal M}_{2,2}$ & Swiss-cheese & 72   & 9    & 3    & 1 & 0     & $\{30, 12\}$ \\
${\cal M}_{2,3}$ & Swiss-cheese & 144 & 2    & 3    & 3 & 3     & $\{32, 42\}$  \\
${\cal M}_{2,4}$ & Swiss-cheese & 144 & 3    & 3    & 3  & 2    & $\{42, 32\}$  \\
${\cal M}_{2,5}$ & Swiss-cheese & 144 & 3    & 3    & 3  & 2    & $\{42, 32\}$  \\
${\cal M}_{2,6}$ & Hybrid & 162 & 0    & 3    & 3 & 0    & $\{36, 36\}$  \\
${\cal M}_{2,7}$ & Swiss-cheese & 164 & 3    & 5    & 5 & 5     & $\{42, 50\}$   \\
${\cal M}_{2,8}$ & K3-fibered & 168 & 8    & 4   & 0 & 0      & $\{56, 24\}$  \\
${\cal M}_{2,9}$ & K3-fibered & 168 & 8    & 4   & 0 & 0       & $\{56, 24\}$   \\
${\cal M}_{2,10}$ & K3-fibered & 168 & 0    & 0    & 4 & 8  & $\{24, 56\}$   \\
${\cal M}_{2,11}$ & K3-fibered & 168 & 0    & 0    & 4 & 2     & $\{24, 44\}$   \\
${\cal M}_{2,12}$ & K3-fibered & 168 & 0    & 0    & 4 & 5     & $\{24, 50\}$    \\
${\cal M}_{2,13}$ & Swiss-cheese & 168 & 3    & 3    & 3 & 1     & $\{42, 34\}$  \\
${\cal M}_{2,14}$ & Hybrid & 168 & 0 & 3    & 7 & 11    & $\{36, 62\}$   \\
${\cal M}_{2,15}$ & Hybrid & 168 &0 & 3 & 5 & 5     & $\{36, 50\}$  \\
${\cal M}_{2,16}$ & Hybrid & 168 & 3 & 6    & 10 & 14    & $\{42, 68\}$   \\
${\cal M}_{2,17}$ & Hybrid & 168 & 6    & 9 & 13 & 17     & $\{48, 74\}$  \\
${\cal M}_{2,18}$ & Swiss-cheese & 176 & 2    & 5    & 5 & 5     & $\{44, 50\}$   \\
${\cal M}_{2,19}$ & Swiss-cheese & 180 & 0    & 3    & 3 & 3     & $\{36, 42\}$   \\
${\cal M}_{2,20}$ & Hard & 186 & 0    & 3 & 7 & 14    & $\{36, 68\}$  \\
${\cal M}_{2,21}$ & Swiss-cheese & 200 & 2    & 6    & 12 & 24    & $\{44, 84\}$   \\
${\cal M}_{2,22}$ & Swiss-cheese & 208 & 1    & 4    & 12 & 36    & $\{34, 96\}$   \\
${\cal M}_{2,23}$ & Swiss-cheese & 208 & 1    & 4    & 12 & 36    & $\{34, 96\}$  \\
${\cal M}_{2,24}$ & Swiss-cheese & 208 & 1    & 4    & 12 & 36    & $\{34, 96\}$   \\
${\cal M}_{2,25}$ & Swiss-cheese & 228 & 0    & 1    & 1 & 1     & $\{24, 34\}$   \\
${\cal M}_{2,26}$ & Swiss-cheese & 236 & 1    & 2    & 2 & 2     & $\{34, 44\}$   \\
${\cal M}_{2,27}$ & Swiss-cheese & 240 & 2    & 7    & 21 & 63    & $\{44, 126\}$   \\
${\cal M}_{2,28}$ & Swiss-cheese & 240 & 2    & 7    & 21 & 63    & $\{44, 126\}$   \\
${\cal M}_{2,29}$ & Swiss-cheese & 240 & 2    & 7    & 21 & 63    & $\{44, 126\}$  \\
${\cal M}_{2,30}$ & K3-fibered & 252  & 4    & 2    & 0 & 0      & $\{52, 24\}$   \\
${\cal M}_{2,31}$ & K3-fibered & 252  & 4    & 2    & 0 & 0      & $\{52, 24\}$   \\
${\cal M}_{2,32}$ & K3-fibered & 252 & 0    & 0    & 2 & 4 &  $\{24, 52\}$  \\
${\cal M}_{2,33}$ & K3-fibered & 252 & 0    & 0    & 2 & 0       & $\{24, 36\}$   \\
${\cal M}_{2,34}$ & K3-fibered & 252 & 0 & 0    & 2 & 2       & $\{24, 44\}$   \\
${\cal M}_{2,35}$ & Swiss-cheese & 252 & 1    & 2    & 4 & 6     & $\{34, 60\}$   \\
${\cal M}_{2,36}$ & Swiss-cheese & 260 & 0    & 2    & 2 & 2     & $\{36, 44\}$  \\
${\cal M}_{2,37}$ & Swiss-cheese & 260 & 0    & 1    & 3 & 9     & $\{24, 66\}$  \\
${\cal M}_{2,38}$ & Swiss-cheese & 284 & 0    & 2    & 4 & 8     & $\{36, 68\}$   \\
${\cal M}_{2,39}$ & Swiss-cheese & 540 & 9    & 3    & 1 & 0    & $\{102, 36\}$ \\
\hline
\end{longtable}
\end{center}

\begin{center}
\renewcommand{\arraystretch}{1.2}
\begin{longtable}{|c||c|c|c|c|c|}
\caption{The GV dependent quantities $n_i$ and $n_{ij}$ for all the 39 CY geometries which appear in the instanton prepotential ${\cal F}_{\rm inst}$ given in \cref{eq:F-inst-nis}.}\\
 \hline
CY \# & $n_1$  & $n_2$  & $n_{11}$           & $n_{12}$ & $n_{22}$           \\
 \hhline{|=#=|=|=|=|=|}
 \endhead
 \label{tab_cydata-h11eq2-2}$1$ & $63   $& $63   $& $\frac{567}{8}    $&$ 2754   $&$ \frac{567}{8}    $\\
$2$ & $54   $& $3    $& $\frac{243}{4}    $&$  -108   $&$ -\frac{45}{8}    $\\
$3$ & $252  $& $6    $& $-\frac{18441}{2} $&$ 7524   $&$ \frac{3}{4}      $\\
$4$ & $6    $& $252  $& $\frac{3}{4}      $&$  7524   $&$ -\frac{18441}{2} $\\
$5$ & $6    $& $252  $& $\frac{3}{4}      $&$ 7524   $&$ -\frac{18441}{2} $\\
$6$ & $189  $& $189  $& $\frac{1701}{8}   $&$  8262   $&$ \frac{1701}{8}   $\\
$7$ & $56   $& $20   $& $-265             $&$ 2635   $&$ \frac{5}{2}      $\\
$8$ & $640  $& $4    $& $10112            $&$  640    $&$ \frac{1}{2}      $\\
$9$ & $640  $& $4    $& $10112            $&$ 640    $&$ \frac{1}{2}      $\\
$10$ & $4    $& $640  $& $\frac{1}{2}     $&$  640    $&$ 10112            $\\
$11$ & $64   $& $640  $& $8                $&$ 6912   $&$ 10112            $\\
$12$ & $16   $& $640  $& $2                $&$  2144   $&$ 10112            $\\
$13$ & $60   $& $56   $& $\frac{15}{2}     $&$ 7164   $&$ -265             $\\
$14$ & $1    $& $186  $& $\frac{1}{8}      $&$  640    $&$ \frac{837}{4}    $\\
$15$ & $18   $& $186  $& $\frac{1}{4}      $&$ 2439   $&$ \frac{837}{4}    $\\
$16$ & $1    $& $20   $& $\frac{1}{8}      $&$  640    $&$ \frac{1}{2}      $\\
$17$ & $1    $& $-2   $& $\frac{1}{8}      $&$ 640    $&$ -\frac{1}{4}     $\\
$18$ & $27   $& $60   $& $-\frac{405}{8}   $&$  2515   $&$ \frac{15}{2}     $\\
$19$ & $27   $& $180  $& $-\frac{405}{8}   $&$ 6804   $&$ \frac{405}{2}    $\\
$20$ & $-2   $& $177  $& $-\frac{1}{4}     $&$  178    $&$ \frac{1593}{8}   $\\
$21$ & $-4   $& $48   $& $-\frac{9}{2}     $&$ 96     $&$ 6                $\\
$22$ & $3    $& $40   $& $-\frac{45}{8}    $&$  -80    $&$ 5                $\\
$23$ & $3    $& $40   $& $-\frac{45}{8}    $&$ -80    $&$ 5                $\\
$24$ & $3    $& $40   $& $-\frac{45}{8}    $&$  -80    $&$ 5                $\\
$25$ & $252  $& $252  $& $-\frac{18441}{2} $&$ 216080 $&$ \frac{519}{2}    $\\
$26$ & $252  $& $28   $& $-\frac{18441}{2} $&$  27824  $&$ \frac{3}{2}      $\\
$27$ & $3    $& $28   $& $-\frac{45}{8}    $&$ -56    $&$ \frac{7}{2}      $\\
$28$ & $3    $& $28   $& $-\frac{45}{8}    $&$  -56    $&$ \frac{7}{2}      $\\
$29$ & $3    $& $28   $& $-\frac{45}{8}    $&$ -56    $&$ \frac{7}{2}      $\\
$30$ & $2496 $& $2    $& $224064           $&$  2496   $&$ \frac{1}{4}      $\\
$31$ & $2496 $& $2    $& $224064           $&$ 2496   $&$ \frac{1}{4}      $\\
$32$ & $2    $& $2496 $& $\frac{1}{4}     $&$  2496   $&$ 224064           $\\
$33$ & $288  $& $2496 $& $288              $&$ 216576 $&$ 224064           $\\
$34$ & $24   $& $2496 $& $1                $&$  24000  $&$ 224064           $\\
$35$ & $1    $& $56   $& $\frac{1}{8}      $ & $ 2496 $  & $-265          $   \\
$36$ & $56   $& $280  $& $-265             $&$  26144  $&$ 295              $\\
$37$ & $3    $& $220  $& $-\frac{45}{8}    $&$ -440   $&$ \frac{575}{2}    $\\
$38$ & $-4   $& $256  $& $-\frac{9}{2}     $&$  512    $&$ 316              $\\
$39$ & $540  $& $3    $& $\frac{1215}{2}   $&$  -1080  $&$ -\frac{45}{8}    $\\ 
 \hline
\end{longtable}
\end{center}
\noindent


\section{Computation of Gopakumar-Vafa invariants}
\label{sec:GVinvarcomputation}

In this appendix, we recall briefly facts about genus $0$ Gopakumar-Vafa (GV) invariants, and the r\^ole they play in string compactifications. Let us consider the topological string A-model, with target space given by a compact CY $3$-fold $\tilde{X}$. Different worldsheet topologies, labeled by their genus $g$, contribute to the topological string partition function differently. In particular, the total free energy for the A-model with target space $\tilde{X}$ can be written as a power series in the string coupling $g_s$ as 
\begin{equation}
    \mathcal{F}_{A}=\sum_{g=0}^\infty \mathcal{F}_g(t)g_s^{2-2g}\coma
\end{equation} 
where $t$ is a collective notation for the complexified K\"ahler moduli. 

Focusing on the case of $g=0$, we have that $\mathcal{F}_g(t)$ takes the following form:
\begin{equation}
    \mathcal{F}_0(t)= -\frac{1}{3!}\kappa^0_{ijk}t^it^jt^k + t^j \int_X c_2(\tilde{X}) \wedge J_j+ \frac{\chi}{2}\zeta(3)+ \frac{1}{(2\pi i)^3} \sum_{d_i} A_{d_i}e^{2\pi i d_i t^i}\fstop
    \label{eq:Fzero_pf}
\end{equation}
In \eqref{eq:Fzero_pf}, $\kappa^0_{ijk}$ are the (classical, geometric) triple intersection numbers of $\tilde{X}$, $c_2(\tilde{X})$ the second Chern class, $\chi$ is the Euler characteristic of $\tilde{X}$, $\zeta(3)$ is the Riemann zeta function evaluated at $3$, $d_i$ are charges labelling homology classes of curves in $H^2(\tilde{X},\mathbb{Z})$ and $A_{d_i}$ are rational numbers called \emph{Gromov-Witten invariants}. 

The last summation in \eqref{eq:Fzero_pf} encodes the non-perturbative part of the free energy of the A-model. Its interpretation is clearer from the dual point of view of M-theory compactified on $\tilde{X}$. In this case, the worldsheet corrections can be encoded as
\begin{equation}
    \mathcal{F}_{\text{inst}}(t^i)=\frac{1}{(2\pi i)^3}\sum_{\beta \in H_2(\tilde{X},\ZZ)} n_\beta \text{Li}_3 \left( q^\beta\right)\coma
\end{equation}
where 
\begin{equation}
    \text{Li}_3 \left( x\right)=\sum_{m=1}^\infty \frac{x^m}{m^3}\coma q^{\beta}= e^{2\pi i d_i t^i}\fstop
\end{equation}
The integer numbers $n_\beta$ are called \emph{Gopakumar-Vafa invariants}. From the M-theory perspective, they count the number of holomorphic curves of genus $g$ within a given homology class $\beta$.\footnote{They can also be interpreted as the number of BPS states obtained by M2-branes wrapping such holomorphic curves (see e.g. \cite{Katz:1999xq}).} 

Being our Calabi-Yau $\tilde{X}$ compact, the computation of the genus zero GV invariants can be performed explicitly using the technique explained in \cite{Hosono:1993qy,Hosono:1994ax} (see also \cite{Alim:2021vhs}). We briefly review such prescription now.\footnote{We briefly mention that, in case $X$ was not compact, the GV invariants could as well be easily computed by the topological vertex method, e.g. \cite{Aganagic:2002qg,Aganagic:2003db} or by a recent approach of \cite{Collinucci:2021ofd,Collinucci:2021wty}} For simplicity of exposition, we focus on CY $3$-folds $\tilde{X}$ which are complete intersections of $k$ polynomial equations a in product of $s$ projective spaces. This family of CY are usually specified by a configuration matrix 
\begin{equation}
	\left[
	\begin{tabular}{c|cccc}
		$\PP^{n_1}$ &   $q_1^1$ & $\cdots$  & $q_k^1 $ \\
		$\PP^{n_2}$  &   $q_1^2$ & $\cdots$  &$ q_k^2$  \\
		$\vdots$ &   $\vdots$ & $\ddots$ & $\vdots$  \\ 
		$\PP^{n_{\tilde{h}^{1,1}}} $&   $q_1^{\tilde{h}^{1,1}} $& $\cdots$ & $q_k^{\tilde{h}^{1,1}} $
	\end{tabular}
	\right] \fstop
\end{equation}
We stress that it is actually possible to extend the prescription for the computation of the genus zero GV invariants to CICYs in an ambient space given by products of weighted projective spaces \cite{Hosono:1994ax,Alim:2021vhs}, or even cases of a single anticanonical hypersurface in a generic toric Fano ambient space \cite{Hosono:1993qy}. Nevertheless, since we just aim to give some illustrative description of the algorithm presented in \cite{Hosono:1993qy,Hosono:1994ax}, we will review it here only in its most simple instance: that of projective CICYs.\footnote{See also \cite{Carta:2021sms} for a more detailed description.} In general, the algorithm needs the classical triple intersection numbers of the CY and the Mori cone generators of its mirror, as we will see in the following paragraphs.

To compute the GV invariants, one needs to compute the quantum corrected triple intersection numbers $\kappa_{ijk}$, which are related to the prepotential $\mathcal{F}$ as
\begin{equation}
\kappa_{ijk}(t) = \de_{t^i}\de_{t^j}\de_{t^k}\mathcal{F}_0(t)\fstop
\label{eq:quantumKwithF}
\end{equation}
By mirror symmetry the $t^i$ correspond to complex structure moduli space of the mirror CY $X$ (see \cite{Hosono:1994av,Hori:2003ic} and references therein.).

The generators of the Mori cone of $X$ are vectors $l^{(i)}$, given by
\begin{equation}
	l^{(i)}=\left(-q_1^{(i)},\ldots -q_{k}^{(i)}\,;\, \ldots, 0,1,\ldots, 1,0,\ldots\right)\equiv\left(\left\{l_{0j}^{(i)}\right\}\,;\,\left\{l_r^{(i)}\right\}\right)\coma
\end{equation}
where $i=1,\ldots, h^{2,1}$; $j=1,\ldots, k$, and the number of $1$'s in $\left\{l_r^{(i)}\right\}$ is equal to $n_i+1$ at a position corresponding to the $\PP^{n_i}$ that has been considered. 

The period vector $\Pi(z)$ for $X$ is a vector with $2h^{2,1}+2$ components. The \emph{fundamental period} is given by
\begin{equation}
	w_0(z)=\sum_{n_1\geq 0}\ldots \sum_{n_{h^{2,1}}\geq 0}c(n) \prod_{i=1}^{h^{2,1}}z_i^{n_i}\coma
\end{equation}
where\footnote{$n!=\Gamma(n+1)$ is the Euler's Gamma function.}
\begin{equation}
	c(n)=\dfrac{\displaystyle\prod_j\Gamma\left(1-\sum_{s=1}^{h^{2,1}}l_{0j}^{(s)}n_s\right)}{\displaystyle\prod_{i}\Gamma\left(1+\sum_{s=1}^{h^{2,1}}l_i^{(s)}n_s\right)}\fstop
\end{equation}

The full period vector $\Pi(z)$ can be defined as~\cite{Hosono:1993qy,Hosono:1994ax}
\begin{equation}
	\renewcommand*{\arraystretch}{2}
	\Pi(z)=\left(\begin{array}{*3{>{\displaystyle}c}p{5cm}} w_0(z) \\ \left.\pderr{\rho_i}w_0(z,\rho)\right|_{\rho=0}\\ \left.\frac{1}{2}\kappa^0_{ijk}\pderr{\rho_j}\pderr{\rho_k}w_0(z,\rho)\right|_{\rho=0}\\
		\left.-\frac{1}{6}\kappa^0_{ijk}\pderr{\rho_i}\pderr{\rho_j}\pderr{\rho_k}w_0(z,\rho)\right|_{\rho=0}
	\end{array}\right)\coma
\end{equation}
where $\kappa_{ijk}^0$ are the classical triple intersection numbers of $X$ and 
\begin{equation}
	w_0(z,\rho)=\sum_{n_1\geq 0}\ldots \sum_{n_{h^{2,1}}\geq 0}c(n+\rho) \prod_{i=1}^{h_{2,1}}z_i^{n_i+\rho_i}\fstop
	\label{eq:genfunperiod}
\end{equation}

We proceed defining the mirror map to be 
\begin{equation}
	t^i(z)=\frac{w_i(z)}{w_0(z)}\coma
	\label{eq:tinwz}
\end{equation}
where
\begin{equation}
	w_i(z)=\sum_{n_1\geq 0}\ldots \sum_{n_{h^{2,1}}\geq 0}\left.\frac{1}{2\pi i}\pderr{\rho_i}c(n+\rho)\right|_{\rho=0}\prod_{i=1}^{h^{2,1}}z_i^{n_i}+w_0(z)\frac{\ln z_i}{2\pi i}\coma
\end{equation}
so that quantum-corrected triple intersection numbers $\kappa_{ijk}$ can be expressed as
\begin{equation}
	\kappa_{ijk}(t)=\de_{t^i}\de_{t^j}\dfrac{\displaystyle \left.\frac{1}{2}\kappa^0_{kab}\pderr{\rho_a}\pderr{\rho_b}w_0(z,\rho)\right|_{\rho=0}}{w_0(z)}(t)\coma
	\label{eq:kappacomp}
\end{equation}
where it is clear that the fraction is computed first as function of the c.s. moduli $z_i$, then, one substitutes the inverse of Eq.~\eqref{eq:tinwz}, and takes the last two derivatives with respect to the K\"ahler moduli $t^i$.

The extraction of the GV is obtained by comparison of \cref{eq:quantumKwithF,eq:kappacomp}. The algorithm, schematically reviewed above, was coded in the Mathematica notebook \texttt{INSTANTON}~\cite{Klemm:2001aaa}.


\section{GV invariants for CY 3-folds with $h^{1,1}=2$}

 We compute the Gopakumar-Vafa invariants for all these 39 CY geometries, which can be directly used for computing the non-perturbative pieces of the prepotential (${\cal F}_{\text{inst}}$) using Eq. \eqref{eq:Finst}.

\begin{table}[H]
	\centering
	\begin{subtable}{\textwidth}
		\centering
		{\footnotesize{
				\begin{tabular}{|c||c|c|c|c|c|}
					\hline
					\diagbox{$d_1$}{$d_2$} &$0$&$1$&$2$&$3$&$4$\\
					\hhline{|=#=|=|=|=|=|}
					$0$&$0$&$63$&$63$&$54$&$63$\\
					$1$&$63$&$2754$&$47628$&$497430$&$3791691$\\
					$2$&$63$&$47628$&$4369464$&$172317699$&$409644224$\\
					$3$&$54$&$497430$&$172317699$&$18654101550$&$1051549169670$\\
					$4$&$63$&$3791691$&$4096442241$&$1051549169670$&$122327286921828$\\
					\hline
				\end{tabular}
		}}
		\caption{CY 1.}
	\end{subtable}\\
	\begin{subtable}{\textwidth}
		\centering
		{\footnotesize{
				\begin{tabular}{|c||c|c|c|c|c|}
					\hline
					\diagbox{$d_1$}{$d_2$} &$0$&$1$&$2$&$3$&$4$\\
					\hhline{|=#=|=|=|=|=|}
					$0$&$0$&$3$&$-6$&$27$&$-192$\\
					$1$&$54$&$-108$&$270$&$-1728$&$15444$\\
					$2$&$54$&$1215$&$-4968$&$45927$&$-539190$\\
					$3$&$72$&$43767$&$41364$&$-641988$&$10612620$\\
					$4$&$54$&$482895$&$-1080432$&$7104780$&$-141250284$\\
					\hline
				\end{tabular}
		}}
		\caption{CY 2.}
	\end{subtable}\\
	\begin{subtable}{\textwidth}
		\centering
		{\footnotesize{
				\begin{tabular}{|c||c|c|c|c|c|}
					\hline
					\diagbox{$d_1$}{$d_2$} &$0$&$1$&$2$&$3$&$4$\\
					\hhline{|=#=|=|=|=|=|}
					$0$&$0$&$6$&$0$&$0$&$0$\\
					$1$&$252$&$7524$&$7524$&$252$&$0$\\
					$2$&$-9252$&$30780$&$5549652$&$16761816$&$5549652$\\
					$3$&$848628$&$-4042560$&$45622680$&$10810105020$&$56089743576$\\
					$4$&$-114265008$&$691458930$&$-6771588480$&$107939555010$&$31014597012048$\\
					\hline
				\end{tabular}
		}}
		\caption{CY 3.}
	\end{subtable}\\
	\begin{subtable}{\textwidth}
		\centering
		{\footnotesize{
				\begin{tabular}{|c||c|c|c|c|c|}
					\hline
					\diagbox{$d_1$}{$d_2$} &$0$&$1$&$2$&$3$&$4$\\
					\hhline{|=#=|=|=|=|=|}
					$0$&$0$&$252$&$-9252$&$848628$&$-114265008$\\
					$1$&$6$&$7524$&$30780$&$-4042560$&$691458930$\\
					$2$&$0$&$7524$&$5549652$&$45622680$&$-6771588480$\\
					$3$&$0$&$252$&$16761816$&$10810105020$&$107939555010$\\
					$4$&$0$&$0$&$5549652$&$56089743576$&$31014597012048$\\
					\hline
				\end{tabular}
		}}
		\caption{CY 4.}
	\end{subtable}
\end{table}
\begin{table}[H]
	\ContinuedFloat
	\begin{subtable}{\textwidth}
		\centering
		{\footnotesize{
				\begin{tabular}{|c||c|c|c|c|c|}
					\hline
					\diagbox{$d_1$}{$d_2$} &$0$&$1$&$2$&$3$&$4$\\
					\hhline{|=#=|=|=|=|=|}
					$0$&$0$&$252$&$-9252$&$848628$&$-114265008$\\
					$1$&$6$&$7524$&$30780$&$-4042560$&$691458930$\\
					$2$&$0$&$7524$&$5549652$&$45622680$&$-6771588480$\\
					$3$&$0$&$252$&$16761816$&$10810105020$&$107939555010$\\
					$4$&$0$&$0$&$5549652$&$56089743576$&$31014597012048$\\
					\hline
				\end{tabular}
		}}
		\caption{CY 5.}
	\end{subtable}\\
	\begin{subtable}{\textwidth}
		\centering
		{\footnotesize{
				\begin{tabular}{|c||c|c|c|c|c|}
					\hline
					\diagbox{$d_1$}{$d_2$} &$0$&$1$&$2$&$3$&$4$\\
					\hhline{|=#=|=|=|=|=|}
					$0$&$0$&$189$&$189$&$162$&$189$\\
					$1$&$189$&$8262$&$142884$&$1492290$&$11375073$\\
					$2$&$189$&$142884$&$13108392$&$516953097$&$12289326723$\\
					$3$&$162$&$1492290$&$516953097$&$55962304650$&$3154647509010$\\
					$4$&$189$&$11375073$&$12289326723$&$3154647509010$&$366981860765484$\\
					\hline
				\end{tabular}
		}}
		\caption{CY 6.}
	\end{subtable}\\
	\begin{subtable}{\textwidth}
		\centering
		{\footnotesize{
				\begin{tabular}{|c||c|c|c|c|c|}
					\hline
					\diagbox{$d_1$}{$d_2$} &$0$&$1$&$2$&$3$&$4$\\
					\hhline{|=#=|=|=|=|=|}
					$0$&$0$&$20$&$0$&$0$&$0$\\
					$1$&$56$&$2635$&$5040$&$190$&$-40$\\
					$2$&$-272$&$2760$&$541930$&$2973660$&$2454600$\\
					$3$&$3240$&$-45440$&$933760$&$277421695$&$2644224240$\\
					$4$&$-58432$&$1001340$&$-18770880$&$563282580$&$208000930200$\\
					\hline
				\end{tabular}
		}}
		\caption{CY 7.}
	\end{subtable}\\
	\begin{subtable}{\textwidth}
		\centering
		{\footnotesize{
				\begin{tabular}{|c||c|c|c|c|c|}
					\hline
					\diagbox{$d_1$}{$d_2$} &$0$&$1$&$2$&$3$&$4$\\
					\hhline{|=#=|=|=|=|=|}
					$0$&$0$&$4$&$0$&$0$&$0$\\
					$1$&$640$&$640$&$0$&$0$&$0$\\
					$2$&$10032$&$72224$&$10032$&$0$&$0$\\
					$3$&$288384$&$7539200$&$7539200$&$288384$&$0$\\
					$4$&$10979984$&$757561520$&$2346819520$&$757561520$&$10979984$\\
					\hline
				\end{tabular}
		}}
		\caption{CY 8.}
	\end{subtable}\\
	\begin{subtable}{\textwidth}
		\centering
		{\footnotesize{
				\begin{tabular}{|c||c|c|c|c|c|}
					\hline
					\diagbox{$d_1$}{$d_2$} &$0$&$1$&$2$&$3$&$4$\\
					\hhline{|=#=|=|=|=|=|}
					$0$&$0$&$4$&$0$&$0$&$0$\\
					$1$&$640$&$640$&$0$&$0$&$0$\\
					$2$&$10032$&$72224$&$10032$&$0$&$0$\\
					$3$&$288384$&$7539200$&$7539200$&$288384$&$0$\\
					$4$&$10979984$&$757561520$&$2346819520$&$757561520$&$10979984$\\
					\hline
				\end{tabular}
		}}
		\caption{CY 9.}
	\end{subtable}\\
	\begin{subtable}{\textwidth}
		\centering
		{\footnotesize{
				\begin{tabular}{|c||c|c|c|c|c|}
					\hline
					\diagbox{$d_1$}{$d_2$} &$0$&$1$&$2$&$3$&$4$\\
					\hhline{|=#=|=|=|=|=|}
					$0$&$0$&$640$&$10032$&$288384$&$10979984$\\
					$1$&$4$&$640$&$72224$&$7539200$&$757561520$\\
					$2$&$0$&$0$&$10032$&$7539200$&$2346819520$\\
					$3$&$0$&$0$&$0$&$288384$&$757561520$\\
					$4$&$0$&$0$&$0$&$0$&$10979984$\\
					\hline
				\end{tabular}
		}}
		\caption{CY 10.}
	\end{subtable}
\end{table}
\begin{table}[H]
	\ContinuedFloat
	\begin{subtable}{\textwidth}
		\centering
		{\footnotesize{
				\begin{tabular}{|c||c|c|c|c|c|}
					\hline
					\diagbox{$d_1$}{$d_2$} &$0$&$1$&$2$&$3$&$4$\\
					\hhline{|=#=|=|=|=|=|}
					$0$&$0$&$640$&$10032$&$288384$&$10979984$\\
					$1$&$64$&$6912$&$742784$&$75933184$&$7518494784$\\
					$2$&$0$&$14400$&$8271360$&$2445747712$&$532817161216$\\
					$3$&$0$&$6912$&$31344000$&$26556152064$&$12305418469184$\\
					$4$&$0$&$640$&$48098560$&$130867460608$&$130700405114112$\\
					\hline
				\end{tabular}
		}}
		\caption{CY 11.}
	\end{subtable}\\
	\begin{subtable}{\textwidth}
		\centering
		{\footnotesize{
				\begin{tabular}{|c||c|c|c|c|c|}
					\hline
					\diagbox{$d_1$}{$d_2$} &$0$&$1$&$2$&$3$&$4$\\
					\hhline{|=#=|=|=|=|=|}
					$0$&$0$&$640$&$10032$&$288384$&$10979984$\\
					$1$&$16$&$2144$&$231888$&$23953120$&$2388434784$\\
					$2$&$0$&$120$&$356368$&$144785584$&$36512550816$\\
					$3$&$0$&$-32$&$14608$&$144051072$&$115675981232$\\
					$4$&$0$&$3$&$-4920$&$5273880$&$85456640608$\\
					\hline
				\end{tabular}
		}}
		\caption{CY 12.}
	\end{subtable}\\
	\begin{subtable}{\textwidth}
		\centering
		{\footnotesize{
				\begin{tabular}{|c||c|c|c|c|c|}
					\hline
					\diagbox{$d_1$}{$d_2$} &$0$&$1$&$2$&$3$&$4$\\
					\hhline{|=#=|=|=|=|=|}
					$0$&$0$&$56$&$-272$&$3240$&$-58432$\\
					$1$&$60$&$7164$&$8280$&$-136320$&$3004020$\\
					$2$&$0$&$55800$&$5295132$&$9912720$&$-194443200$\\
					$3$&$0$&$105160$&$129026480$&$10230627060$&$22327525780$\\
					$4$&$0$&$55800$&$948373920$&$441508016160$&$29055345533568$\\
					\hline
				\end{tabular}
		}}
		\caption{CY 13.}
	\end{subtable}\\
	\begin{subtable}{\textwidth}
		\centering
		{\footnotesize{
				\begin{tabular}{|c||c|c|c|c|c|}
					\hline
					\diagbox{$d_1$}{$d_2$} &$0$&$1$&$2$&$3$&$4$\\
					\hhline{|=#=|=|=|=|=|}
					$0$&$0$&$186$&$186$&$168$&$186$\\
					$1$&$1$&$640$&$22295$&$324012$&$3017570$\\
					$2$&$0$&$0$&$10032$&$2367222$&$138486166$\\
					$3$&$0$&$0$&$0$&$288384$&$239874519$\\
					$4$&$0$&$0$&$0$&$0$&$10979984$\\
					\hline
				\end{tabular}
		}}
		\caption{CY 14.}
	\end{subtable}\\
	\begin{subtable}{\textwidth}
		\centering
		{\footnotesize{
				\begin{tabular}{|c||c|c|c|c|c|}
					\hline
					\diagbox{$d_1$}{$d_2$} &$0$&$1$&$2$&$3$&$4$\\
					\hhline{|=#=|=|=|=|=|}
					$0$&$0$&$186$&$186$&$168$&$186$\\
					$1$&$18$&$2439$&$58032$&$709281$&$5931612$\\
					$2$&$-2$&$442$&$480662$&$39212788$&$1402803900$\\
					$3$&$0$&$-512$&$109488$&$232354413$&$35519961168$\\
					$4$&$0$&$768$&$-119142$&$60915700$&$165368736700$\\
					\hline
				\end{tabular}
		}}
		\caption{CY 15.}
	\end{subtable}\\
	\begin{subtable}{\textwidth}
		\centering
		{\footnotesize{
				\begin{tabular}{|c||c|c|c|c|c|}
					\hline
					\diagbox{$d_1$}{$d_2$} &$0$&$1$&$2$&$3$&$4$\\
					\hhline{|=#=|=|=|=|=|}
					$0$&$0$&$20$&$-2$&$0$&$0$\\
					$1$&$1$&$640$&$6474$&$1152$&$-1023$\\
					$2$&$0$&$0$&$10032$&$733560$&$4458084$\\
					$3$&$0$&$0$&$0$&$288384$&$75445435$\\
					$4$&$0$&$0$&$0$&$0$&$10979984$\\
					\hline
				\end{tabular}
		}}
		\caption{CY 16.}
	\end{subtable}
\end{table}
\begin{table}[H]\ContinuedFloat
	\centering
	\begin{subtable}{\textwidth}
		\centering
		{\footnotesize{
				\begin{tabular}{|c||c|c|c|c|c|}
					\hline
					\diagbox{$d_1$}{$d_2$} &$0$&$1$&$2$&$3$&$4$\\
					\hhline{|=#=|=|=|=|=|}
					$0$&$0$&$-2$&$0$&$0$&$0$\\
					$1$&$1$&$640$&$641$&$4$&$5$\\
					$2$&$0$&$0$&$10032$&$208126$&$8734$\\
					$3$&$0$&$0$&$0$&$288384$&$23177356$\\
					$4$&$0$&$0$&$0$&$0$&$10979984$\\
					\hline
				\end{tabular}
		}}
		\caption{CY 17.}
	\end{subtable}\\
	\begin{subtable}{\textwidth}
		\centering
		{\footnotesize{
				\begin{tabular}{|c||c|c|c|c|c|}
					\hline
					\diagbox{$d_1$}{$d_2$} &$0$&$1$&$2$&$3$&$4$\\
					\hhline{|=#=|=|=|=|=|}
					$0$&$0$&$60$&$0$&$0$&$0$\\
					$1$&$27$&$2515$&$12210$&$12210$&$2515$\\
					$2$&$-54$&$1620$&$525370$&$7643520$&$31493340$\\
					$3$&$243$&$-9840$&$478410$&$266348155$&$6864059040$\\
					$4$&$-1728$&$82620$&$-3726120$&$282482940$&$198108021600$\\
					\hline
				\end{tabular}
		}}
		\caption{CY 18.}
	\end{subtable}\\
	\begin{subtable}{\textwidth}
		\centering
		{\footnotesize{
				\begin{tabular}{|c||c|c|c|c|c|}
					\hline
					\diagbox{$d_1$}{$d_2$} &$0$&$1$&$2$&$3$&$4$\\
					\hhline{|=#=|=|=|=|=|}
					$0$&$0$&$180$&$180$&$180$&$180$\\
					$1$&$27$&$6804$&$138510$&$1478520$&$11337165$\\
					$2$&$-54$&$4860$&$5103972$&$336441600$&$9824455980$\\
					$3$&$243$&$-29520$&$5034150$&$9738919980$&$1154124236340$\\
					$4$&$-1728$&$247860$&$-37757340$&$11148032160$&$27397662877008$\\
					\hline
				\end{tabular}
		}}
		\caption{CY 19.}
	\end{subtable}\\
	\begin{subtable}{\textwidth}
		\centering
		{\footnotesize{
				\begin{tabular}{|c||c|c|c|c|c|}
					\hline
					\diagbox{$d_1$}{$d_2$} &$0$&$1$&$2$&$3$&$4$\\
					\hhline{|=#=|=|=|=|=|}
					$0$&$0$&$177$&$177$&$186$&$177$\\
					$1$&$-2$&$178$&$20291$&$317172$&$2998628$\\
					$2$&$0$&$3$&$-177$&$332040$&$73458379$\\
					$3$&$0$&$5$&$-708$&$44790$&$794368$\\
					$4$&$0$&$7$&$-1068$&$75225$&$-4468169$\\
					\hline
				\end{tabular}
		}}
		\caption{CY 20.}
	\end{subtable}\\
	\begin{subtable}{\textwidth}
		\centering
		{\footnotesize{
				\begin{tabular}{|c||c|c|c|c|c|}
					\hline
					\diagbox{$d_1$}{$d_2$} &$0$&$1$&$2$&$3$&$4$\\
					\hhline{|=#=|=|=|=|=|}
					$0$&$0$&$48$&$0$&$0$&$0$\\
					$1$&$-4$&$96$&$6756$&$15808$&$6756$\\
					$2$&$-4$&$144$&$-2256$&$48912$&$5296692$\\
					$3$&$-12$&$480$&$-9024$&$151200$&$-2228328$\\
					$4$&$-48$&$2352$&$-54432$&$947680$&$-13844832$\\
					\hline
				\end{tabular}
		}}
		\caption{CY 21.}
	\end{subtable}\\
	\begin{subtable}{\textwidth}
		\centering
		{\footnotesize{
				\begin{tabular}{|c||c|c|c|c|c|}
					\hline
					\diagbox{$d_1$}{$d_2$} &$0$&$1$&$2$&$3$&$4$\\
					\hhline{|=#=|=|=|=|=|}
					$0$&$0$&$40$&$0$&$0$&$0$\\
					$1$&$3$&$-80$&$780$&$54192$&$121410$\\
					$2$&$-6$&$200$&$-3120$&$29640$&$-425600$\\
					$3$&$27$&$-1280$&$27580$&$-365040$&$3953900$\\
					$4$&$-192$&$11440$&$-316400$&$5457240$&$-70079040$\\
					\hline
				\end{tabular}
		}}
		\caption{CY 22.}
	\end{subtable}
\end{table}
\begin{table}[H]
	\ContinuedFloat
		\begin{subtable}{\textwidth}
		\centering
		{\footnotesize{
				\begin{tabular}{|c||c|c|c|c|c|}
					\hline
					\diagbox{$d_1$}{$d_2$} &$0$&$1$&$2$&$3$&$4$\\
					\hhline{|=#=|=|=|=|=|}
					$0$&$0$&$40$&$0$&$0$&$0$\\
					$1$&$3$&$-80$&$780$&$54192$&$121410$\\
					$2$&$-6$&$200$&$-3120$&$29640$&$-425600$\\
					$3$&$27$&$-1280$&$27580$&$-365040$&$3953900$\\
					$4$&$-192$&$11440$&$-316400$&$5457240$&$-70079040$\\
					\hline
				\end{tabular}
		}}
		\caption{CY 23.}
	\end{subtable}\\
	\begin{subtable}{\textwidth}
		\centering
		{\footnotesize{
				\begin{tabular}{|c||c|c|c|c|c|}
					\hline
					\diagbox{$d_1$}{$d_2$} &$0$&$1$&$2$&$3$&$4$\\
					\hhline{|=#=|=|=|=|=|}
					$0$&$0$&$40$&$0$&$0$&$0$\\
					$1$&$3$&$-80$&$780$&$54192$&$121410$\\
					$2$&$-6$&$200$&$-3120$&$29640$&$-425600$\\
					$3$&$27$&$-1280$&$27580$&$-365040$&$3953900$\\
					$4$&$-192$&$11440$&$-316400$&$5457240$&$-70079040$\\
					\hline
				\end{tabular}
		}}
		\caption{CY 24.}
	\end{subtable}\\
	\begin{subtable}{\textwidth}
		\centering
		{\footnotesize{
				\begin{tabular}{|c||c|c|c|c|c|}
					\hline
					\diagbox{$d_1$}{$d_2$} &$0$&$1$&$2$&$3$&$4$\\
					\hhline{|=#=|=|=|=|=|}
					$0$&$0$&$252$&$228$&$252$&$228$\\
					$1$&$252$&$216080$&$19024344$&$692373600$&$14948021796$\\
					$2$&$-9252$&$1292760$&$11018901520$&$3184161458112$&$365266217922840$\\
					$3$&$848628$&$-169787520$&$111419296080$&$1505294251736400$&$780554765497523112$\\
					$4$&$-114265008$&$29041275060$&$-16093658577240$&$17751139250586480$&$303947661674939596160$\\
					\hline
				\end{tabular}
		}}
		\caption{CY 25.}
	\end{subtable}\\
	\begin{subtable}{\textwidth}
		\centering
		{\footnotesize{
				\begin{tabular}{|c||c|c|c|c|c|}
					\hline
					\diagbox{$d_1$}{$d_2$} &$0$&$1$&$2$&$3$&$4$\\
					\hhline{|=#=|=|=|=|=|}
					$0$&$0$&$28$&$-2$&$0$&$0$\\
					$1$&$252$&$27824$&$173000$&$36016$&$-16132$\\
					$2$&$-9252$&$143640$&$117759512$&$2195372472$&$7532433400$\\
					$3$&$848628$&$-18865280$&$1060457520$&$1282755656240$&$41865486354648$\\
					$4$&$-114265008$&$3226808340$&$-157440192340$&$13673441420900$&$20578685883966672$\\
					\hline
				\end{tabular}
		}}
		\caption{CY 26.}
	\end{subtable}\\
	\begin{subtable}{\textwidth}
		\centering
		{\footnotesize{
				\begin{tabular}{|c||c|c|c|c|c|}
					\hline
					\diagbox{$d_1$}{$d_2$} &$0$&$1$&$2$&$3$&$4$\\
					\hhline{|=#=|=|=|=|=|}
					$0$&$0$&$28$&$0$&$0$&$0$\\
					$1$&$3$&$-56$&$378$&$14427$&$14427$\\
					$2$&$-6$&$140$&$-1512$&$9828$&$-69804$\\
					$3$&$27$&$-896$&$13426$&$-122472$&$837900$\\
					$4$&$-192$&$8008$&$-154280$&$1841868$&$-15811488$\\
					\hline
				\end{tabular}
		}}
		\caption{CY 27.}
	\end{subtable}\\
	\begin{subtable}{\textwidth}
		\centering
		{\footnotesize{
				\begin{tabular}{|c||c|c|c|c|c|}
					\hline
					\diagbox{$d_1$}{$d_2$} &$0$&$1$&$2$&$3$&$4$\\
					\hhline{|=#=|=|=|=|=|}
					$0$&$0$&$28$&$0$&$0$&$0$\\
					$1$&$3$&$-56$&$378$&$14427$&$14427$\\
					$2$&$-6$&$140$&$-1512$&$9828$&$-69804$\\
					$3$&$27$&$-896$&$13426$&$-122472$&$837900$\\
					$4$&$-192$&$8008$&$-154280$&$1841868$&$-15811488$\\
					\hline
				\end{tabular}
		}}
		\caption{CY 28.}
	\end{subtable}
\end{table}
\begin{table}[H]
	\ContinuedFloat
	\begin{subtable}{\textwidth}
		\centering
		{\footnotesize{
				\begin{tabular}{|c||c|c|c|c|c|}
					\hline
					\diagbox{$d_1$}{$d_2$} &$0$&$1$&$2$&$3$&$4$\\
					\hhline{|=#=|=|=|=|=|}
					$0$&$0$&$28$&$0$&$0$&$0$\\
					$1$&$3$&$-56$&$378$&$14427$&$14427$\\
					$2$&$-6$&$140$&$-1512$&$9828$&$-69804$\\
					$3$&$27$&$-896$&$13426$&$-122472$&$837900$\\
					$4$&$-192$&$8008$&$-154280$&$1841868$&$-15811488$\\
					\hline
				\end{tabular}
		}}
		\caption{CY 29.}
	\end{subtable}\\
	\begin{subtable}{\textwidth}
		\centering
		{\footnotesize{
				\begin{tabular}{|c||c|c|c|c|c|}
					\hline
					\diagbox{$d_1$}{$d_2$} &$0$&$1$&$2$&$3$&$4$\\
					\hhline{|=#=|=|=|=|=|}
					$0$&$0$&$2$&$0$&$0$&$0$\\
					$1$&$2496$&$2496$&$0$&$0$&$0$\\
					$2$&$223752$&$1941264$&$223752$&$0$&$0$\\
					$3$&$38637504$&$1327392512$&$1327392512$&$38637504$&$0$\\
					$4$&$9100224984$&$861202986072$&$2859010142112$&$861202986072$&$9100224984$\\
					\hline
				\end{tabular}
		}}
		\caption{CY 30.}
	\end{subtable}\\
	\begin{subtable}{\textwidth}
		\centering
		{\footnotesize{
				\begin{tabular}{|c||c|c|c|c|c|}
					\hline
					\diagbox{$d_1$}{$d_2$} &$0$&$1$&$2$&$3$&$4$\\
					\hhline{|=#=|=|=|=|=|}
					$0$&$0$&$2$&$0$&$0$&$0$\\
					$1$&$2496$&$2496$&$0$&$0$&$0$\\
					$2$&$223752$&$1941264$&$223752$&$0$&$0$\\
					$3$&$38637504$&$1327392512$&$1327392512$&$38637504$&$0$\\
					$4$&$9100224984$&$861202986072$&$2859010142112$&$861202986072$&$9100224984$\\
					\hline
				\end{tabular}
		}}
		\caption{CY 31.}
	\end{subtable}\\
	\begin{subtable}{\textwidth}
		\centering
		{\footnotesize{
				\begin{tabular}{|c||c|c|c|c|c|}
					\hline
					\diagbox{$d_1$}{$d_2$} &$0$&$1$&$2$&$3$&$4$\\
					\hhline{|=#=|=|=|=|=|}
					$0$&$0$&$2496$&$223752$&$38637504$&$9100224984$\\
					$1$&$2$&$2496$&$1941264$&$1327392512$&$861202986072$\\
					$2$&$0$&$0$&$223752$&$1327392512$&$2859010142112$\\
					$3$&$0$&$0$&$0$&$38637504$&$861202986072$\\
					$4$&$0$&$0$&$0$&$0$&$9100224984$\\
					\hline
				\end{tabular}
		}}
		\caption{CY 32.}
	\end{subtable}\\
	\begin{subtable}{\textwidth}
		\centering
		{\footnotesize{
				\begin{tabular}{|c||c|c|c|c|c|}
					\hline
					\diagbox{$d_1$}{$d_2$} &$0$&$1$&$2$&$3$&$4$\\
					\hhline{|=#=|=|=|=|=|}
					$0$&$0$&$2496$&$223752$&$38637504$&$9100224984$\\
					$1$&$288$&$216576$&$152031744$&$100021045248$&$63330228232704$\\
					$2$&$252$&$6391296$&$19638646848$&$34832157566976$&$47042083144050624$\\
					$3$&$288$&$104994816$&$1180450842624$&$4962537351009792$&$13025847457256417280$\\
					$4$&$252$&$1209337344$&$43199009739072$&$401057938191181824$&$1937385878589532624320$\\
					\hline
				\end{tabular}
		}}
		\caption{CY 33.}
	\end{subtable}\\
	\begin{subtable}{\textwidth}
		\centering
		{\footnotesize{
				\begin{tabular}{|c||c|c|c|c|c|}
					\hline
					\diagbox{$d_1$}{$d_2$} &$0$&$1$&$2$&$3$&$4$\\
					\hhline{|=#=|=|=|=|=|}
					$0$&$0$&$2496$&$223752$&$38637504$&$9100224984$\\
					$1$&$24$&$24000$&$17244192$&$11552340480$&$7397501182992$\\
					$2$&$-2$&$4544$&$92555600$&$239152764672$&$387426434941992$\\
					$3$&$0$&$-4096$&$22945824$&$911674812096$&$4444162191527440$\\
					$4$&$0$&$6144$&$-12818168$&$283693717248$&$13387851546648736$\\
					\hline
				\end{tabular}
		}}
		\caption{CY 34.}
	\end{subtable}
\end{table}
\begin{table}[H]
	\ContinuedFloat
		\begin{subtable}{\textwidth}
		\centering
		{\footnotesize{
				\begin{tabular}{|c||c|c|c|c|c|}
					\hline
					\diagbox{$d_1$}{$d_2$} &$0$&$1$&$2$&$3$&$4$\\
					\hhline{|=#=|=|=|=|=|}
					$0$&$0$&$56$&$-272$&$3240$&$-58432$\\
					$1$&$1$&$2496$&$201380$&$339904$&$-5551734$\\
					$2$&$0$&$0$&$223752$&$149435776$&$10344027040$\\
					$3$&$0$&$0$&$0$&$38637504$&$99139408274$\\
					$4$&$0$&$0$&$0$&$0$&$9100224984$\\
					\hline
				\end{tabular}
		}}
		\caption{CY 35.}
	\end{subtable}\\
	\begin{subtable}{\textwidth}
		\centering
		{\footnotesize{
				\begin{tabular}{|c||c|c|c|c|c|}
					\hline
					\diagbox{$d_1$}{$d_2$} &$0$&$1$&$2$&$3$&$4$\\
					\hhline{|=#=|=|=|=|=|}
					$0$&$0$&$280$&$260$&$280$&$260$\\
					$1$&$56$&$26144$&$1285680$&$27317440$&$369165320$\\
					$2$&$-272$&$38640$&$112034912$&$17397477760$&$1117302570720$\\
					$3$&$3240$&$-636160$&$225004960$&$1204150282400$&$335794977815760$\\
					$4$&$-58432$&$14018760$&$-4378961880$&$2806670899040$&$19093623519550272$\\
					\hline
				\end{tabular}
		}}
		\caption{CY 36.}
	\end{subtable}\\
	\begin{subtable}{\textwidth}
		\centering
		{\footnotesize{
				\begin{tabular}{|c||c|c|c|c|c|}
					\hline
					\diagbox{$d_1$}{$d_2$} &$0$&$1$&$2$&$3$&$4$\\
					\hhline{|=#=|=|=|=|=|}
					$0$&$0$&$220$&$260$&$220$&$260$\\
					$1$&$3$&$-440$&$23050$&$15075856$&$650382435$\\
					$2$&$-6$&$1100$&$-92720$&$4571160$&$-1442208140$\\
					$3$&$27$&$-7040$&$822850$&$-57853400$&$5991277900$\\
					$4$&$-192$&$62920$&$-9471700$&$876629160$&$-76811833000$\\
					\hline
				\end{tabular}
		}}
		\caption{CY 37.}
	\end{subtable}\\
	\begin{subtable}{\textwidth}
		\centering
		{\footnotesize{
				\begin{tabular}{|c||c|c|c|c|c|}
					\hline
					\diagbox{$d_1$}{$d_2$} &$0$&$1$&$2$&$3$&$4$\\
					\hhline{|=#=|=|=|=|=|}
					$0$&$0$&$256$&$284$&$256$&$284$\\
					$1$&$-4$&$512$&$199696$&$6311936$&$104656856$\\
					$2$&$-4$&$768$&$-62440$&$15098880$&$11018163216$\\
					$3$&$-12$&$2560$&$-252032$&$33786368$&$-2740916912$\\
					$4$&$-48$&$12544$&$-1526508$&$177694720$&$-15447666984$\\
					\hline
				\end{tabular}
		}}
		\caption{CY 38.}
	\end{subtable}\\
	\begin{subtable}{\textwidth}
		\centering
		{\footnotesize{
				\begin{tabular}{|c||c|c|c|c|c|}
					\hline
					\diagbox{$d_1$}{$d_2$} &$0$&$1$&$2$&$3$&$4$\\
					\hhline{|=#=|=|=|=|=|}
					$0$&$0$&$3$&$-6$&$27$&$-192$\\
					$1$&$540$&$-1080$&$2700$&$-17280$&$154440$\\
					$2$&$540$&$143370$&$-574560$&$5051970$&$-57879900$\\
					$3$&$540$&$204071184$&$74810520$&$-913383000$&$13593850920$\\
					$4$&$540$&$21772947555$&$-49933059660$&$224108858700$&$-2953943334360$\\
					\hline
				\end{tabular}
		}}
		\caption{CY 39.}
	\end{subtable}
	\caption{GV invariants for the CY with $h^{1,1}=2$.}
	\label{tab:GVh11=2}
\end{table}


\newpage

\bibliographystyle{JHEP}
\bibliography{reference}

\end{document}